\documentclass[twocolumn]{aastex62}

\usepackage[bottom]{footmisc}
\usepackage{amsmath}
\usepackage{todo}
\received{February, 2019}
\revised{--}
\accepted{--}
\submitjournal{ApJ}

\newcommand{\cca}{Center for Computational Astrophysics, Flatiron Institute, Simons Foundation, 162 Fifth Avenue, New York, NY 10010, USA}

\shorttitle{The structure of the MW--LMC DM halo}
\shortauthors{Garavito-Camargo et al.}

\begin{document}

\title{Quantifying the impact of the Large Magellanic Cloud on the structure of the Milky Way's dark matter halo using Basis Function Expansions}

\correspondingauthor{Nicol\'as Garavito-Camargo}
\email{jngaravitoc@email.arizona.edu}

\author[0000-0001-7107-1744]{Nicol\'as Garavito-Camargo}
\affil{Steward Observatory, University of Arizona, 933 North Cherry Avenue,Tucson, AZ 85721, USA.}

\author[0000-0003-0715-2173]{Gurtina Besla}
\affiliation{Steward Observatory, University of Arizona, 933 North Cherry Avenue,Tucson, AZ 85721, USA.}

\author[0000-0003-3922-7336]{Chervin F.P Laporte}
\altaffiliation{CITA National Fellow}
\affiliation{Kavli Institute for the Physics and Mathematics of the Universe (WPI), The University of Tokyo Institutes for Advanced Study (UTIAS), The University of Tokyo, Chiba 277-8583, Japan}
\affiliation{Department of Physics and Astronomy, University of Victoria, 3800 Finnerty Road, Victoria, B.C., V8P 4HN, Canada}

\author[0000-0003-0872-7098]{Adrian~M.~Price-Whelan}
\affiliation{\cca}


\author[0000-0002-6993-0826]{Emily C. Cunningham} 
\affiliation{\cca}

\author[0000-0001-6244-6727]{Kathryn V. Johnston}
\affiliation{Department of Astronomy, Columbia University, New York, NY 10027, USA.}
\affiliation{\cca}

\author{Martin Weinberg}
\affiliation{Department of Astronomy, University of Massachusetts, Amherst, MA 01003-9305, USA}

\author{Facundo A. G\'omez}
\affiliation{Instituto de Investigaci\'on Multidisciplinar en Ciencia y Tecnolog\'ia, Universidad de La Serena, Ra\'ul Bitr\'an 1305, La Serena, Chile.}
\affiliation{Departamento de F\'isica y Astronom\'ia, Universidad de La Serena, Av. Juan Cisternas 1200 N, La Serena, Chile.}




\begin{abstract}

Indications of disequilibrium  
throughout the Milky Way (MW) highlight the need for compact,
flexible, non-parametric descriptions of phase--space distributions
of galaxies. We present a new representation of the 
current Dark Matter (DM) distribution and potential derived from N-body simulations of the 
MW and Large Magellanic Cloud (LMC) system using Basis Function Expansions (BFEs).
We incorporate methods to maximize the physical signal
in the representation.
As a result, the simulations of 10$^8$ DM particles representing the distorted MW(MW+LMC) system can be described by $\sim$236(2067) coefficients.
We find that the LMC induces asymmetric perturbations (odd $l,m$)
to the MW's halo, which are inconsistent with oblate, prolate, or triaxial halos.
Furthermore, the energy in high-order even modes ($l,m>2$) is similar to average triaxial halos found in cosmological simulations. As such, the response of the MW's halo to the LMC must be accounted for in order to recover the imprints of its assembly history.   

The LMC causes the outer halo ($>30$ kpc) to shift from the disk center of mass (COM) by $\sim15-25$ kpc at present day, manifesting as a dipole in the BFE and in the radial velocities of halo stars. The shift depends on the LMC's infall mass, the distortion of the LMC's halo and the MW halo response.
Within 30 kpc, halo tracers are expected to orbit the COM of the MW's disk, regardless of LMC infall mass. 

The LMC's halo is also distorted by MW tides, we discuss the implications for its mass loss and the subsequent effects on current Magellanic satellites.

\end{abstract}


\section{Introduction}

Current and upcoming surveys (e.g. SDSS, PS-1, Gaia, H3, S5, HSC surveys, DESI, WEAVE, LSST, 4MOST, PFS, etc) will soon map the Milky Way's (MW's) stellar 
halo with unprecedented depth and precision.

We are thus poised to utilize the phase space properties of the stellar halo and substructures 
within it to make significant progress on two key problems that the fields of near-field Cosmology and Galactic Archaeology have been building towards over the last few decades: 1) constraining the distribution of dark matter (DM) throughout our Galaxy \citep[e.g.][]{Johnston99, Springel08, Bonaca14, Sanderson15, Malhan19, Reino20}; and 2) reconstructing the cosmic assembly history of the Galaxy \citep[e.g.][]{Johnston96,Johnston98,Helmi99,Helmi01, Ibata01, Bullock05,Wang11,Vera-Ciro13,Deason16, Helmi20, Naidu20, Naidu21}.

The same surveys that allow unprecedented scales in terms of numbers of stars and volume over which the MW has been
mapped, also reveal exquisite details of our Galaxy that defy simple descriptions. In particular, they challenge some
fundamental assumptions that prior interpretive work has typically made (e.g. that the Galaxy is in equilibrium and has
a smooth phase-space distribution), highlighting the limitations of the theoretical and analytical tools currently at
our disposal.

In particular, it is now clear that in order to constrain the DM distribution of the MW, we must account for
the recent infall of the Large Magellanic Cloud (LMC, $< 2$ Gyr ago, \cite{Besla07, Kallivayalil13}).

The expected infall mass of the LMC is $\sim 10^{11}$ M$_\odot$ \citep{Besla10,Besla12, Besla15, Boylan-Kolchin11, Patel17,Penarrubia16, Erkal18b}. Corresponding to roughly to $10\%$ to $20\%$ of the total mass of the MW. As such,
the LMC will affect: the structure of the MW's DM halo \citep[][ this paper]{Weinberg98a, Laporte18a,garavito-camargo19a, tamfal20}, the kinematics of the stellar halo \citep[][]{garavito-camargo19a, petersen20,Cunningham20, Petersen21, Erkal20c}, the orbital dynamics of satellites and globular 
clusters \citep[e.g:][]{Patel20,Erkal20b, Garrow20},
the dynamics of stellar streams \citep{Vera-Ciro13, Gomez15, Erkal18a, Koposov18, Erkal18b, Shipp19}, the structure of the MW's disk \citep{Weinberg98b, Laporte18, Laporte18a}, 
 the kinematics of hypervelocity stars \citep{Boubert17, Kenyon18, Erkal19a, Boubert20}, and even the dynamics of the solar neighborhood \citep{Besla19,Gardner20, Hinkel20}. 

In this paper, we present a framework in which to
understand and analyze the present-day gravitational potential and DM density distribution of the
combined MW and LMC (MW--LMC) system, using Basis Function Expansions (BFEs). We focus on the current state of the MW--LMC system in order to explain both the framework and to 
present new predictions for the present day structure of the DM halo density and gravitational 
potential. In subsequent papers, we will present the time-evolution of the MW--LMC models, 
enabling the community to model the dynamics of substructure 
in a time evolving MW--LMC potential.
The simulations utilized in this study are those presented in \cite[][hereafter G19]{garavito-camargo19a}, which account for the tidal field of the LMC, the deformation and mass loss of the LMC's halo since infall, 
and the MW halo’s response to the LMC's passage, including the formation of a
DM dynamical friction wake trailing the LMC.

One promising way to constrain the structure and potential of the MW's DM halo is by modelling the orbits of stellar streams. At radii larger than 20 kpc, the majority of estimates for the shape of the MW’s DM halo come from modeling the Sagittarius (Sgr) Stream \citep{Ibata01, Helmi04, Johnston05, Law10, Deg13, Vera-Ciro13, Dierickx17, Vasiliev20}. \cite{Law10} present a best-fit halo model that is triaxial with the major axis perpendicular to the disk.
They also explored the impact of a static LMC on the Sgr. Stream, but limited their studies to a maximal LMC mass of $6\times10^{10}$M$_\odot$. Building on this work, \cite{Vera-Ciro13} argued that the shape of the MW's halo is changing as a function of radius, as expected in
cosmological simulations. They advocated for a static model in which the inner MW's halo shape is spherical and becomes prolate in the outer regions, with a major axis towards the LMC, which they model as a massive infalling halo ($10^{11}$M$_\odot$). Indeed, \cite{Gomez15} have illustrated that the orbit of the Sgr. dSph is affected by the LMC, which will change the phase space properties of the Sgr. Stream \citep{Laporte18}. Recently, \cite{Vasiliev20} showed that the inclusion of the LMC changes the dynamics of the Sgr. dSph and Stream due to both direct torques and the response of the MW halo. 

These studies of Sgr. indicate that stellar streams probe asymmetries in the MW's DM halo owing to the LMC {\it in
addition to} structure arising from the cosmological assembly history of the MW 
\citep[e.g:][]{Wechsler02, Ludlow13,Correa15, Wang20, Vera-Ciro11, Prada19, Drakos19}
or changes in halo shape expected from different DM particle models
\citep[e.g:][]{Yoshida20, Peter13, Bose16}. Without a framework to quantify the LMC's impact on the halo, we cannot
disentangle these processes. In this study we will quantify the primary effects of the orbit of the LMC through the DM
halo of the MW, the dynamical friction wake and the collective response. 
We also account for the distortions in the LMC's DM halo owing to the tidal field of the MW. Some DM particles will no
longer be bound to the LMC, forming a distorted distribution of LMC DM debris. Note that the majority of studies of the
impact of the LMC on the kinematics of substructure in the stellar halo have not accounted for the mass loss or
distortions of the LMC halo \citep{Vera-Ciro13, Gomez15, Erkal18b, Patel20, petersen20}. We will explicitly quantify
the importance of the LMC debris to the structure of the halo.

In this study we will utilize the 8 high-resolution simulations of the MW--LMC system presented in G19: 4 LMC models, ranging from 8-25 $\times 10^{10}$ M$_\odot$, and 2 MW models with different anisotropy profiles (isotropic and radially biased). We then analytically describe the complex DM distribution of the combined system using BFEs.  

BFEs have been used in the literature to solve Poisson’s equation \citep[see the review by][]{Sellwood97} 
and perform N-body simulations \citep[e.g][]{Hernquist92,Johnston95,Bullock05,Choi09, petersen20}. Given their 
high accuracy in describing asymmetric DM halos, BFEs have been used to 
characterize potentials and density fields from cosmological simulations \citep{Lowing11, Sanders20}. 
These analytic, reconstructed potentials have also been used to simulate the evolution of substructure, such 
as stellar streams, in complex DM halos \citep[e.g][]{Lowing11, Dai18}, the dynamics of satellite galaxies
\citep{Choi09, Sanders20}, and bars in disks \citep[e.g][]{Petersen19-disk}. 

Typically, BFEs can involve, of order, $10^3-10^4$ terms. 
We identify and remove terms that represent noise based on the methodology outlined in \cite{Weinberg96}, 
thus reducing the expansion by factors of 10-100. Our goal is to use the remaining coefficients (which
actually contain physical information) to quantify the halo response of the MW to the passage of the LMC and 
understand the magnitude of this perturbation versus standard idealized prolate, oblate or triaxial halos.

This paper is organized as follows: in section \ref{sec:sims_review} we summarize the 
computational methods. In section \ref{sec:mwlmc_bfe}, we present the BFE for the MW and LMC at
the present day, including the unbound DM debris from the LMC. In section \ref{sec:power} we discuss
how the coefficients in a BFE can provide intuition about the MW's halo shape. We then compare our 
results to that of oblate, prolate and triaxial halos in section \ref{sec:power_shape}. We discuss 
the resulting shape of the MW's DM halo in section~\ref{sec:mw_com}. We show how our results scale 
as a function of LMC mass and the anisotropy profile of the MW's halo in section~\ref{sec:lmc_mass}.
We discuss how BFEs can be applied to other areas of astrophysics in section~\ref{sec:bfe_applications}.
We present our conclusions in section \ref{sec:conclusions}.

\section{Computational Methods:}

N--body simulations are a powerful computational tool to 
study how galaxies form and evolve \citep[e.g][]{White78, Barnes92}. In the case of the MW, many cosmological 
N--body simulations have shown how a MW-like galaxy
grows across cosmic time \citep[e.g.][]{Brook04, Springel08, Guedes11, Aumer13, Grand17, Agertz20}. 
However, identifying the contribution of a particular component
of the galaxy to the total gravitational potential is very challenging
as these components can be asymmetric and evolve with time. In the current era, 
with full 6D phase space information for many MW halo stars, globular clusters and satellites at our disposal, 
we may have sufficient information to constrain 
the DM distribution of the MW. However, we do not presently have a framework 
to understand and quantify the asymmetries and time-evolving components of the MW's
halo. Owing to its recent infall into our halo and large infall mass, 
the LMC is currently the largest perturber of the MW's DM distribution, inducing large scale 
($>$50 kpc) perturbations across the entire halo (G19).

In order to quantify and understand the time-evolving perturbations caused by the 
LMC to the MW halo, we analyze the constrained N--body simulations presented 
in G19 
using BFEs.
In \S \ref{sec:sims_review} we summarize the main properties of 
these simulations. \S \ref{sec:HOBFE} reviews BFE, where
we focus on a particular BFE, the Hernquist expansion \citep{Hernquist92}.

\subsection{N--body simulations}\label{sec:sims_review}

The suite of N--body simulations used in this work were
presented in G19  

The suite consists of eight
high-resolution simulations, with dark matter particle mass $m_p=1.5\times 10^{4} M_{\odot}$\footnote{Note, that 
this value was incorrectly quoted as $4\times10^4$ M$_{\odot}$ in Table 1 of G19},
run with Gadget-3 \citep{Springel08}. We use two
MW models, each with the same Hernquist DM halo density profile 
with virial mass $M_{vir}=1.2 \times 10^{12}$M$_{\odot}$, an
exponential disk of mass $M_d = 5.78
\times 10^{10}$M$_{\odot}$, and a bulge of mass $M_b = 0.7 \times
10^{10}$M$_{\odot}$. 
We adopt two different with different halo distribution functions, where one is
isotropic and the other is radially biased. These are represented
by anisotropy parameters, $\beta(r)=0$ and
$\beta(r)=-0.5-0.2 \frac{d\rm{ln}\rho(r)}{d\rm{ln}r}$,
respectively. 

For the LMC, we have four
models with different halo masses but with the same enclosed mass
within 9 kpc, fixed by the rotation curve of the LMC
\citep{vandermarel14}. This observational constraint implies that the concentration of the halo is set for each LMC mass model and that the inner mass profile is similar for each LMC mass model. This has important consequences to the halo response, as we will illustrate.

Both the MW and the LMC halos were initialized 
with GalIC \citep{Yurin14} with a Hernquist DM halo, but matched to an
NFW halo in the inner parts following \citet{vandermarel14}.
Table \ref{tab:sims_summary} summarizes the
main properties of these simulations. 
We take simulation \#7 as our fiducial simulation since the MW halo model (radially biased) 
and LMC mass are consistent with current accepted values \citep[e.g.,][]{Moster13,Cunningham18b}.

The DM halos of the MW and the LMC
were evolved separately in isolation for 2 Gyrs in to order to guarantee energetic 
equilibrium in the halos.
The LMC was placed at the virial radius of the MW halo (ICs values are reported in Table 8 of G19). 
The simulations were run for $\sim$2 Gyr, and follow the evolution of 
the LMC on its first infall into the MW. 
At the present time the position and velocity vectors of the simulated LMC are
within 2$\sigma$ of the observed values reported in \cite{Kallivayalil13}.

\begin{table*}
  \centering
  \begin{tabular}{c c c c }
  \hline
  \hline
    \textbf{Simulation:} & \textbf{LMC halo mass at infall} $[\times 10^{10} \rm{M}_{\odot}]$ & \textbf{MW kinematics} & Name \\
    \hline
    \#1 & 0.8  & Isotropic ($\beta_{DM}=0$) & MW model 1 + LMC 1\\
    \#2 & 1.0  & Isotropic ($\beta_{DM}=0$) & MW model 1 + LMC 2\\
    \#3 & 1.8  & Isotropic ($\beta_{DM}=0$) &  MW model 1 + LMC 3\\
    \#4 & 2.5  & Isotropic ($\beta_{DM}=0$) &  MW model 1 + LMC 4\\
    \#5 & 0.8  & Radially biased ($\beta_{DM}=-0.5-0.2\alpha(r)$) &  MW model 2 + LMC 1\\
    \#6 & 1.0  & Radially biased ($\beta_{DM}=-0.5-0.2\alpha(r)$) & MW model 2 + LMC 2\\
    \textbf{\#7} & \textbf{1.8}  & \textbf{Radially biased ($\beta_{DM}=-0.5-0.2\alpha(r)$)} &  \textbf{MW model 2 + LMC 3}\\
    \#8 & 2.5  & Radially biased ($\beta_{DM}=-0.5-0.2\alpha(r)$) &  MW model 2 + LMC 4\\
    \hline
    \hline
  \end{tabular}
  \caption{Summary of the N--body simulations used in this study. The virial mass of the MW DM halo is $1.2\times 10^{12}M_{\odot}$, corresponding to a Hernquist halo mass of $1.57\times 10^{12}M_{\odot}$ and a Hernquist scale length of $a=40.8$ kpc. $\alpha= \frac{dln \rho(r)}{dln r}$. We take simulation \#7 as our fiducial simulation. LMC halo mass refers to the Hernquist mass adopted for the DM profile. Note that these LMC halo masses were incorrectly listed as the virial mass in Table 2 in G19.}
  \label{tab:sims_summary}
\end{table*}

\subsection{The Hernquist Basis Function Expansion:}\label{sec:HOBFE}

    To analyze the present-day snapshot from our N--body simulations we use BFEs. Several BFE expansions have been developed in the last three decades. For DM halos, \cite{Clutton-Brock73} built a BFE whose zeroth-order basis is the Plummer profile \citep{Plummer11}. Similarly, \cite{Hernquist92} built a BFE based on 
    the Hernquist profile \citep{Hernquist90}; given its accuracy, this expansion has been widely used in the literature \citep[e.g][]{Johnston01,Johnston02, Johnston02b, Sanders20}.
    Recently,
    \cite{Lilley18a} presented an analytic flexible BFE expansion based on the NFW profile \citep{NFW}, and presented a more general

    family of double-power law expansions parametrized with two parameters 
    in \citep{Lilley18b}. 
    Another option is to solve the Poisson equation numerically. which is a particular form of the Sturm-Liouville equation \citep{Weinberg99}. For example, in \cite{Dai18}, the NFW basis was solved numerically for the \textit{Eris} cosmologically simulated halo \citep{Guedes11}. 
    
    In our case, the Hernquist BFE is a natural choice since our simulations were initialized with Hernquist DM halos. Furthermore, \citep{Sanders20} demonstrate
    a higher performance and accuracy of the Hernquist BFE with respect to the other expansions. 
    
    In this section we summarize the main equations of the Hernquist expansion. For completeness, the derivation can be found in Appendix~\ref{sec:scf_derivation}. For a detailed and comprehensive derivation of this expansion we refer the reader to sections 2.2 and 3.1 in \cite{Hernquist92} or section 2.1 in \cite{Lowing11}.
    
    The density and potential for a system described by particles is represented by the following expansions:

    \begin{equation}\label{eq:rho_bfe}
     \begin{split}
      \rho(r, \theta, \phi) =  \sum_{n}^{n_{max}} \sum_l^{l_{max}} \sum_m^l  Y_{lm}(\theta) \rho_{nl}(r)
      (S_{nlm} \cos{m\phi}\\
      + T_{nlm} \sin{m\phi}
    \end{split}
   \end{equation}

   \begin{equation}\label{eq:phi_bfe}
   \begin{split}
     \Phi(r, \theta, \phi) = \sum_{n}^{n_{max}} \sum_l^{l_{max}} \sum_m^l Y_{lm}(\theta) \Phi_{nl}(r)
     (S_{nlm} \cos{m\phi} \\
     + T_{nlm} \sin{m\phi}
   \end{split}
   \end{equation}
   
   Where the coefficients $S_{nlm}$ and $T_{nlm}$ are defined as: 
   
   \begin{equation}\label{eq:coeff}
    \begin{aligned}
      S_{nlm} = \dfrac{(2-\delta_{m0})}{I_{nl}} \sum_k^{N} m_k
             \Phi_{nl}(r_k)Y_{lm}(\theta_k) \cos{m\phi_k} \\
             T_{nlm} = \dfrac{(2-\delta_{m0})}{I_{nl}} \sum_k^N m_k 
             \Phi_{nl}(r_k)Y_{lm}(\theta_k) \sin{m\phi_k} 
    \end{aligned}
    \end{equation}
    
    The expressions for $\rho_{nl}$ and $\Phi_{nl}$ can be found in Appendix \ref{sec:scf_derivation}. The normalization factor $I_{nl}$ is defined as:

    \begin{equation}\label{eq:Inl}
      I_{nl} = -K_{nl}
      \dfrac{4\pi}{2^{8l+6}}\dfrac{\Gamma(n+4l+3)}{n!(n+2l+3/2)[\Gamma(2l+3/2)]^2}
    \end{equation}

    And $K_{nl}$ is defined as: 
    
    \begin{equation}
    K_{nl}=\dfrac{1}{2}n(n+4l+3) +(l+1)(2l+1)
    \end{equation}
    
    Python public implementations of the \cite{Hernquist92} BFE can be found in the galactic dynamics Python-libraries 
    \verb+Gala+ \citep{gala, Price-Whelan:2017} \footnote{\href{http://gala.adrian.pw/en/latest/examples/Arbitrary-density-SCF.html}{http://gala.adrian.pw/en/latest/examples/Arbitrary-density-SCF.html}}, and \verb+Galpy+ \citep{Galpy}.
    An example of how this BFE decomposes the density and potential for a prolate, oblate, and triaxial halo can be found in Appendix \ref{sec:ideal_halos}. In this paper we used a modified version of the \verb+Gala+ library, which implements a new noise reduction method developed in this work (Appendix~\ref{sec:bfe_length})
    and is also adjusted for reading N--body snapshots.

  A challenge with BFEs is choosing the order of the expansion, i.e. $n_{max}$ and $l_{max}$. This choice is not obvious and depends on the number of particles and morphology of the system. \cite{Weinberg96} developed a method to truncate the expansion by studying the noise generated by the discrete nature of the simulation. In Appendix~\ref{sec:bfe_length}, we discuss the nature of the noise in BFE and how  the noise is reduced by truncating the expansion.

\section{Results: BFE of the MW and the LMC System}\label{sec:mwlmc_bfe}



In this section we present a combined BFE that represents the present-day MW--LMC system. 
A similar approach has been used previously to simulate mergers of massive galaxies
\citep[e.g.,][]{VanAlbada77, Villumsen82, Villusmen83}, we follow these approaches to analyze our N-body simulations. We analyze the MW--LMC system using three BFEs. 
In order to compute each of the expansions we organize the DM particles 
from the MW and the LMC into three different components, as defined below.


\begin{itemize}

    \item \textit{MW's DM halo:} All the DM particles that initially comprised the MW's DM halo 
      before the LMC's infall. The DM dynamical friction wake and the Collective Response are present in 
      this component, but the LMC's DM debris is not.

    \item \textit{LMC:} All LMC DM particles that are gravitationally 
    bound to the LMC at the present time. The LMC's halo has been tidally distorted by the 
    MW, requiring a BFE to characterize its asphericity and evolution.
    
    \item \textit{LMC's DM debris:} In our simulations the LMC has transferred a large fraction of its infall
    DM halo mass 
    to the MW's DM halo due to tidal forces. We define this DM debris as all the LMC's DM 
    particles that are presently gravitationally unbound to the LMC. The LMC's DM debris forms 
    the largest DM flow of particles in the MW at the present time. \citep[see also][]{Besla19}. 
    Note, that the mass of LMC's debris depends on the initial conditions, such as
    the mass profile of the LMC's DM halo
    and total mass (see section~\ref{sec:lmc_mass}). 
    

\end{itemize}

\begin{figure*}[h]
    \centering
    \includegraphics[scale=0.45]{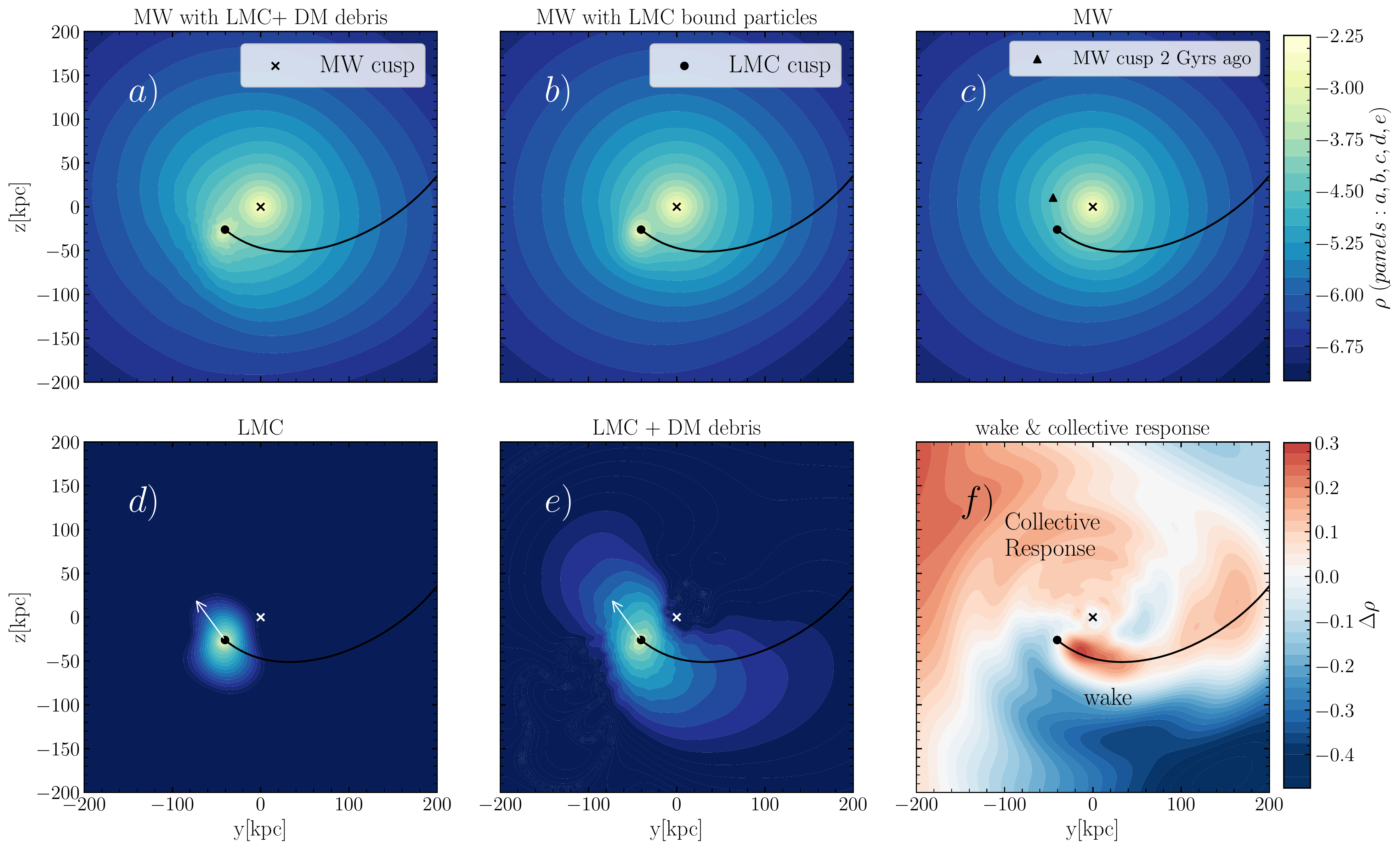}
    
    \caption{MW and LMC projected DM density reconstruction created using BFEs for the present-day snapshot of the MW--LMC simulation \#7 described in Section \ref{sec:sims_review}. The densities are computed in the $x=0$ Galactocentric plane in a slab, 10 kpc in thickness. Panel $a$ 
    shows the combined density field of the MW and the LMC, computed using two BFEs: one centered on the MW 
    (panel $c$; no LMC particles) 
    and one on the LMC (panel $e$; including LMC debris and bound particles, see Sections \ref{sec:mwlmc_bfe} and \ref{sec:LMC_DM_debris} for details).
    Panel $b$ shows the density field of the MW without 
    the LMC's DM debris, i.e. panel $c$, but also including the BFE for the LMC's bound particles (panel $d$).  
     In panel $c$ (the density field of the MW halo with no LMC particles),
 the original location of the MW cusp at the start of the simulation is marked by the black triangle. The present day cusp location of the MW(LMC) is marked by the x(circle) in all panels.
    All the panels are normalized 
    to the same color bar. BFEs enable the characterization of the LMC's direct contributions to the density field and accurately disentangle this contribution from the perturbations it induces in the halo. 
    Panel $f$ shows the MW halo 
    density field (panel $c$) computed as a density contrast $\delta \rho$, 
    as defined in equation~\ref{eq:denscontrast}.}
    \label{fig:bfe_mwlmc}
\end{figure*}

In the following, we describe three BFEs constructed to study the above three 
components: 1) the MW's DM halo 2) the LMC (particles bound to the LMC); 3) 
all the LMC's particles including its DM debris;
The MW--LMC system is described fully by the expansion of the MW combined with the expansion of the LMC and its DM debris. 
The density field for each of these BFEs is shown in Figure \ref{fig:bfe_mwlmc}. 

A summary of the main characteristics of the 
expansions used in this work can be found in table \ref{tab:BFE_summary}. 
In the following sections we discuss the details of how each of these 
expansions are computed. The resulting density field for various combinations of the LMC, LMC Debris and MW DM halo are summarized in Figure~\ref{fig:bfe_mwlmc}.

\begin{table*}[h]
    \centering
    
    \begin{tabular}{c c c c c c}
        \hline
        Component & subscript & Number of coefficients &  $r_s$ [kpc] & $\Gamma_{opt}$ &  $n_{max}, l_{max} $\\
        \hline
        MW's DM halo & MW  & 236 & 40.85 & 5 & 20, 20\\
        LMC's bound particles + DM debris &  LMC + DM debris & 1831 & 12-25 & 2 & 30, 30 \\
        LMC's bound particles & LMC & 20 & 12-25 & 8 & 20, 20\\
        \hline
        \hline
    \end{tabular}
    \caption{Summary of the computed BFE. A Noise subtraction was carried out following the procedure described in Section \ref{sec:bfe_length}. The Hernquist scale-length ($r_s$) is shown in column 3, and the optimal `signal-to-noise' in column 4.}
    \label{tab:BFE_summary}
\end{table*}

\subsection{The MW's DM halo} \label{sec:MWDMBFE}
Here we utilize BFEs for MW particles alone to describe the MW's DM halo response due
to the passage of the LMC.
Panel $c)$ in Figure \ref{fig:bfe_mwlmc} shows the BFE for the present-day MW's DM halo without
the LMC particles.  Panel $f)$
shows the density contrast for the
MW's DM halo, 
computed as:

\begin{equation}
\delta \rho = \dfrac{\rho}{\rho_{000}} - 1,
\label{eq:denscontrast}
\end{equation}

where $\rho_{000}$ corresponds to the density computed with only the first term in the expansion.

Panel $f)$ clearly illustrates 
the halo response.  

\begin{figure*}
    \centering
    \includegraphics[scale=0.11]{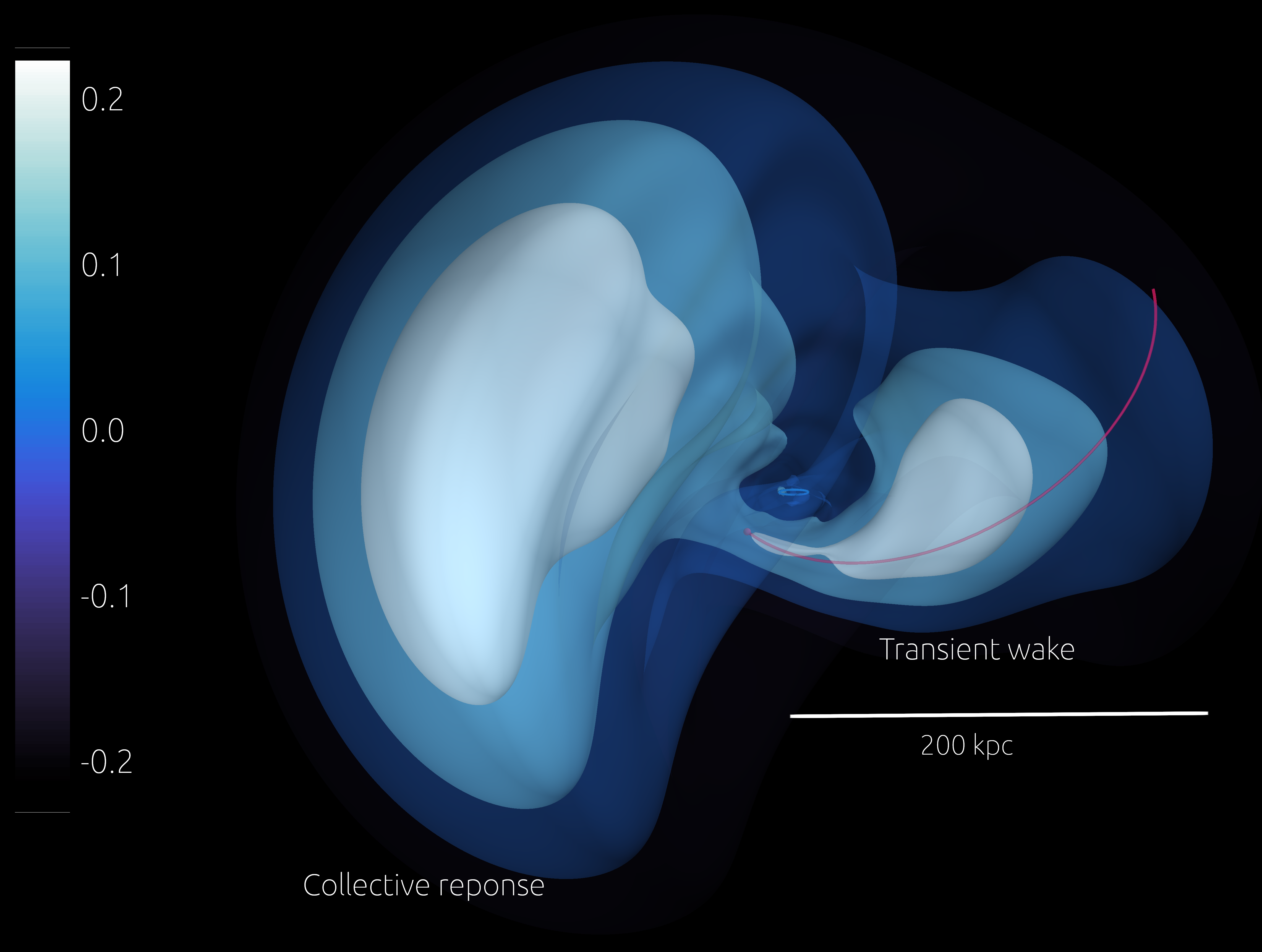}
    \caption{The LMC-induced DM dynamical friction wake and Collective Response in the MW DM halo at the present day, in the Galactocentric YZ plane. The density contours are computed using the BFE for the MW's DM halo. The colorbar shows the density contrast as defined in Equation~\ref{eq:denscontrast}. White contours represent the overdensities, while the darker blue contours show the underdensities. The dynamical friction wake is a large-scale structure ranging from $\sim$ 50 kpc, near the LMC (red circle), out to the edge of the halo. The Collective Response is the larger overdensity that appears predominantly north of the MW disk (the latter is marked by the central blue ellipse). The Collective Response also appears to the south of the MW disk, at large distances. The red line marks the past passage of the LMC, which tracks the location of the dynamical friction wake. A 3d animated rendering of the density field of the MW illustrating the halo response to the LMC's passage, can be found here \href{https://vimeo.com/546207117}{https://vimeo.com/5462071170}}
    \label{fig:3d_Wake}
\end{figure*}

A collisionless self-gravitating system responds to a perturbation through excitation
of specific modes \citep[e.g:][]{Kalnajs77, Weinberg93}. For the collisionless
Boltzmann equation, the spectrum of solutions is represented by discrete and
continuous modes. Overall, the discrete modes are sustained by resonances
\citep[e.g:][]{Weinberg94} and damp slowly over time (the Collective Response) while
the continuous modes \citep{Vandervoort03} are non-resonant and damp quickly (the
dynamical friction DM Wake). In the case of a satellite sinking in a larger host, the
host response is commonly known as the "wake" which encompasses both discrete and
continuous modes. However, given their markedly distinct and striking morphologies we
make a distinction between the two as follows:

\textit{LMC's Dynamical Friction Wake:} The wake is the overdensity of DM particles
trailing the LMC as illustrated in Figure~\ref{fig:3d_Wake}. The LMC excites modes that will phase-mix on time scales shorter than an orbital time. These modes can be thought of
as wave-packets that are excited by the perturbation. Classically, the dynamical friction wake is
described as the local scattering of DM halo particles as a massive perturber travels
through the medium of a larger host. In the case of a homogeneous and infinite
background medium, this leads to the classic dynamical friction equation derived in
\citep{Chandrasekhar43}. \cite{Mulder83} and \cite{Weinberg86} showed that dynamical
friction was the result of the wake formation.  Here we define the LMC’s dynamical
friction wake as the non-resonant continuous modal response of the halo to the
passage of the LMC. As such, the dynamical friction wake will damp quickly after a
few crossing times.

\textit{Collective Response:}  

The Collective Response is shown in the upper left of Figure~\ref{fig:3d_Wake}. It is a large overdensity in the halo, mainly in the Northern hemisphere. It has been recognized that satellite-halo interactions lead to a barycenter excursion \citep{Weinberg89, ChoiPhDT}. The amplitude of the barycenter excursion (or displacement) depends on the mass and orbit of the satellite. For a galaxy, the barycenter displacement is complex because the system is not a point mass and therefore the response is described by the excitation of its $l=1$ weakly damped mode \citep{Weinberg89}. In contrast to the dynamical friction wake, the Collective Response is the result of the excitation of discrete modes (e.g. $l=1, m=1$), which are sustained by resonances and self-gravity. These perturbations will damp slowly over time-scales longer than the orbital time. Here we define the Collective Response as the discrete modal response of the halo to the passage of the LMC predominantly seen in the ($l=1$ mode). As such, the Collective Response will persist over many crossing times.

In the remainder of this section, we will study the combined effect of both the dynamical friction wake and the Collective Response using the MW BFE (no LMC particles). 

A 3D animated rendering of the
density contrast (equation \ref{eq:denscontrast}), illustrating the halo response to the LMC's passage, can be found here \href{https://vimeo.com/546207117}{https://vimeo.com/546207117}. 
The present-day frame from the animation for the density contrast (equation \ref{eq:denscontrast}) is shown in Figure \ref{fig:3d_Wake}.

We do not include an expansion for the baryonic components of the MW (disk and bulge), 
although they are included in the N-body simulation as live components
(see Section \ref{sec:sims_review}). Consequently, the MW's DM halo and the LMC do feel the gravitational potential of the MW disk and bulge.

\subsection{The LMC: Defining bound particles with BFEs}\label{sec:lmc_body}

We use BFEs to compute the gravitational potential $\Phi$ of the LMC using all the LMC's DM particles. The expansion is centered on the LMC, where the cusp of the LMC was
computed using the shrinking sphere algorithm described in \cite{Power03}. 
We compute the expansion up to order
$n_{max}=l_{max}=20$. The gravitational potential is computed using only the 
coefficients with $\Gamma>$5. We keep the halo
scale length fixed during the iterative procedure since the inner region of the LMC's halo does 
not change significantly during its orbit. We test this by fitting the scale length in
each iteration to the LMC's DM particle distribution.

We define particles as bound to the LMC if their kinetic energy (KE) is less than the gravitational potential energy ($\Phi$). This is
similar to the iterative procedure used in \cite{Johnston99}. In practice we use the energy per 
particle mass and hence a particle is bound if: $\dfrac{1}{2}v_k^2 < \Phi_{BFE}$, where $\Phi_{BFE}$ is 
computed using equation \ref{eq:phi} at the location of each particle, $k$. The velocity $v_k$ is computed with respect to the LMC's cusp.

We iterate the computation of the bound particles for every particle 
until we reach 1\% convergence in the number of bound particles. The mean number of iterations required 
is $\sim$ 5. 

Panel $d)$ in Figure \ref{fig:bfe_mwlmc} shows the BFE reconstruction of the density field of the bound LMC particles at the present time for the fiducial model. The LMC DM halo exhibits an S-shape produced by the tidal field of the MW. This is most pronounced in the galactocentric y-z plane (coincident with the orbital plane of the LMC).

\begin{figure}[h!]
    \centering
    \includegraphics[scale=0.5]{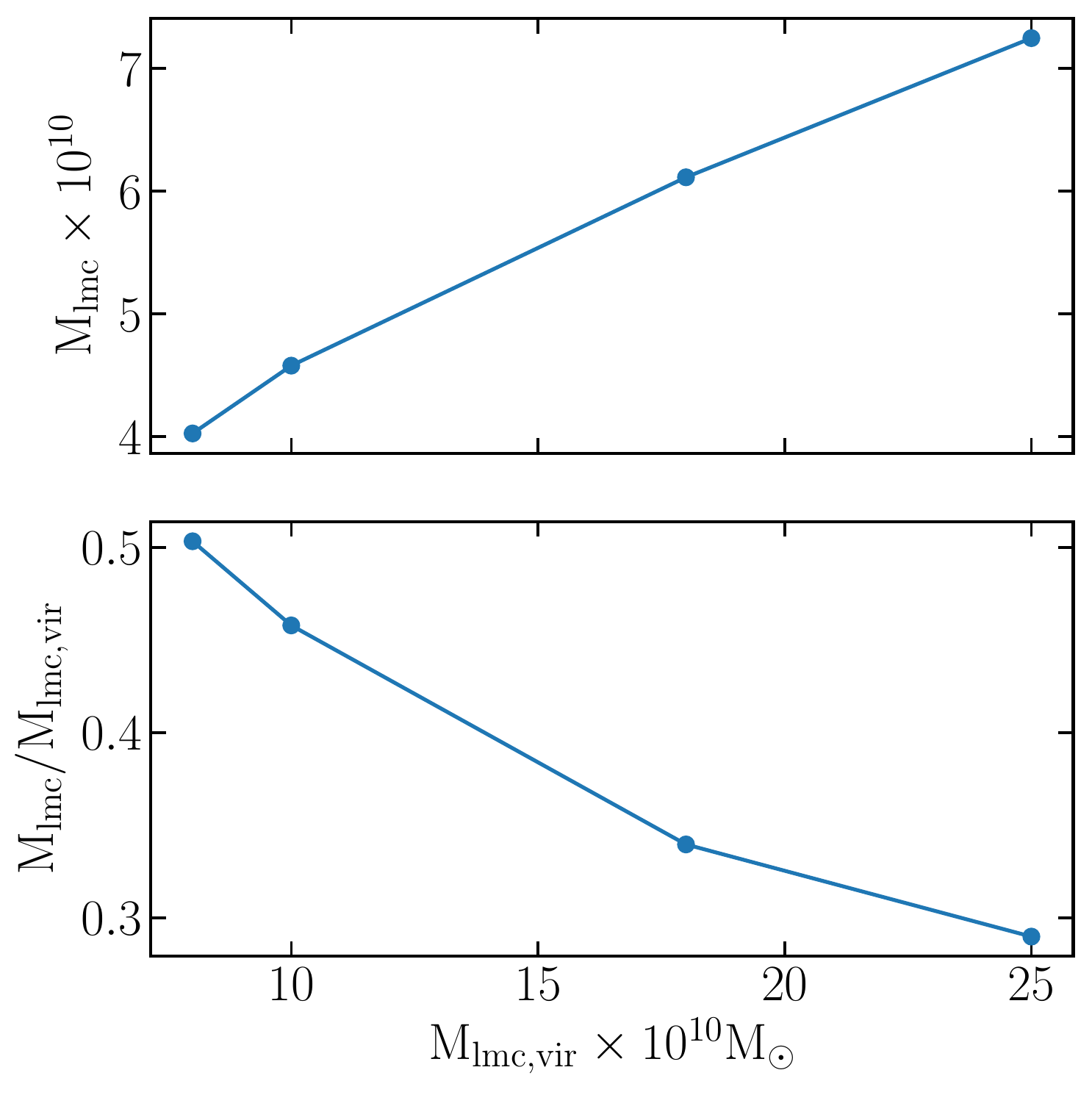}
    \caption{\textit{Top panel:} Present-day bound mass of the LMC, $M_{\rm LMC}$, as a function of the LMC virial mass at infall,
    $M_{\rm LMC, vir}$, for the four LMC halo masses explored in this study
    ([8, 10, 18, 25]$\times 10^{10}$ $M_\odot$).  \textit{Bottom panel:} Ratio of $M_{\rm LMC}$ to $M_{\rm LMC,vir}$ as a function of $M_{\rm LMC,vir}$. 
    The DM halo of each LMC is initially modeled with a Hernquist profile with a concentration constrained by the observed rotation curve \citep{vandermarel14}. The resulting Hernquist scale length, $r_s$, ranges from 10--25 kpc, where higher mass LMC halos have lower concentrations and larger $r_s$ than lower mass LMC halos. The higher mass LMC halos thus have more mass at larger distances. This material is more easily captured by the MW, explaining why 
    the ratio of present-day bound mass to the infall mass decreases with increasing LMC infall mass.}
    \label{fig:LMC_bound_mass}
\end{figure}

The total bound mass of the LMC at the present-day ($M_{\rm{LMC}}$) depends on both the 
orbit of the LMC and on the DM halo profile of the LMC. 
Figure \ref{fig:LMC_bound_mass} illustrates the bound mass of the LMC at the present day, $M_{\rm{LMC}}$, as a function 
of the LMC's initial virial infall mass  ($M_{\rm{LMC}, vir}$), for the four LMC models used in this study (all Hernquist halos).

Bound $M_{\rm{LMC}}$ ranges from $\sim 4 \times 10^{10}$ (50\% of the infall mass of the lowest mass LMC model) up to $\sim 7 \times 10^{10} M_{\odot}$ (30\% of the infall mass of the highest mass LMC model).
The most massive LMC models 
lose a larger fraction of mass compared to the less massive models. This is expected 
because the concentration

of these halo models are different. 
Each LMC's DM halo is initially represented by a Hernquist profile with 
a scale radius chosen to reproduce the observed rotation curve of the LMC (see G19).
As such, low mass models have a higher halo concentration compared to high mass models. For
these low concentration models, the DM particles in the outskirts of the LMC's DM halo are less bound to the LMC
than in the higher concentration models (low mass models).  As a result, the
ratio of bound to infall virial mass is lower in the higher mass models (see lower panel of
Figure \ref{fig:LMC_bound_mass}).

\begin{figure}
    \includegraphics[scale=0.4]{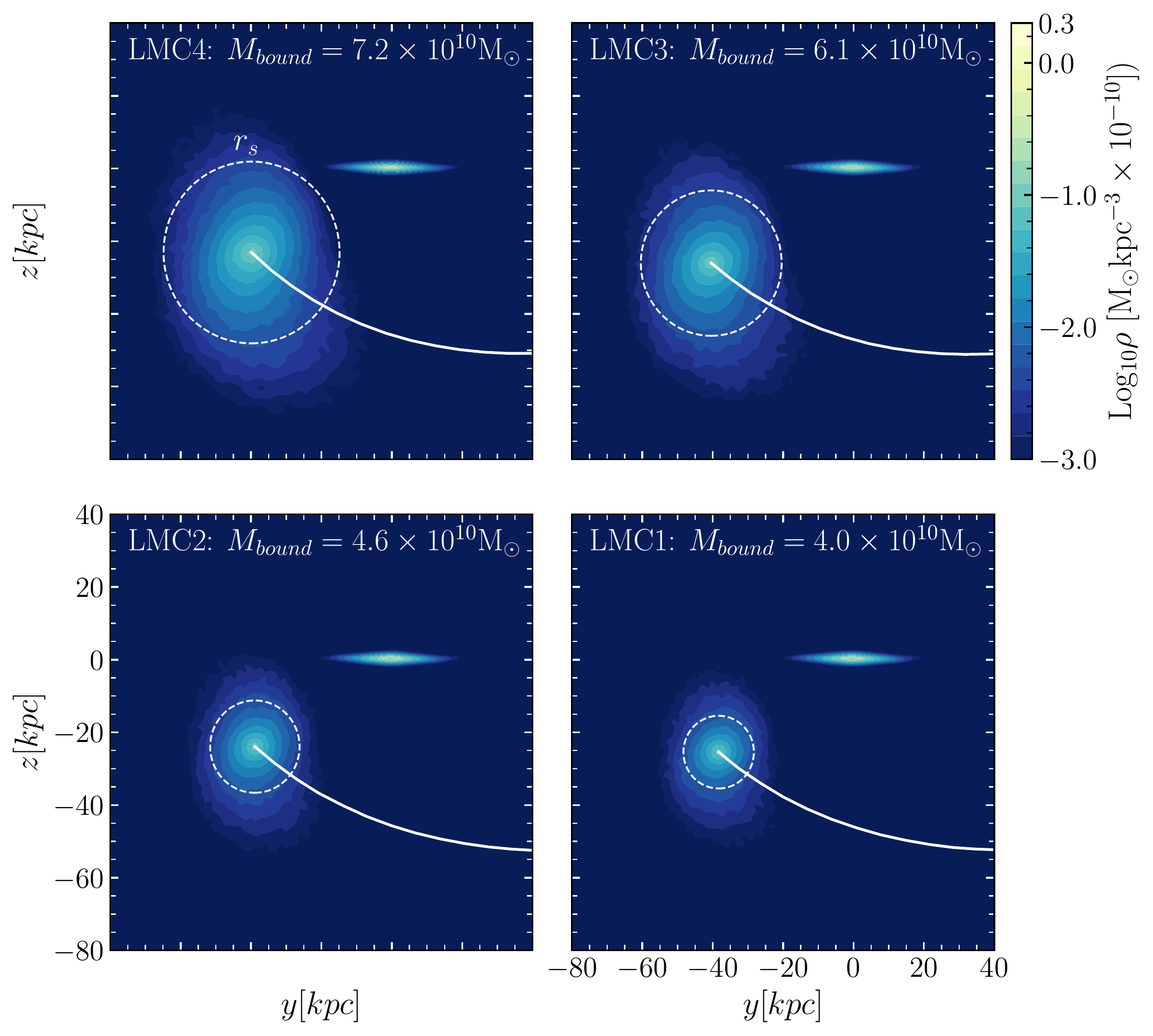}
    \caption{Projected density distribution of the LMC's bound particles in the Galactocentric
    $y-z$ plane. The white circles illustrate the LMC's DM halo scale length, $r_s$. The MW's
    disk particles are shown as a reference. The scale length of the LMC's DM halo can be as close as 30 kpc from the Galactic center. Note that particles from the LMC's unbound debris do overlap with the disk of the MW \citep[][see panel $e)$ of Figure~\ref{fig:bfe_mwlmc}]{Besla19}.}\label{fig:LMC_dmhalo}
\end{figure}

We illustrate the extent of the bound LMC's DM halo in Figure \ref{fig:LMC_dmhalo}
for all of our four models, in the Galactocentric $y-z$ plane. As a reference, the MW's disk particles are also plotted.
The present-day bound LMC's DM halo is very extended, where the edge of the halo scale length (white circle) can be as close as $\sim$30 kpc
from the Galactic center. 

\subsection{The LMC's DM debris}\label{sec:LMC_DM_debris}

Over the last 2 Gyrs, the LMC has lost more than 50\% of its pre-infall mass (see 
lower panel in Figure \ref{fig:LMC_bound_mass}). This DM debris extends over
large distances throughout the DM halo. The LMC's DM debris can impact the shape of the MW's halo (see Section \ref{sec:mw_com}) and can
extend the reach of direct detection experiments to lower WIMP mass \citep{Besla19}.

The mass of the LMC's DM debris is found using the same iterative method described in  
section \ref{sec:lmc_body}, but now requiring that the kinetic energy exceeds the potential energy, 
thus defining unbound LMC particles.

\begin{figure}
    \centering
    \includegraphics[scale=0.6]{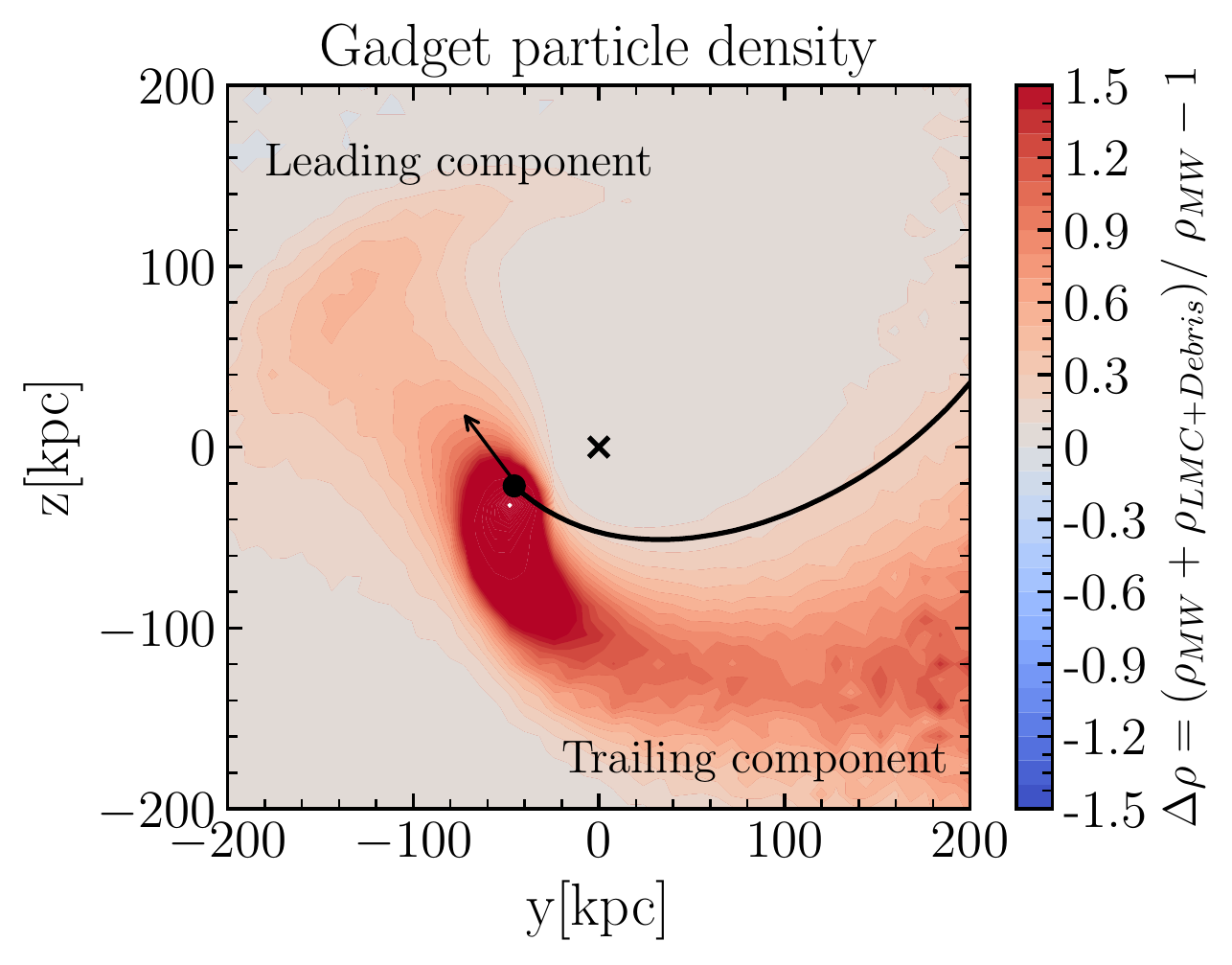}
    \includegraphics[scale=0.6]{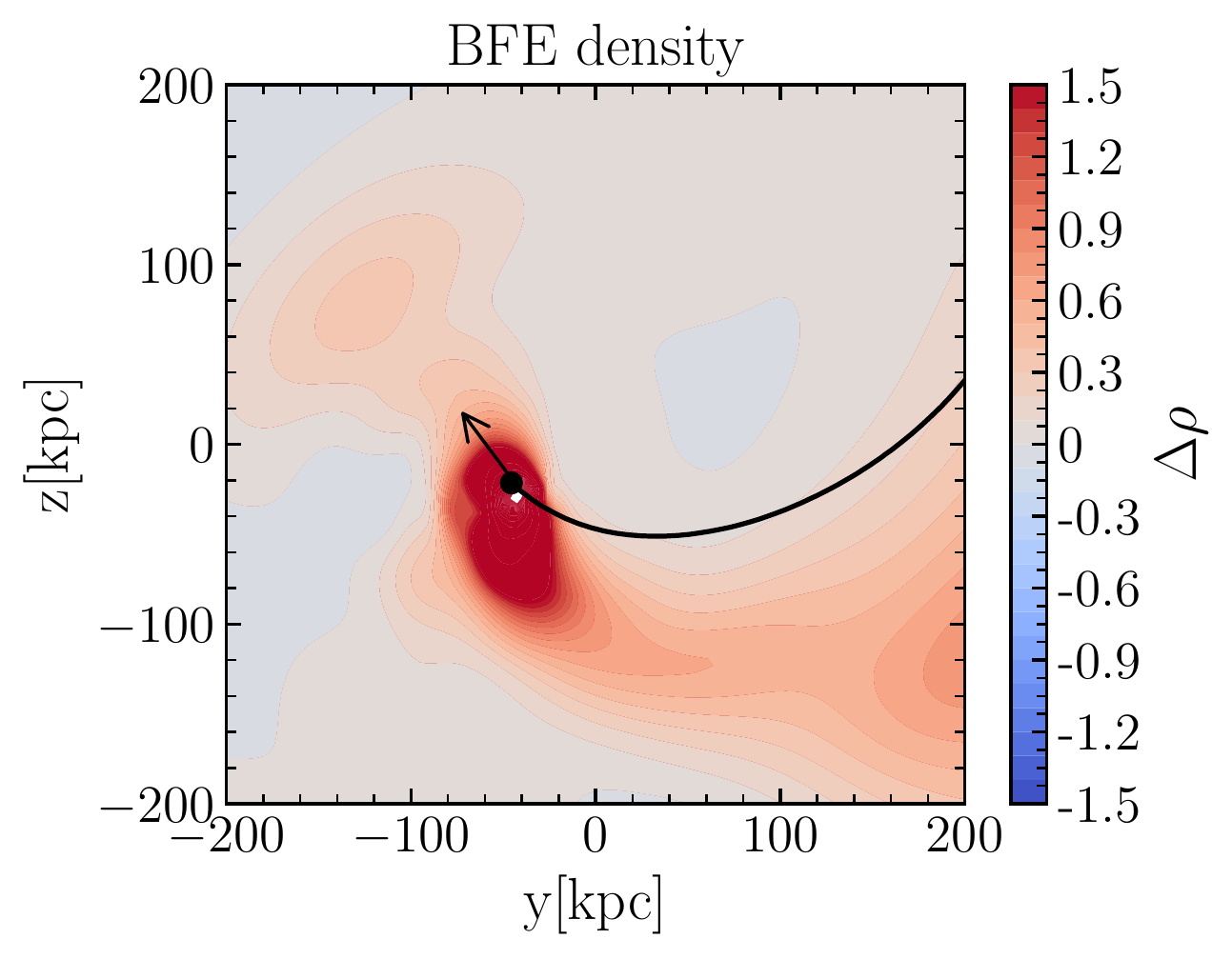}
    \caption{Density contrast of the LMC DM debris, with respect to the MW (panel $c$ in Figure \ref{fig:density_contours}). The density contrast is computed using equation~\ref{eq:denscontrast}, where the denominator is now the MW density including the halo response (not just the monopole).    The debris is illustrated within a 10 kpc slab in $y-z$ Galactocentric plane, centered at $x=0$ kpc. The bound LMC was removed by masking high density regions (density contrasts $>$ 1.5), enabling us to isolate the structure of the debris.
    The LMC DM debris has both trailing and leading components that extend over a large region of the MW's DM halo. In the leading component, 
    a shell structure is starting to form as the debris particles, now bound to the MW, reach apocenter in their orbits. The trailing component appears to have a higher density relative to the background owing to the formation of the dynamical friction wake, which creates underdensities in the MW DM distribution in the volume encompassing the trailing LMC DM debris.  \textit{Top panel} shows the density computed from the particle data. 
    \textit{Bottom panel} shows the density computed with the BFE. 
    The BFE reproduces 
    the amplitude and morphology of the distribution of LMC debris from the particle data (top panel).} 
    \label{fig:debris_density}
\end{figure}

We utilize one high-order ($n_{max}=l_{max}=30$) expansion to capture  
all LMC particles, i.e. both bound and unbound particles. 
The BFE for the bound LMC only requires 20 terms. However, including the highly distorted debris, 1831 terms and a signal to noise threshold of 2 (see Appendix~\ref{sec:bfe_length}) were needed to reproduce the density field computed from the LMC debris particles (top vs. bottom panels in Figure~\ref{fig:debris_density}).

Through comparison with the BFE for the bound particles (previous section), we can identify the properties of the LMC's DM debris. The debris is found to be a much more extended asymmetric structure, as illustrated in panel $e)$ in
Figure \ref{fig:bfe_mwlmc} for the fiducial model (relative to panel $d$). 

In Figure~\ref{fig:debris_density}, we illustrate the density contrast of the LMC debris with respect to the MW (panel $c$ of Figure \ref{fig:bfe_mwlmc}). This is computed using equation~\ref{eq:denscontrast}, where the denominator is now the MW density (not just the monopole).
We isolate the structure of the LMC debris by masking high density regions in the LMC particle distribution (density contrast $>$1.5), which correspond to the bound LMC.
Two components comprise the LMC’s DM debris: trailing and leading,

1) \textit{Trailing component:} LMC DM particles that extend from the LMC out to $-200$ kpc in $\hat{z}$ and beyond 200 kpc in $\hat{y}$ (Galactocentric). The DM density enhancement of the trailing debris relative  to the MW (which includes the halo response, i.e. the wake)
is $\sim$100\% for the fiducial model. Meaning that in 
the regions of the halo traced by the debris,
the mass in the debris
equals the local DM mass of the MW in that same volume. 

This high ratio is aided by the fact that the dynamical friction wake causes a significant decrease in the MW’s DM content in that region, by $\sim 30\%-40\%$ (panel $f$ in Figure \ref{fig:density_contours}).

2) \textit{Leading component:} LMC DM particles 
in the north galactic hemisphere, extending from the LMC out to 150 kpc in $\hat{z}$ and $\hat{y}$. The leading component of the LMC debris overlaps with the Collective Response (panel $d$ vs. $e$ in Figure \ref{fig:density_contours}). The shape of the leading component is 
asymmetric, resembling the shell structure typically formed through mergers of satellites on radial orbits \citep{Piran87, Heisler90}.
The dynamics of tidal debris from satellite galaxies has been extensively studied  
\citep[e.g][]{Choi09, Sanderson10, Amorisco15, Hendel15, Drakos20}. In general the 
resulting morphology of tidal debris is governed by the satellite's orbit, internal rotation, mass, and internal structure. However, for massive satellites, the self gravity of the debris, the satellite's gravitational influence on the debris, and the host halo's response to the satellite, are also important. All of these processes are captured in our simulations. 

In order to fully constrain and predict the morphology of the LMC's debris, simulations that cover a more complete parameter space are needed. 
For example, varying the density profile of the LMC and including perturbations and mass loss from the SMC. 
We conclude that, in our simulations, the mass of the LMC's DM debris is non-negligible and can dominate the density field of the DM around the MW in some regions of the southern hemisphere. 

The distribution of the debris is highly asymmetric and evolves in time - it 
cannot be well-represented by the infall of a static LMC halo. We expect that the distribution of the LMC DM debris can affect the dynamics of stellar streams, GCs, and satellite galaxies; this will be studied in subsequent papers.

\subsection{The Combined BFE for the MW and LMC System}

With the three optimal BFEs computed for each component of the system 
(MW, LMC, LMC+Debris) 
we can now quantify the density, potential, acceleration and shape of the combined MW--LMC
system in a compact and accurate manner.
Specifically, we combine the BFE for the MW (section \ref{sec:MWDMBFE}) with the BFE for the LMC+Debris (section \ref{sec:LMC_DM_debris}). The combined expansion is seen in panel $a$ in Figure \ref{fig:density_contours}). 

For the fiducial simulation, the combined expansion 
consists of 236 
coefficients for the MW and 1831 for the LMC+debris (see Table~\ref{tab:BFE_summary} and previous section). Each combined expansion contains the information of the entire 100 million particle simulation.

\begin{figure*}[h]
\centering
  \includegraphics[scale=0.55]{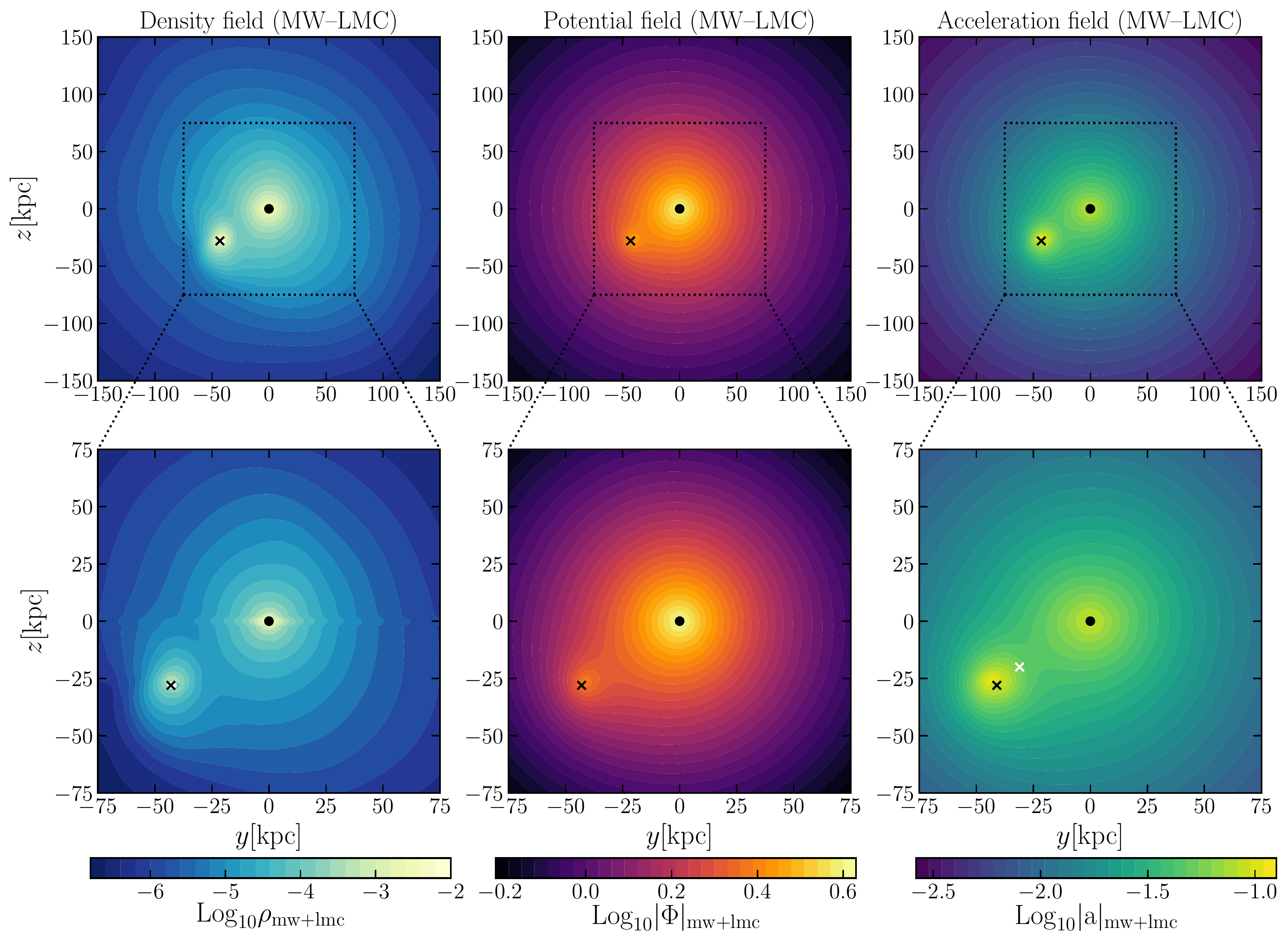}
 
  \caption{MW--LMC reconstructed density (left panels), potential (middle panels) and acceleration (right panels) fields of the present day combined MW+LMC system (including halo response and LMC debris), 
  for the fiducial simulation \#7. Top panels show the DM halo out to 150 kpc, while the bottom panels shows the inner regions, out to 75 kpc. 
  The top left panel is the same as panel $a$ in Figure \ref{fig:density_contours}. 
  The BFE as computed using 
  2067 coefficients selected with the procedure outlined in  
  Section \ref{sec:bfe_length}. We also included the potential of the MW disk and bulge analytically. The black dot marks the COM of the MW disk. 
  The black x marks the COM of the LMC halo. The white x in the bottom right panel marks the local minimum in the acceleration field for the system. 
  The present-day, combined MW--LMC DM potential and acceleration field is clearly distorted due to the response of the halo (Collective Response and dynamical friction wake), the LMC halo itself and its DM debris.}\label{fig:bfemwlmc}
\end{figure*}

Figure \ref{fig:bfemwlmc}, 

illustrates 
the projected density (left panel), acceleration (middle panel), and potential (right panel)
fields for the combined MW--LMC system computed with the BFE. Each row shows a different size-scale of the system,  from a radius of 75 kpc (top panel)
 to 150 kpc (bottom panel). 
Note that, to create these combined potentials, the center of each of the expansions is different. The expansion of the MW  

is centered on the MW COM, while the expansion of the LMC+Debris is centered on the LMC's COM.

The inner region ($<$35 kpc) of the MW's DM halo is governed 
by the presence of the disk and the bulge. The contribution from the 
MW disk to plotted the density, potential and acceleration fields, was added by fitting the simulated MW disk from the
N-body simulations with a Miyamoto-Nagai profile 
\citep{Miyamoto-Nagai}. The bulge was also
added analytically following the Hernquist profile used in the initial conditions from G19. 
While it is possible to 
use a BFE for the bulge and the disk, respectively \citep{Petersen19-disk}, in this study 
we are primarily concerned with the perturbations at large distances, 
where the deviations from the analytic representations of the disk and bulge are negligible. 

An interesting feature observed in the simulated acceleration field is the local minimum value  
between the MW disk and the LMC.
This minimum is located at $\vec{r}=(-2.5 \hat{x}, -31 \hat{y}, -20\hat{z})$ kpc (see white cross in bottom right panel of Figure \ref{fig:bfemwlmc}), which is along the 
separation vector between the LMC and MW. The position corresponds to ($l$=280, $b=-$32) in galactic coordinates.
The location of this minimum does not vary among our 8 simulations since the enclosed mass
within 50 kpc of the LMC is similar in all models. 
The combined acceleration field provides a snapshot of the non-axisymmetric accelerations experienced by objects in orbit about the MW at the present-day.

Outside 35 kpc, the perturbations from the LMC become apparent. 
Formed by the combined contribution of the MW's DM halo response, the bound LMC  
and the LMC's DM debris, the perturbation is anisotropic. 
The DM distribution outside 35 kpc is elongated in the $\hat{y}$ 
direction, which is in the direction of the DM dynamical friction wake and the DM debris. 
Hence, the DM halo shape/distribution contains information about the location and amplitude of the DM dynamical friction wake.

Ultimately, our goal is to quantify the perturbation caused by the LMC in the MW's DM halo 
using the BFEs, we will discuss this in Section \ref{sec:power}. We will discuss how these results change as a function of the MW anisotropy profile and LMC 
mass in section \ref{sec:lmc_mass}.

\section{Results: Understanding the BFE of the MW--LMC DM halo}\label{sec:power}We start by identifying the most energetic coefficients that dominate the BFE 
of the combined MW--LMC DM halo in Section~\ref{sec:energy_coeff} and \ref{sec:viz_response}. We further build insight into how the shape of the halo is related to the energy in each expansion term and compare these results to idealized prolate, oblate, and triaxial halos to quantify how well the DM distribution of MW--LMC system is described by simplified halo models 
(Section \ref{sec:power_shape}). We finally connect how the shape of the MW's halo is impacted by the MW's barycenter motion in Section \ref{sec:mw_com}.

\subsection{Gravitational energy using the BFE coefficients}\label{sec:energy_coeff}

One of the advantages of using BFEs is the ability to decompose the halo response
into modes whose amplitudes are the coefficients of the expansion. 
The contribution to the total gravitational potential energy $U_{nlm}$ of each mode 
can be computed with the corresponding coefficients. We quantify the halo response to the 
passage of the LMC in terms of the energy in each expansion term. The total gravitational potential $U$ 
for the basis expansion is defined:

\begin{equation}\label{eq:energy1}
  U = \int \rho({\bf{r}}) \Phi({\bf{r}}) d \textbf{r} 
\end{equation}

\noindent Using the orthonormal properties of the BFE, the gravitational potential energy may be expressed in terms of the BFE coefficients as:

\begin{equation}\label{eq:U}
  U = \sum_{nlm} U_{nlm} = \sum_{nlm} I_{nl} \left(S_{nlm}^2 + T_{nlm}^2 \right) 
\end{equation}

Note that equation \ref{eq:U} has units of energy, the coefficients $S$ and $T$ are
unit-less, but the units comes from the normalization factor $I_{nl}$ (see Eq. \ref{eq:Inl}). 
The coefficients used in the energy calculation are already smoothed, as described 
in Section \ref{sec:bfe_length}. The most energetic coefficients represent 
the modes that contribute the most to the total gravitational potential energy
of the system.

\begin{figure*}[h]
    \centering
    \includegraphics[scale=0.55]{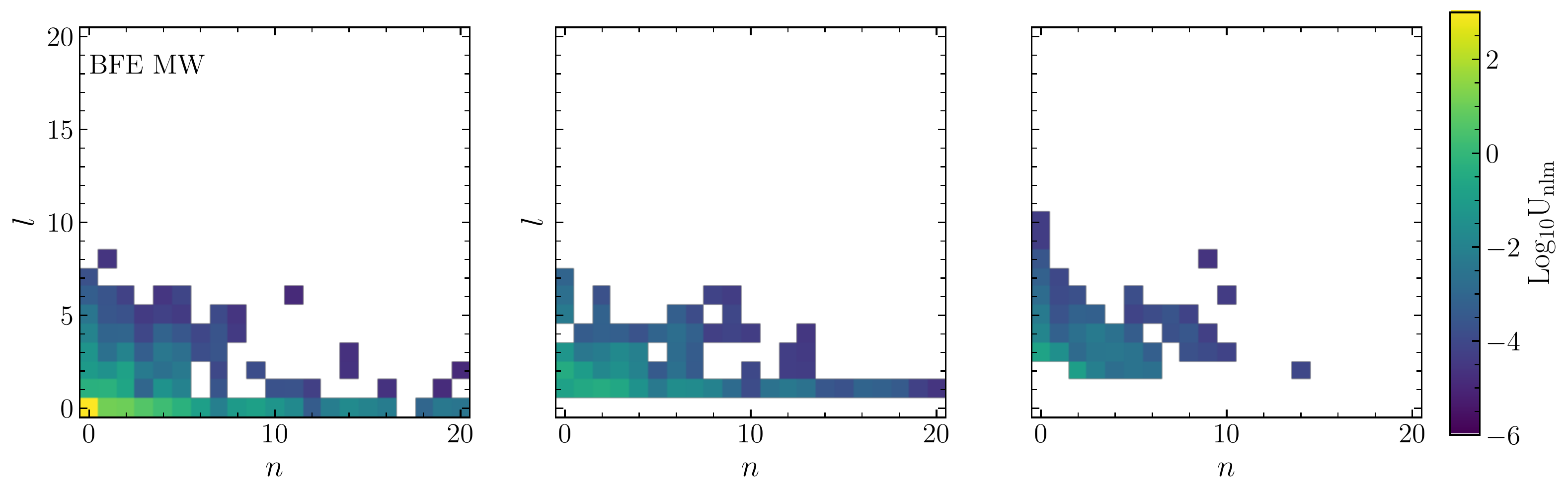}
    \caption{Gravitational energy $U_{nlm}$ of the most energetic coefficients for 
    the BFE of the MW's DM halo, for the fiducial simulation \#7.
    Coefficients are defined by their radial modes, $n$, vs. their angular mode $l$, with
    the dominant $m$ modes indicated in each column. Note that by definition of the BFE, there are no coefficients with $l<m$.
    The most energetic term is 
    the monopole, indicating that overall the LMC has not dramatically altered the symmetry of the original dark matter halo.
    However, there are energetic terms present from angular modes, indicating that asymmetric 
    structures do exist, even if they do not dominate. In contrast to expectations for
    idealized, symmetric halos (oblate, prolate or triaxial, see Section
    \ref{sec:power_shape}), the LMC induces 
    contributions from dipole terms ($l=1$) and high order odd $l$ terms (e.g., $l=3, 5$).}
    \label{fig:mw_energy}
\end{figure*}

Figure \ref{fig:mw_energy} shows the gravitational potential energy, $U_{nlm}$, of 
all coefficients decomposed in $n$ and $l$, with $m$ indicated per column.
These coefficients were computed for 
our fiducial simulation \#7. The main characteristics of the MW BFEs are summarized as follows:

\begin{enumerate}
    \item  Globally, the most energetic term 
is the monopole, which contributes $99$\% of the total gravitational potential energy in both cases. 
This term is the spherically symmetric Hernquist DM halo. 
    \item The radial modes, $n>0$, contribute up
to order $\sim$ 20, for the $m=l=0$ case.
    \item The angular, $l$, modes
appear up to order $l=14$, but most of the energy in those terms is contained within
order $l<6$. The angular, $m$, modes exhibit contributions up to $m=8$ and $m=12$; however,
most of the energy is contained in the $m=0, 1, 2$ modes 
(shown in Figure~\ref{fig:mw_energy}). 

\end{enumerate}

Although globally the halo energetics are reasonably described by the original, 
unperturbed Hernquist DM halo potential, the existence of energetic
angular $l,m$ modes prove the existence of asymmetries induced by the LMC's 
passage within the halo. 
Furthermore, both the halo Collective Response and the dynamical friction wake  
are localized perturbations that will have a non-negligible 
impact on the  density (G19) and shape of the halo (see Section \ref{sec:mw_com}). 

Studying the effect of the LMC's DM debris is also crucial to interpret the shape of the MW's halo. Since we have separate expansions for the LMC+DM debris and the MW,
we cannot self-consistently quantify the increase in energy in the MW BFE from the LMC's debris. 
However, we will study the effects of the debris on the density field of the MW-LMC system in section~\ref{sec:mw_com}.

\subsection{Visualizing the modes that dominate the halo response}\label{sec:viz_response}

\begin{figure*}[h]
    \centering
    \includegraphics[scale=0.6]{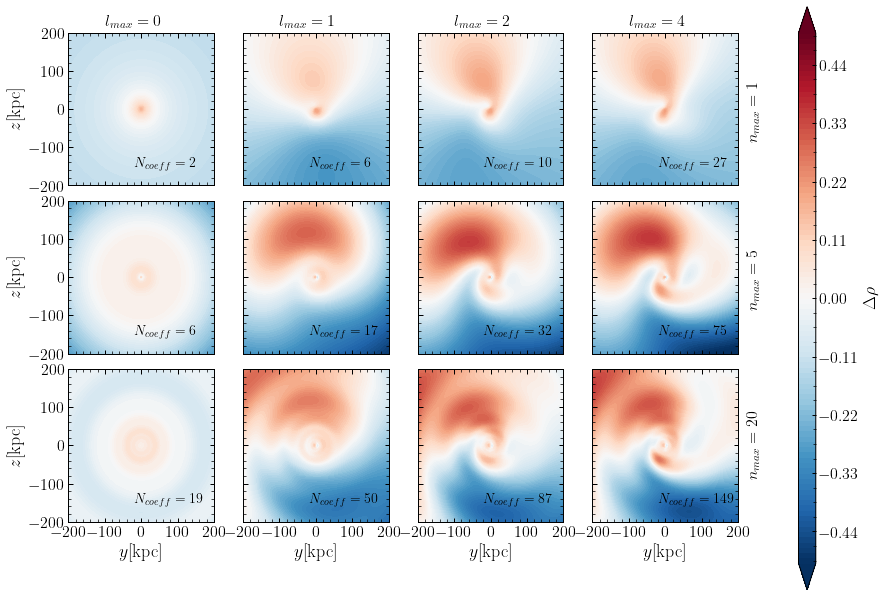}
    \caption{Projected density field of the MW's halo response (Collective Response and dynamical friction wake) reconstructed using different sets of terms of 
    the expansion. Color scale indicates the density relative to the monopole. Each column shows the 
    reconstruction up to a different maximal order in $l$ and $m$, 
    from $l_{max}=m_{max}=0$ (left column) up to $l_{max}=m_{max}=4$ (right column). In each row
    the maximal order of the radial $n$ term is increased.  
    Angular terms build the asymmetric structure of the halo response. The terms with odd $l$ and $m$  
    contribute to the asymmetry in the density field between the north and south (i.e the Collective Response). 
    Radial terms build the radial scale of the response (bottom row). The level of asymmetry needed to 
    reveal the dynamical friction wake is reached at $l_{max}=4$ (right column).}
    \label{fig:wake_l_terms}
\end{figure*}

In this section we qualitatively describe the terms that most contribute to 
the halo response 
(no LMC BFE). Figure \ref{fig:wake_l_terms} shows the projected density field of
the halo response (Collective Response and dynamical friction wake), dissected using different sets of coefficients, ranked by their
order. Each column shows the reconstruction up to a different maximal order in $l$ and $m$, from $l_{max}=m_{max}=0$ 
(left column) up to $l_{max}=m_{max}=4$ (right column). In each row the maximal order of the radial $n$ term is increased. This means that the upper left panel only contains two coefficients, $n=0$ and $n=1$, while the bottom right panel has the most terms, up to $n_{max}=20$ and $l_{max}=4$. Below we list the the salient features for each columns.

{\bf The left column} illustrates 
 spherically symmetric density fields. As the order in $n$ increases, each term contributes to a perturbation at a characteristic radius. 
The combination of these $n$-modes creates a more complex perturbation that sets the radial scale of the halo response. For example, in the lower left panel, the 

overdensities appear mainly beyond 50 kpc, which coincides with where the dynamical friction wake starts to pick up. 

{\bf The middle and right columns:} As we increase the order in $l$ and $m$, we see deviations from spherically
symmetric perturbations. For $l_{max}=m_{max}=1$ we clearly see a dipole in the density field, corresponding to 
the Collective Response. The amplitude of the $l_{max}=2$ mode is responsible for applying torques to the MW's disk \citep[e.g][]{Weinberg98a, Gomez13, Gomez17, Laporte18, Laporte18a}. The bottom row illustrates that the radial structure of the dipole is 
defined by the $n$ terms. The level of asymmetry needed to reveal the dynamical friction wake is reached at $l_{max}=m_{max}$=4. 

Distinguishing between the Collective Response and the dynamical friction wake is not straightforward, 
but this exercise illustrates the order of the modes that contribute to these structures. 
Dissecting all the dynamical features in the simulations would require additional analysis of
the coefficients and its time evolution.

\subsection{The shape of the MW's DM halo in comparison to standard halo shapes}\label{sec:power_shape}

In our simulations we start with an idealized spherical 
halo. However, cosmologically, halos are not expected to be spherical \citep[e.g][]{White79, Zemp11, Vera-Ciro11, Chua19, Emami20}. 

In this section, we use the BFE to analyze the shape of the MW--LMC halo shape and compare it 
with MW--like halos proposed in the literature. More details about the BFE of triaxial halos can be found
in Appendix \ref{sec:ideal_halos}.

We compute BFEs for simulations of oblate, prolate, triaxial halos whose main properties are summarized in table 
\ref{tab:ideal_halos}. The triaxial halo model is the mean result for MW--like halos found in the Illustris cosmological simulation \citep{Chua19}. In addition, we include the shape measurement derived by \cite{Law10} from fitting the Sgr stream 
These halos are characterized by the three axes describing the shape of ellipsoids:
$a$ is the major, $b$ the intermediate, and $c$ the minor axis.
We then utilize axis ratios to define the ellipsoid: $s_{\rho}=c/a$, the minor to 
major axis ratio; and $q_{\rho}=b/a$, the intermediate to major axis ratio. The
triaxiality of the halo can be defined in terms of $s$ and $q$ as:

\begin{equation}\label{eq:T}
    T_{\rm{qs}} = \dfrac{1-q_{\rho}^2}{1-s_{\rho}^2}.
\end{equation}

If $T_{\rm{qs}}\geq 0.67$, the halo is considered to be {\it prolate}. If $0.67\geq T_{\rm{qs}} \geq 0.33$ 
the halo is {\it triaxial}, and if $T_{\rm{qs}}\leq 0.33$ the halo is {\it oblate}.

\begin{table}[h]
    \centering
    \begin{tabular}{c c c }
        \hline
         Name &  $s_{\rho}, q_{\rho}$ & Triaxiality ($T_{\rm{qs}}$)\\
         \hline
         \hline
         Oblate  & (0.5, 1) & 0 \\
         Prolate & (0.5, 0.5) &  1 \\
         Triaxial (Chua+19) & (0.2, 0.8) & 0.527 \\
         Law-Majewski (LM+10) & (0.44, 0.97) & 0.17 \\
         \hline
    \end{tabular}
    \caption{Properties of simulated idealized asymmetric halos. All the halos have $\sim 2\times 10^6$ particles, and were initialized with a Hernquist density profile with the 
    same scale length $r_s=40.85$ kpc of the simulated MW. All the halos were built by deforming the spherical Hernquist halo by the listed values of $s_{\rho}$ and $q_{\rho}$ (in density), the minor to major and intermediate to major axis ratio respectively. 
     The triaxial halo values represent the mean shape of MW--like halos found in the Illustris simulation by \cite{Chua19}. Finally, we include the MW halo shape derived in \cite{Law10} (LM+10) based on the Sgr. stream.}
    \label{tab:ideal_halos} 
\end{table}

\begin{figure*}
    \centering
    \includegraphics[scale=0.6]{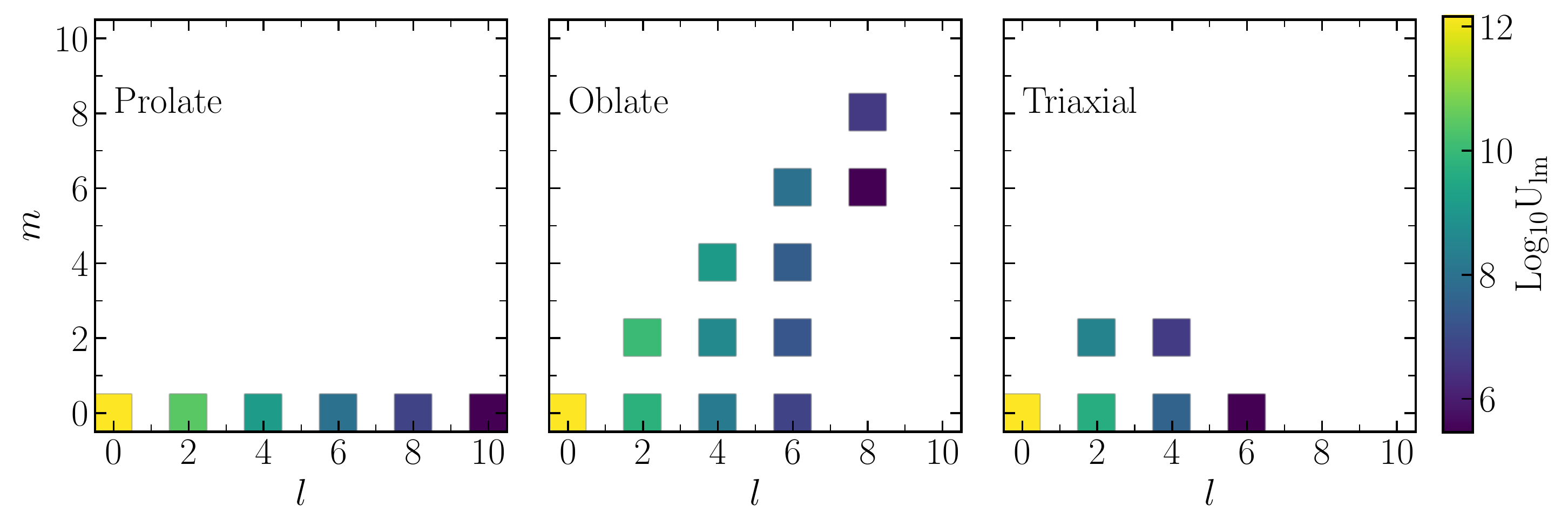}
    \caption{Energy contribution of each $l,m$ mode, summed over all $n$-modes, 
     for idealized and cosmologically motivated aspherical halos: prolate (left), 
     oblate (middle) and triaxial (right).
     The triaxial halo is the average halo shape for halos in the Illustris simulation \citep{Chua19}. Properties of these halos are summarized in in 
     Table \ref{tab:ideal_halos}. 
     In all cases, most of the energy is in
     the $l=0$ mode, comprising $\sim$ 90 \% of the energy, followed
     by the $l=2$ (quadrupole) and the $l=4$ modes.
     Oblate halos consist of higher order $m$ modes, which do not appear in prolate halos.
     For all halos, odd $l$ modes do not appear since
     these modes are not radially symmetric. 
     The existence of odd modes would signify divergence from these standard halos.}

     \label{fig:halos_m_l}
\end{figure*}

Figure \ref{fig:halos_m_l}, summarizes our
findings for the BFE for our simulated oblate, prolate and triaxial halos. Each square in the $l-m$ grid shows the energy corresponding to the sum
of all the $n$-modes with the same $l$ and $m$. That is, $U_{lm}=\sum_{n}
U_{nlm}$. 

Prolate halos only have $m=0$ modes. Corresponding to the `zonal' spherical harmonics
that do not depend on longitude and whose lobes are perpendicular to the
plane of the MW's disk. Oblate halos on the other hand, have a major contribution from the
$|m|=l$ modes. These are represented by the ``sectoral" spherical harmonics, whose lobes are parallel to the MW's disk. The BFE for the triaxial halo is a mixture of that of the prolate and oblate halos. 

The sign of the coefficients can also be used as an indicator of the halo shape. Oblate halos 
have negative values for the $l=2^i_{odd}$ modes. See Appendix ~\ref{sec:ideal_halos} for further details. 
Note that, in all cases, as the axis ratios increase, 
higher-order terms will be needed to properly characterize the structure. The main characteristic of these idealized halos
is that the odd $l$ and $m$ modes do not appear in the expansion, as they are not radially symmetric. As such, the existence of odd modes would signify divergence from these standard halos.

With this intuition in mind, we can now move forward and analyze how a more complicated,
asymmetric structure, such as the DM dynamical friction wake produced by the LMC, will manifest in the BFE space.

\begin{figure}[h]
    \centering
    \includegraphics[scale=0.6]{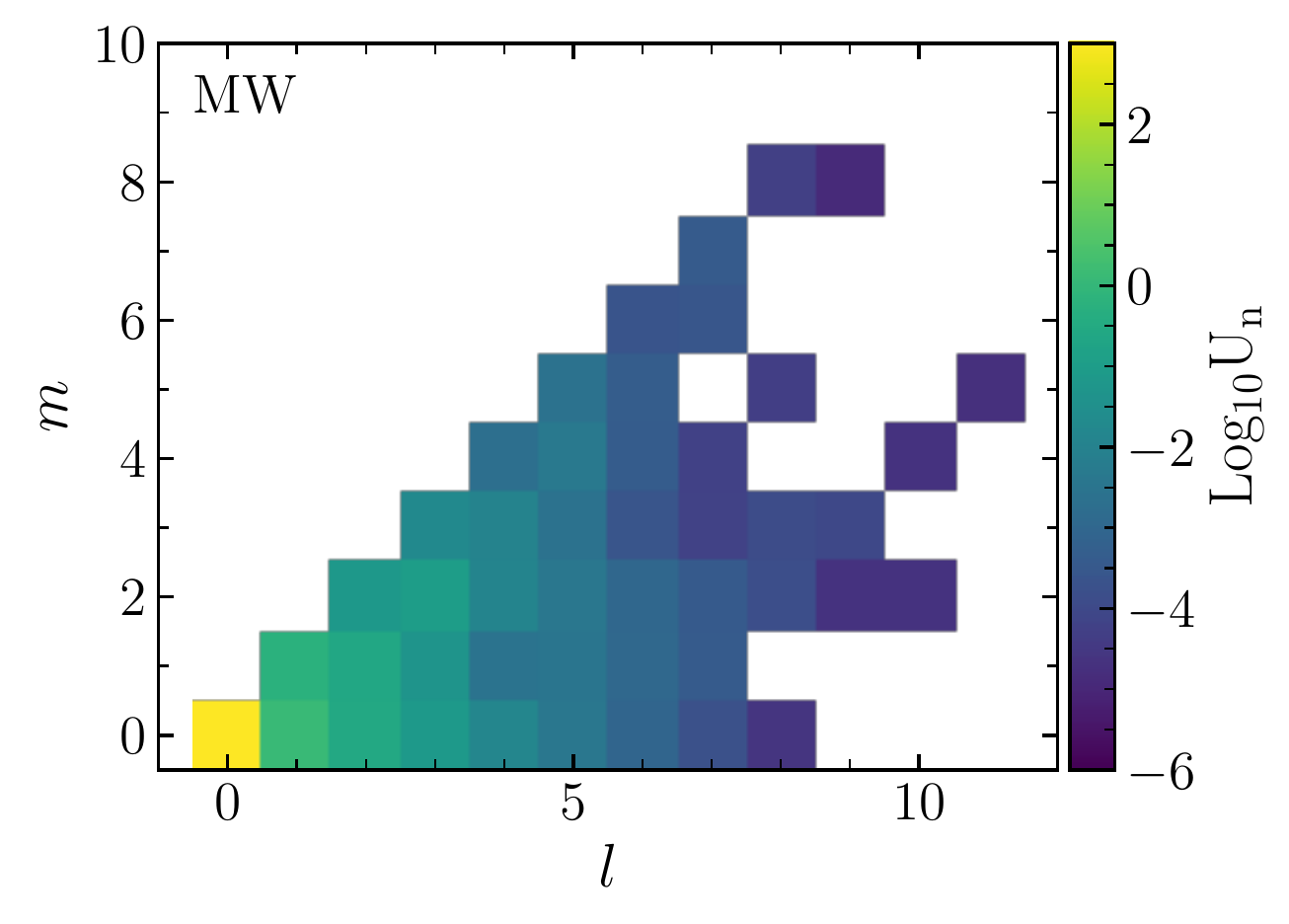}
    \caption{ Similar to Figure~\ref{fig:halos_m_l}, but for the present day MW DM distribution after the passage of the LMC. Results are shown for the fiducial model (simulation \#7). Bound LMC particles are not included. 
 The largest contributions to the energy are from the
$l=0, 1, 2, 3, 4$ terms. This is in contrast to expectations for ideal halos, which do not exhibit odd terms (Figure~\ref{fig:halos_m_l}). These trends are consistent among all the LMC
mass models (see Section \ref{sec:lmc_mass}) and MW models.}
    \label{fig:mw_m_l}
\end{figure}

\begin{figure}[h]
    \centering
    \includegraphics[scale=0.55]{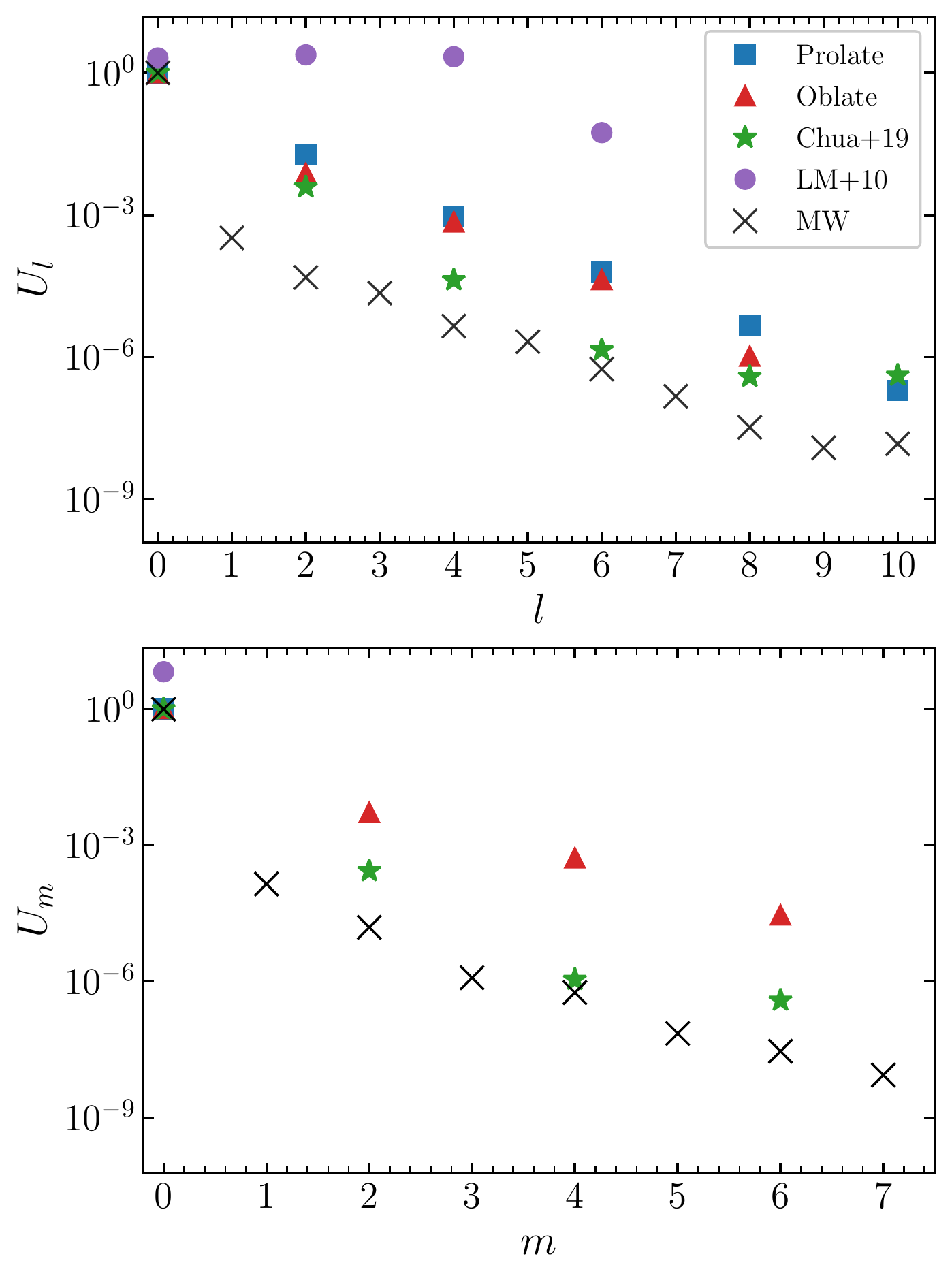}
    \caption{Energy in the $l$ and $m$ modes for the MW BFE (x markers), compared to the prolate (blue squares), and oblate (red triangles). The mean triaxial halos found in Illustris for MW--like galaxies are shown as green stars \citep[][Chua+19]{Chua19}. The MW's DM halo derived by \cite{Law10} (LM+10) is shown with purple circles. Overall, the energy in the $l$ and $m$ modes is higher in the prolate, oblate, and triaxial halos than seen in the simulated MW,
    although these models were chosen to be somewhat extreme. 
    Regardless, the existence of odd terms in the BFE for the MW halo response is inconsistent with these axi-symmetric models. 
    For the Chua+19 halo, the energy in the $l=4, 6$ modes is similar to the MW, suggesting it would be challenging to decompose the MW halo response from the cosmological assembly history of the MW. The LM+10 halo has very high values of energy specially between $l=2-6$, very distinct to those found for the MW halo in presence of the LMC. But we note that our analysis here does not include the LMC BFE. These results show how BFE can be used to explore the shape of DM halos, and how the axis ratios of the halos are proportional to $U_l$ and $U_m$.}
    \label{fig:mw_m_l_2}
\end{figure}

In Figure \ref{fig:mw_m_l}, we plot the energy contribution of each $l,m$ mode, summed over all $n$-modes, for the MW BFE (section~\ref{sec:MWDMBFE}). 

The MW halo responds to the presence of the LMC, but the LMC BFE is not included in this analysis.

Comparing Figure \ref{fig:halos_m_l} with Figure \ref{fig:mw_m_l}, we find   
that the most energetic coefficients in the perturbed MW halo
are distinct from those observed in idealized or cosmologically motivated halos.

In the MW BFE, the monopole term is the dominant mode, followed by the $l=1$ mode, as also found by \cite{tamfal20} and in the kinematics of the halo particle by \cite{Cunningham20}. This
corresponds to the dipole response, which
is induced by the barycenter motion (see \ref{sec:mw_com}). In other words, we find strong contributions from 
radially asymmetric modes (odd $l$-modes), which are completely absent in
the idealized halos shown in Figure~\ref{fig:halos_m_l}. The LMC's impact on the MW halo is not expected to be distorted in a manner consistent with a prolate, oblate or triaxial DM distribution.

In Figure~\ref{fig:mw_m_l_2}, we plot the energy as a function of $l$ and $m$ modes, respectively, to better illustrate the difference between our simulated MW and idealized halos. 

Overall, the energy of the even $l$ and $m$ terms is higher for 
the chosen oblate, prolate, and triaxial halos than in the MW, although these idealized halos were chosen to be somewhat extreme.

There is pronounced difference between the MW BFE and the LM+10 halo, particularly between $l=2-6$.   
This implies that the MW halo response to the LMC alone is not sufficient to generate the
deformations measured by \citep{Law10}.  Also, 
it is still unknown if the response to the passage of the LMC of an initial triaxial MW 
can explain the resulting halo shape of \cite{Law10}. In addition, the direct torque from
the LMC on the Sgr. stream \citep{Vera-Ciro13, Gomez15} has to be taken into account to 
properly interpret the measured shape of \cite{Law10}, as demonstrated in 
\cite{Vasiliev20}. Note that we have not included the LMC BFE in this analysis.

Evidence of the direct torque from the LMC in stellar streams has been detected in the 
Orphan stream \citep{Erkal18b}, where a highly flattened, prolate

MW DM halo was preferred in order to reproduce the morphology of the stream using N-body
simulations.
However, our analysis indicates that the halo response cannot be captured by such static,
axi-symmetric models.
This reinforces the point that efforts to measure halo shapes using streams must to take 
into account both the direct torque from the LMC as well as the halo response, as 
recently shown in \cite{Vasiliev20} for the Sag. Stream.

Interestingly, the average halo from Illustris measured by 
\cite{Chua19} (green stars) is triaxial with power in the $l=4,6$ terms consistent with 
that of our simulated MW. This indicates that it may be challenging to disentangle the 
impact of the LMC from the structure induced by the cosmological assembly history of the 
MW. The quadrupole terms ($l$ and $m$) may serve as a discriminant of the halo 
triaxiality and effect from the LMC. A boost in the quadrupole relative to the higher 
order terms may signal underlying triaxiality. The slope of the energy as a function of 
$l$ and $m$ is correlated with the degree of triaxiality ($T$), where the slope flattens 
as the axis ratios decrease. Also note that the power increases for more highly-triaxial 
halos, as shown by the black stars.

This analysis illustrates that it is not straightforward to disentangle the halo response
due to a passage of a massive satellite from the halo's intrinsic/initial shape in
simulations. A full diagnostic will require 
a proper assessment of the covariances of the coefficients to quantify the halo response 
similar to \cite{Ghil02, Darling2019}.

\subsection{Quantifying the barycenter motion with the MW's halo shape: 
Inner vs. Outer}\label{sec:mw_com}

We now turn our attention towards understanding the origin of the MW's shape in our 
simulations. The present-day shape of the density field of the MW's DM halo is sensitive 
to the recent infall of the LMC. The following physical processes take place over the 
last 2 Gyrs:

1) \textit{Density asymmetries} occur owing to the presence of the LMC (bound DM halo and unbound DM debris) and the dynamical friction wake; 

2) \textit{Barycenter motion:} Direct torques from the LMC and dynamical friction wake move the barycenter of the MW--LMC system. Since the dynamical times of the outer halo of the MW are longer than those of the inner halo, the inner halo will respond faster to the presence of the LMC. Consequently, the COM of the inner halo shifts with respect to the outer halo as a function of time. As a result, the barycenter of the MW--LMC system not only changes over time, but also as a function of Galactocentric distance.
Because of the barycenter displacement, the density in the northern hemisphere is larger than that in the south, at a given Galactocentric radius ($>$ 30 kpc, measured from the disk COM). The barycenter motion is a manifestation of the Collective Response.

As a consequence of both processes, observers in the MW's disk should observe an apparent reflex motion, as has been recently measured \citep{Petersen21, Erkal20c}. 

In this section, we quantify the contribution of the density asymmetries and the barycenter motion to the shape of the MW's DM halo density distribution at the present day.

\begin{figure}
    \centering
    \includegraphics[scale=0.6]{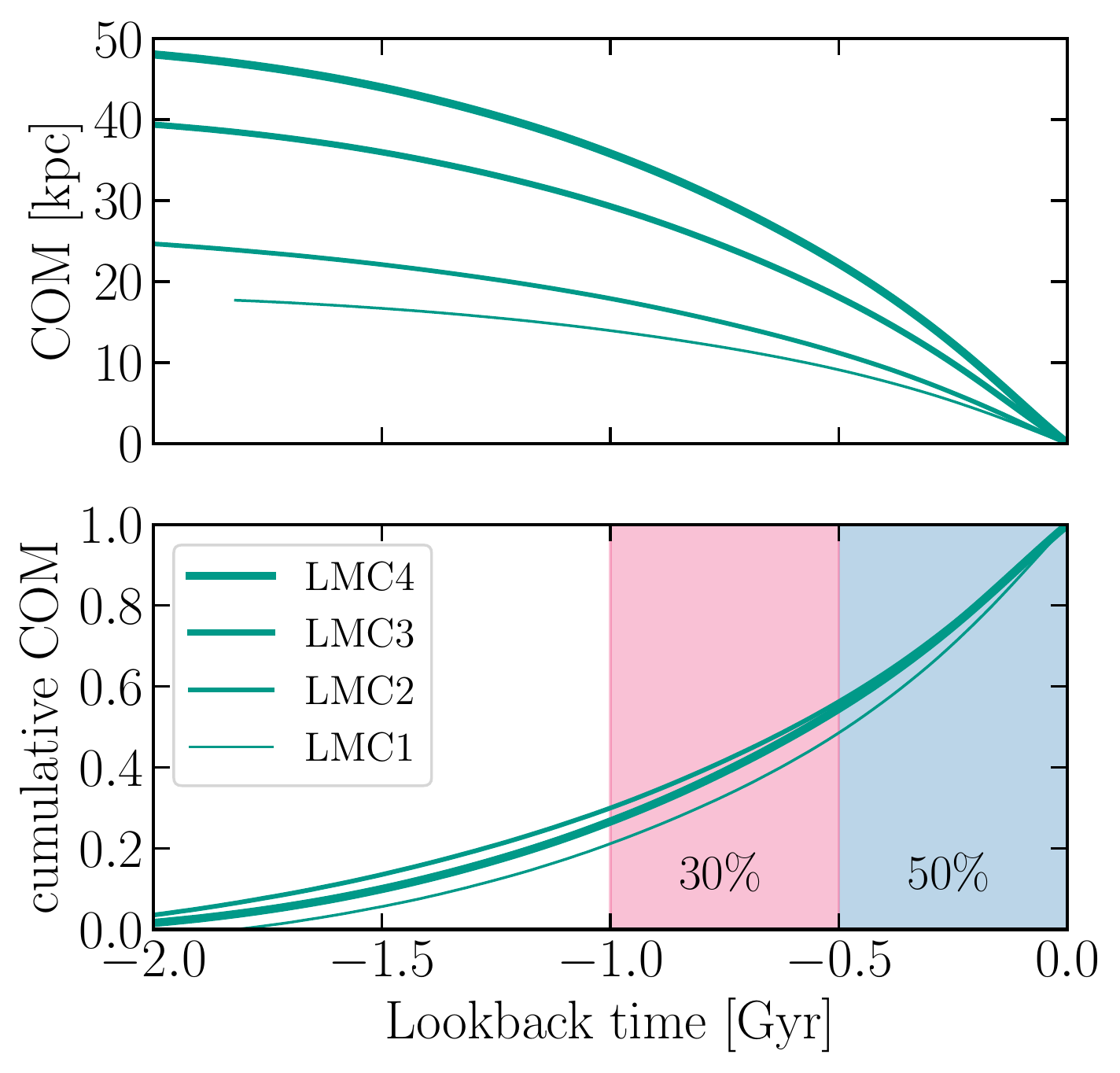}
    \caption{Time evolution of the position of the COM of the MW's disk (Gyr ago). 
    \textit{Top panel:} Distance to the COM of the MW's disk
    relative to the present-day location for the anisotropic MW models (sims 5-8). Line thickness indicates the LMC's mass at infall, where LMC3 is the fiducial model. Over the last 2 Gyrs, the MW disk moves from 20 to 50 kpc, depending on the LMC mass. This is similar to findings in \cite{Gomez15}, but accounts for the halo response and tidal deformation of the LMC. \textit{Bottom panel:} Cumulative COM motion of the MW disk as a function of lookback time. 80$\%$ of the 
    MW's COM movement takes place in the last Gyr and 50$\%$ in the last 0.5 Gyr.}
    \label{fig:mw_com}
\end{figure}

As stated earlier, the COM of the MW's disk has moved
as a function of time owing to the infall of the LMC \citep[e.g][]{Gomez15, petersen20}, but most of this motion has occurred recently. 
This is illustrated in the top panel of Figure~\ref{fig:mw_com}, which shows the change in COM position of the MW's disk with respect to its present day location, as a function of lookback time. 
The bottom panel shows the cumulative COM motion 
as a function of lookback time. From both of these plots we conclude that the COM of the MW's disk 
moves more than 50\% of the total displacement in the last 0.5 Gyrs, which corresponds to a distance of 20 kpc 
for the fiducial LMC mass model (LMC3), in agreement with previous studies \citep{Gomez15, Erkal18b, petersen20}. 
The rapid motion of the disk COM over such a short time-scale implies that the impact on specific objects in orbit about the MW disk will depend on their radius and orbital period.

The entire halo of the MW will not respond as a rigid body to the COM motion of the disk.
Instead, the motion of the halo will vary as a function of radius. The amplitude of the 
corresponding barycenter motion of the MW's halo depends on the infall mass 
of the LMC, its orbit, and MW's halo response.

The inner regions of the halo will move with the disk, but the outer
regions will lag, and hence are displaced from the disk COM. Our goal is
to distinguish the radius of this inner halo and characterize the relative COM motion of the outer halo.

Focusing on the present day halo, we quantify the inner
halo as those regions that follows the COM displacement of the disk. We compute the COM of isodensity contours of the halo using the BFE, as a function of distance. We further distinguish the effects of the LMC and the LMC's DM debris in these calculations by 
comparing results using the BFE with or without those particles (see section \ref{sec:mwlmc_bfe}).

\begin{figure*}
    \centering
    \includegraphics[scale=0.6]{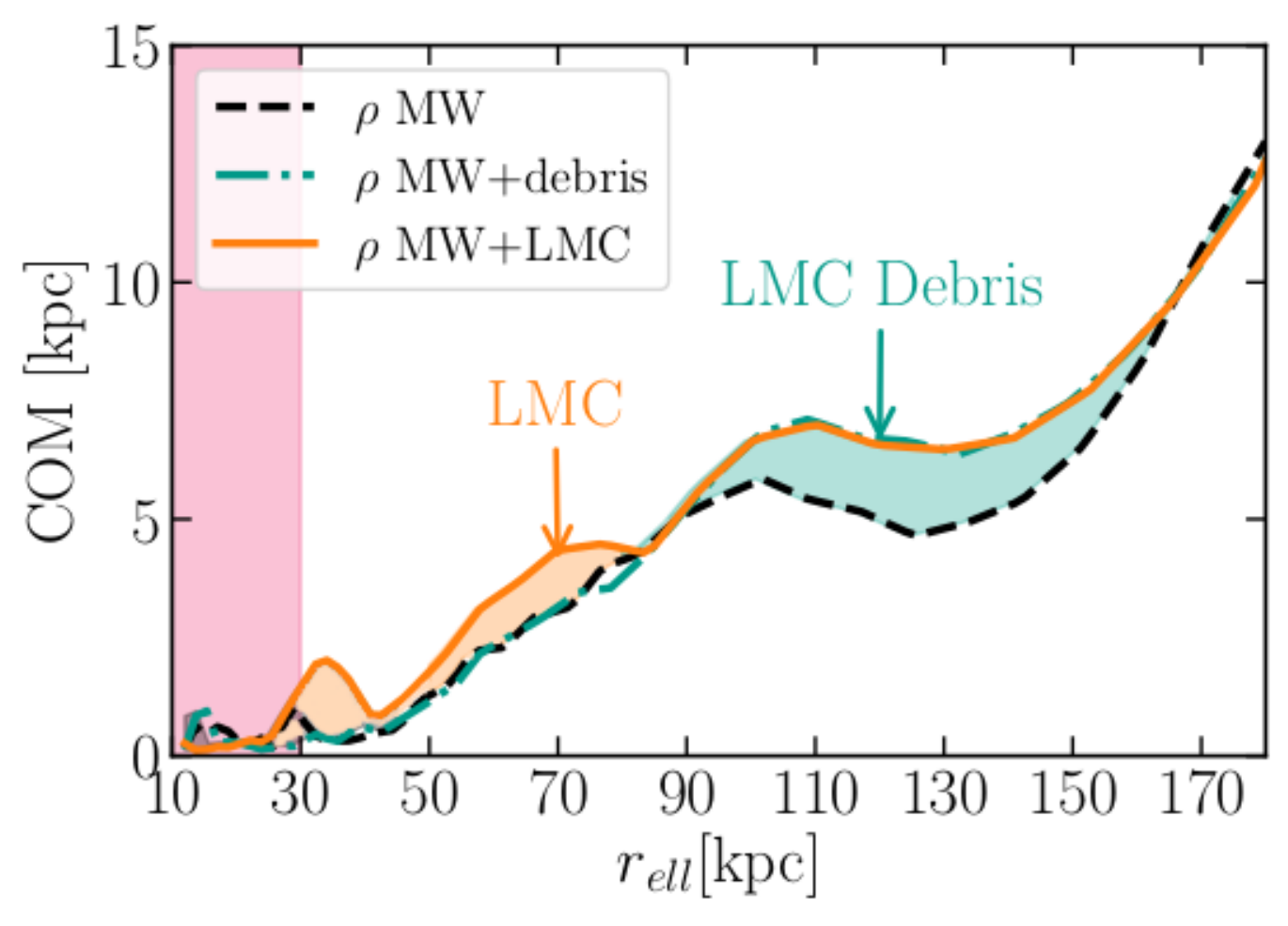}
    \includegraphics[scale=0.6]{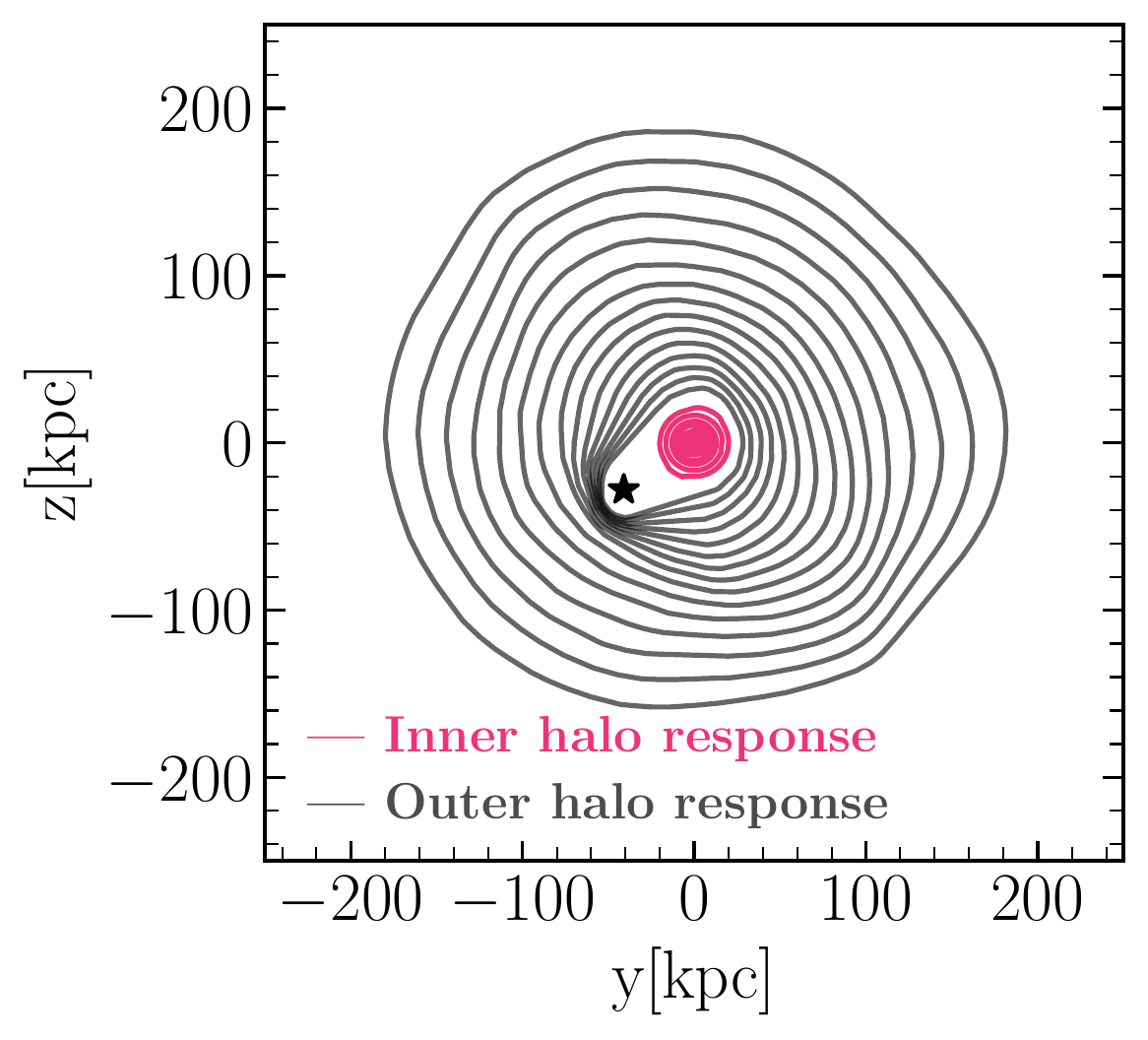}
    \caption{\textit{Left panel:} COM of the MW (black dashed), MW--Debris (blue dashed-dotted), and the full combined MW--LMC system (solid orange)
    as a function of distance. Distance here refers to the ellipsoidal radius of the contour ($r_{ell}$). Density contours were computed using
     the corresponding BFE, where the COM refers to the COM of the corresponding shell (10 kpc thickness). 
    We define an inner halo ($<$30 kpc) that responds coherently to the motion of the disk COM (pink shaded region; see right panel). The COM of the outer halo ($>$30 kpc) is displaced from the disk COM owing multiple effects. The halo response impacts the COM at radii larger than 30 kpc (black dashed line) and is the dominant effect. The LMC bound particles impact the DM mass distribution, and thus the COM, from 50-90 kpc (orange region). Finally, the LMC's DM debris impacts the DM mass distribution at $>$90 kpc (shaded blue region).
     \textit{Right panel:} Density contours showing the shape of
    the inner vs. the outer halo. At 30 kpc, the halo density contours are maximally impacted by the LMC, defining the maximal radius for the inner halo.}
    \label{fig:contours_com}
\end{figure*}

The left panel of Figure \ref{fig:contours_com} shows the resulting COM position of isodensity contours within the present day halo of the MW, as a function of distance, $r_{ell}$, which is the ellipsoid radius:

\begin{equation}\label{eq:rell}
    r_{ell}^2 = x^2 + \dfrac{y^2}{q^2} + \dfrac{z^2}{s^2},
\end{equation}

In the right panel of Figure~\ref{fig:contours_com}, we provide a visualisation of isodensity contours in the combined MW--LMC system.

The left panel of Figure~\ref{fig:contours_com} illustrates that
the COM of the MW halo is 
not coincident with that of the MW disk at all radii. 
Within $r_{ell}\sim$30 kpc, the MW halo moves coherently 
with the MW's disk COM (pink regions in both left and right panels). 

However, beyond $r_{ell}\sim$30 kpc, isodensity contours are distorted by the LMC (right panel).
Consequently, the halo COM position begins to diverge from that of the inner halo. These outer radii are affected by different factors. When MW particles alone are considered (black dashed lines), the offset reflects 
the halo response, including the barycenter motion. 
The inclusion of bound LMC particles increases the displacement from $r_{ell} =$ 30-40 kpc and 50-90 kpc (orange shaded region). The inclusion of unbound debris particles from the LMC augments the COM displacement at $r_{ell}$ larger than 90 kpc (blue shaded region). The orange line denotes the COM displacement for the entire system combined. 
Although we have computed the COM within isodensity shells, the measured displacement should also reflect the behavior of the orbital barycenter of the MW--LMC system.

\begin{figure}
    \centering
    \includegraphics[scale=0.6]{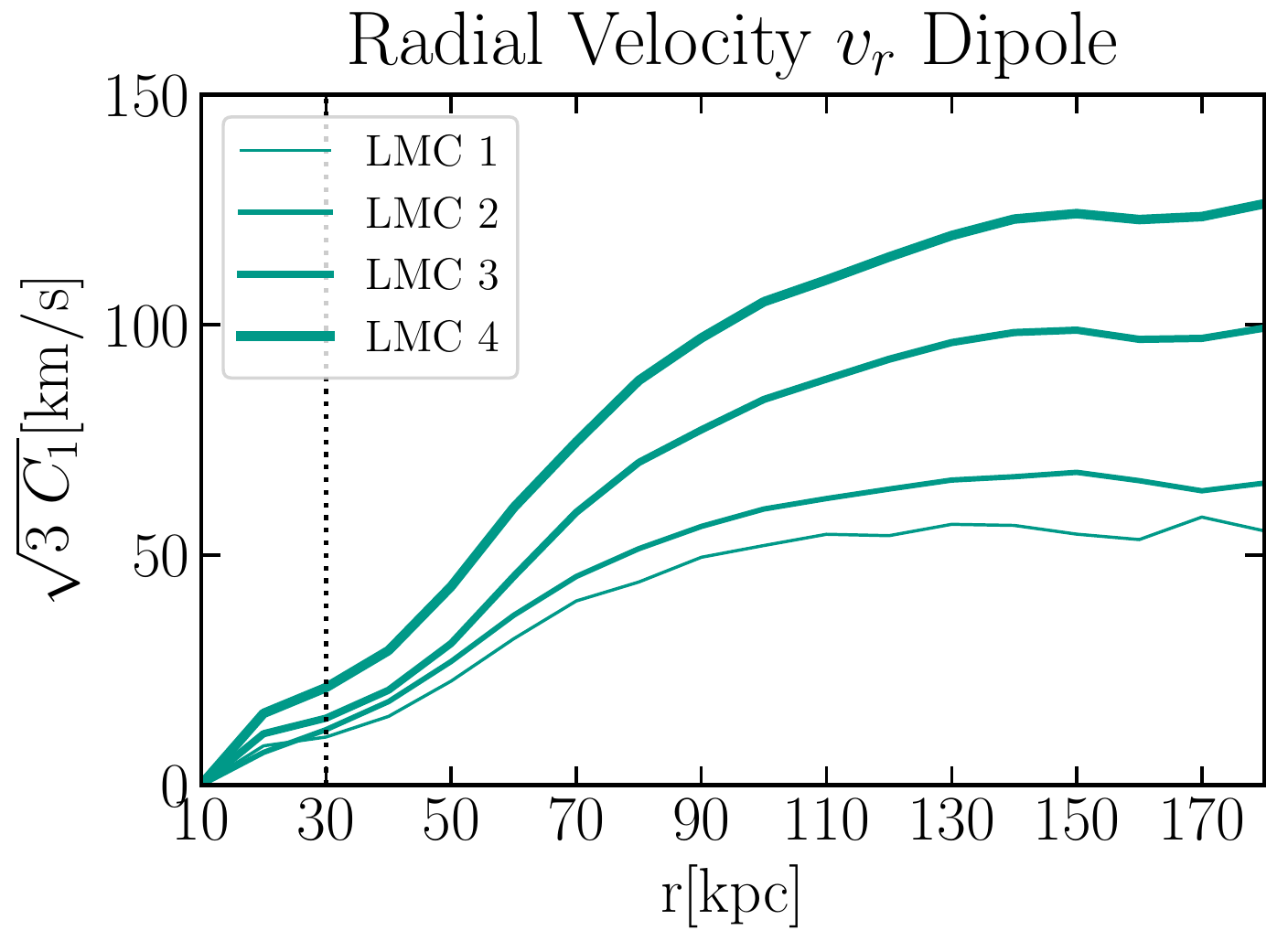}
    \caption{Amplitude of the $\ell=1$ modes (i.e., the dipole modes) in the spherical harmonic expansion of the radial velocity field, for the simulations with an anisotropic MW model (simulations 5-8 in Table~\ref{tab:sims_summary}). We compute the spherical harmonic expansion of the radial velocity field as in \cite{Cunningham20}, using the software package healpy \citep{zonca19}, based on the Healpix scheme \citep{gorski05}. Power was computed using only the MW’s DM halo particles centered on the MW’s disk, where the speeds are impacted by both the halo response and the LMC. We find that the velocity of the dipole begins to increase at$\sim$30 kpc (dashed line), and then increases substantially after 50 kpc for all the LMC mass models, consistent with our findings. This signal is generated by the COM motion of the inner halo ($< 30$ kpc), as illustrated in Figure~\ref{fig:contours_com}.}
    \label{fig:vr_dipole}
\end{figure}

The barycenter motion manifests to observers as a reflex motion in velocities \cite{Petersen21, Erkal20c}. In particular, it induces a dipole pattern in the radial velocities and net vertical velocity $v_z$ of stars/DM in the outer halo ($>30$ kpc). The northern hemisphere appears to be redshifted
while the southern is blueshifted. We quantify this dipole pattern by computing a spherical harmonic expansion; the magnitude of the contribution from the $\ell=1$ modes (i.e., the dipole modes), as a function of radius, is shown in Figure \ref{fig:vr_dipole}. The power in the dipole
increases steeply after 30 kpc, consistent with our findings with the COM analysis. 

While the barycenter motion and the reflex motion can be estimated to first order using simple sourcing
of rigid potentials as done in earlier works \citep[e.g:][]{Gomez15, Erkal18b}, these models
cannot capture the higher order variations as a function of distance which arise from the complex
marriage between contributions from tidal stripping and response of a halo a sinking perturber,
which give rise to different behaviors in the inner and outer halo of the Galaxy.

\section{Discussion}

Here we discuss: 1) The possible observable effects of the MW's barycenter motion in Section~\ref{sec:reflex_motion_observables}; 2) the impact of uncertainties in LMC mass and the velocity anisotropy profile of the MW's halo to our results (Section~\ref{sec:lmc_mass}); 3) the consequences of our results for the present-day satellites of the LMC (Section~\ref{sec:lmc_satellites}); and 4) the applications of BFEs in astrophysics in Section~\ref{sec:bfe_applications}. 

\subsection{Possible observable effects of the MW's barycenter motion}\label{sec:reflex_motion_observables}

Here we outline specific cases where the barycenter motion of the MW must be accounted for. The barycenter motion manifests as {\it both} a spatial displacement and a reflex velocity in the phase space properties of stars in the outer halo ($>$ 30 kpc), relative to an observer in the disk. The barycenter motion also varies as a function of radius and LMC infall mass (see Figures~\ref{fig:mw_com} and  
\ref{fig:vr_dipole}).

\textbf{Models of the Sagittarius stream:} Models of the Sagittarius stream must include the 
gravitational influence of the LMC. Early works by \cite{Law10, Vera-Ciro13} showed that the LMC can have an important contribution 
to the shape of the potential. \cite{Gomez15} showed that the LMC is perturbing the orbit
of Sagittarius and hence its tidal debris will not lie in the plane of the orbit. 

Based on our analysis we expect that stars within the Sagittarius stream will exhibit the barycenter motion of the inner halo (both in terms of the displacement and reflex motion). The Sagittarius stream is unique in that it exists in both the inner, intermediate and outer halos, and thus may show interesting phase space properties at distances larger than 30 kpc vs. at smaller radii.

\textbf{Stellar streams:} We expect the above to be true of any stream that crosses the inner and outer halos. Note that for any stream outside of 30 kpc, the COM of the MW+LMC system is changing as a function of radius (see Figure~\ref{fig:contours_com}).

\textbf{Mass estimates of the Milky Way:} Mass estimates of the MW that use the kinematics of outer halo tracers,
e.g \cite{Watkins10}, will be impacted by the barycenter motion. The reflex motion has already been shown to impact mass estimates derived from the phase space properties of halo stars using, e.g., Jeans modeling \citep{Erkal20b, Deason21}. We argue here that the barycenter displacement (see Figure~\ref{fig:contours_com}) must also be accounted for to accurately assess the errors in such measurements. Note that the halo response to the LMC (DM dynamical friction wake and Collective Response) as well as the LMC debris will result in asymmetric perturbations to the kinematics on the sky (G19) that will impact mass measurements differently depending on the angular location of the tracer on the sky. As shown in G19, their Figure 16, the anisotropy parameter varies across the sky \citep[see also][]{Erkal20b}. 

\textbf{Radial velocities of the outer halo}: The kinematics of the outer halo are 
predicted to exhibit a dipole in 
radial velocities \citep{garavito-camargo19a, petersen20, Cunningham20, Petersen21, Erkal20c}. Where the northern
galactic hemisphere is moving away, while the southern hemisphere is moving towards us. Here we further stress that the dipole should vary as a function of Galactocentric radius and should be maximized at large distances (110-150 kpc, depending on the mass of the LMC; see Figure~\ref{fig:vr_dipole}).

\subsection{Impact of the mass of the LMC and anisotropy of the MW halo}\label{sec:lmc_mass}

Here we explore how our results scale as a 
as a function of LMC mass and anisotropy of the MW halo. All results until now have largely focused 
on our fiducial model (sim \#7). We have computed the BFE for 4 different LMC models [$8-25 \times 10^{11}$ M$_{\odot}$]. 
In two different MW halo models (isotropic and radially biased), as outlined in Table \ref{tab:sims_summary}.  
Our first goal is to understand the changes in the BFE, and thus the structure of the halo response.

\begin{figure}
    \centering
    \includegraphics[scale=0.6]{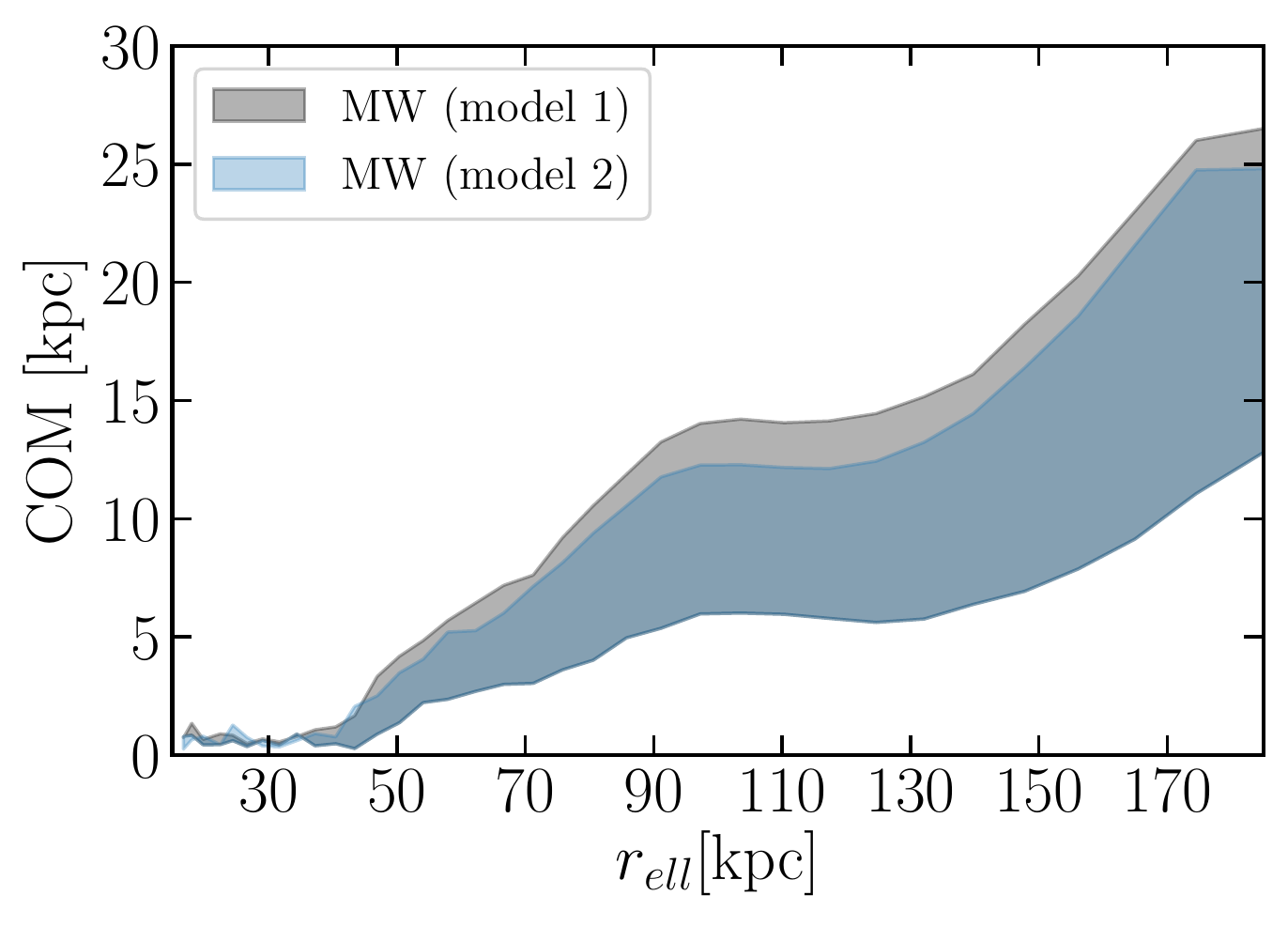}
    \caption{The COM of the MW's DM distribution is computed as a function of $r_{ell}$ using eight BFEs that span both MW models (isotropic/model 1 or radially biased/model 2 anisotropy profiles; grey and blue, respectively) and all LMC mass models ($1-4$, denoted by the width of the shaded regions). There is little difference found between models 1 or 2. In the inner halo ($<30-50$ kpc) the COM motion is consistent for all models, while 
    the impact of the LMC mass is stronger in the outer regions of the halo.}
    \label{fig:com_lmc_mass}
\end{figure}

We start by studying the effect of the LMC mass and MW model on the displacement of the COM of the inner halo with respect to the outer regions, which manifests as a reflex motion. This is shown in 
Figure \ref{fig:com_lmc_mass}. Both MW models lead to similar results (blue vs gray shaded regions). The width of the shaded line represents results from all four LMC models.
The assumed LMC mass affects the amplitude of the COM displacement only at radii $>$ 30-50 kpc. 
The larger impact of the assumed LMC mass on the COM displacement of the outer regions of the MW halo is expected as the bound mass of the LMC was larger at infall vs. at present day (see Figure \ref{fig:LMC_bound_mass}). This is consistent with the results of the radial velocity dipole in Figure~\ref{fig:vr_dipole}. 

In our study we identify a Galactocentric radius of 30 kpc as the boundary between the inner and outer halo, defined as where both the COM of the halo is no longer coincident with the disk center and where the velocity dipole is appreciable (Figures \ref{fig:contours_com} and \ref{fig:vr_dipole}). We find that this value of 30 kpc is independent of LMC infall mass. This is because although the LMC models have different infall masses, they are all constrained to have similar inner mass profiles owing to constraints from the observed rotation curve (\S~\ref{sec:sims_review} and G19). 

\begin{figure}
    \centering
    \includegraphics[scale=0.6]{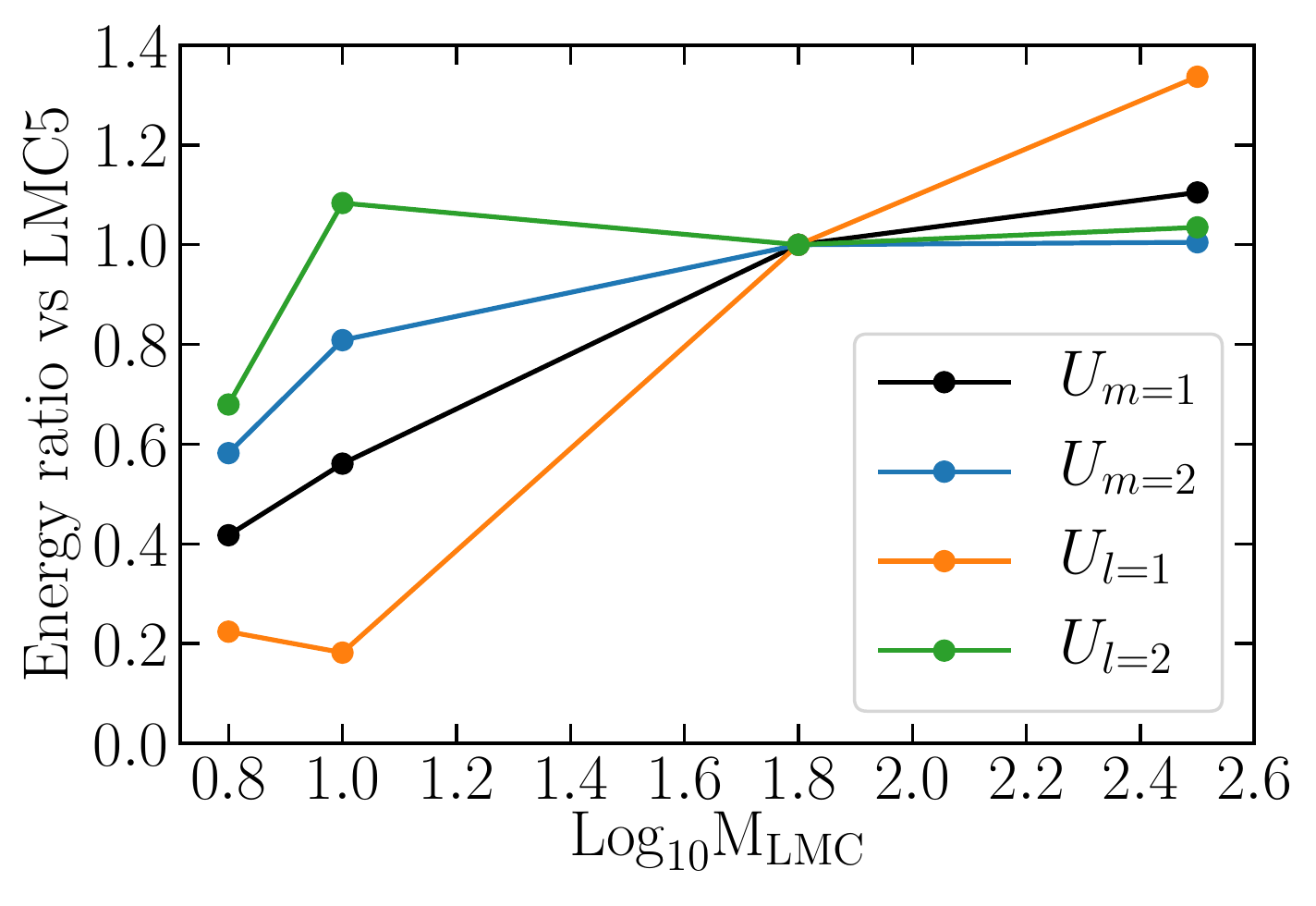}
    \caption{ Gravitational energy of the $m=1, 2$ \& $l=1,2$ modes with respect to the monopole as a function of LMC mass for the MW BFE. The $l=1, m=1$ mode contains information about the dipole induced by the LMC, and hence of the barycenter motion of the MW. The power of these dipoles modes increase as a function of LMC mass, this is in agreement with the increasing amplitude of the displacement of MW's COM as a function of the LMC mass. The quadruple nodes do not depend as strongly on the LMC mass.}
    \label{fig:Um1}
\end{figure}

We now study the total gravitational energy (U) in the coefficients 
as a function of LMC mass. We use the BFEs of the MW to compute the gravitational energy in different modes as a function of LMC mass, as illustrated in Figure~\ref{fig:Um1}. We find that the $l=1, m=1$ modes are the most impacted by LMC mass. These modes represent the amplitude of the barycenter motion (see middle column of Figure~\ref{fig:wake_l_terms}). The figure shows that the ratio of the energy of the $l=1, m=1$, relative to the fiducial LMC mass model, increases as a function of LMC mass. The increase in amplitude in these terms is not linear, in particular beyond our fiducial LMC3 model, the increase is not as dramatic. The minimum value of the dipole is 20\% of that of the fiducial model. For the quadrupole modes, $m=2, l=2$, the change in amplitude with mass is less pronounced.

\subsection{Bound satellites of the LMC}\label{sec:lmc_satellites}

Using the BFE of the simulated LMC we can estimate which of the observed MW satellite galaxies
are bound to the LMC at the present time, accounting for the 
gravitational potential of the dynamical friction wake, the Collective Response, the LMC and the LMC's DM debris.
We define whether a satellite is bound or unbound using the criteria outlined in \ref{sec:lmc_body}.

\begin{figure} 
    \centering
    \includegraphics[scale=0.6]{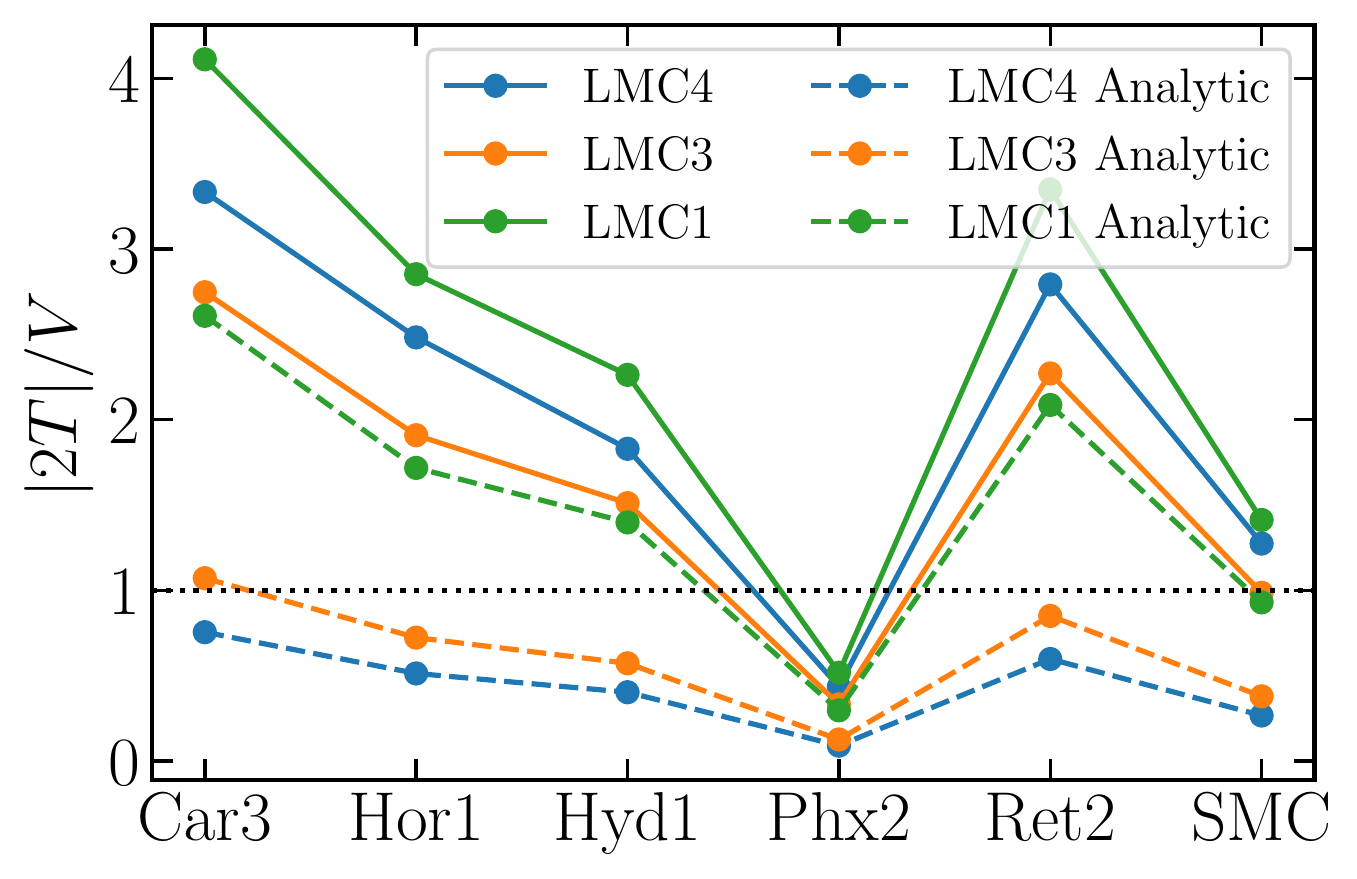}
    \caption{Ratio of the kinetic to potential energy ($|2T|/V$) of proposed LMC satellite galaxies from \citep{Patel20} as a function of LMC infall mass. Satellites are bound if the ratio is less than 1.  The ratio is computed using the BFE for the simulated LMC (solid lines) and for an analytic, rigid LMC halo that represents the LMC at infall (dashed lines). Using the BFE, only Phoenix2 is currently bound to the LMC. The SMC is barely bound only if the infall mass of the LMC is 1.8 $\times 10^{11}$ M$_{\odot}$ (i.e., our fiducial model LMC3). In contrast to the analytic models, which indicate that the strength of the binding energy increases with LMC mass, we find that LMC4 is unable to capture satellites other than Phx2. 
    This results from the lower concentration of this halo required to match the observed LMC rotation curve, making it more susceptible to mass loss from MW tides. 
    Ignoring the mass loss of the LMC will thus overestimate the number of presently bound satellites of the LMC, but these galaxies may still have been bound to the LMC at some point in the past.}
    \label{fig:lmc_satellites}
\end{figure}

The solid lines in Figure \ref{fig:lmc_satellites} illustrate our main findings for 
six of the proposed Magellanic satellites in the literature \citep[e.g,][]{Jethwa16, Sales17, Erkal20b,Patel20}, 
as a function of LMC mass. 

Note that we have ignored the explicit gravitational potential of the SMC, which is instead implicitly accounted for in the range of LMC masses explored.
The dashed lines show the corresponding analytic 
calculation. 
assuming a rigid spherical LMC with the same mass and scale lengths as those used in the 
initial conditions for the G19 simulations (i.e no mass loss over time). We find 
that, at the present time, the only bound ultra-faint dwarf satellite of the LMC is Phoenix 2.  

The SMC is barely bound to the LMC only in our fiducial model, LMC3. Increasing the 
mass of the LMC (e.g. LMC4) does not improve the LMC's ability to hold on to the SMC. 
This is because the LMC models were constructed to match the rotation curve, which 
required decreasing the concentration of the LMC halo as the infall mass increased. 
In the simulations, this makes it easier to unbind material (see section~\ref{sec:LMC_DM_debris}), which is not captured in the analytic models.

\begin{figure*}
    \centering
    \includegraphics[scale=0.65]{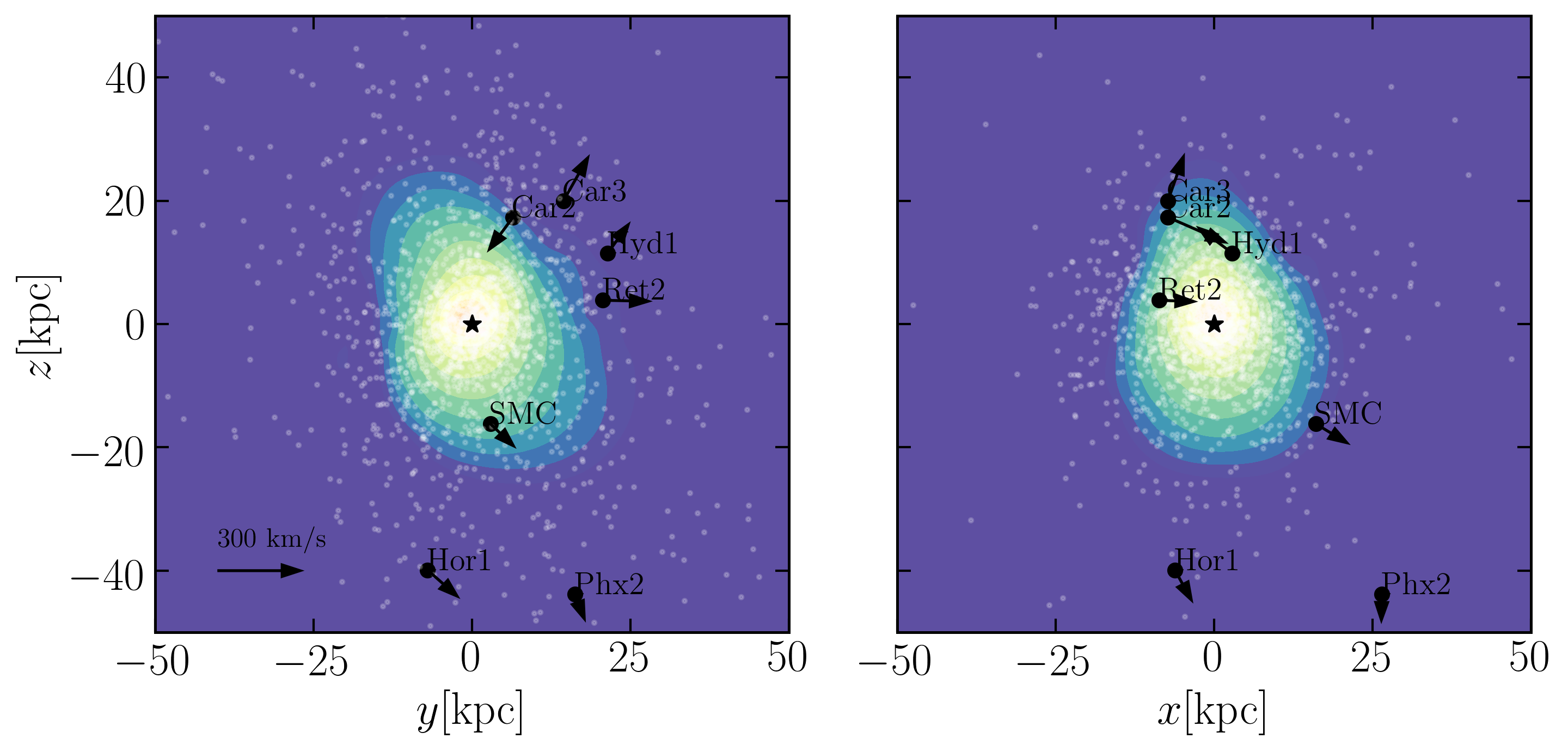}
    \caption{Projected positions and velocity vectors of the most bound satellites of the LMC, plotted in the LMC's rest frame. The sizes of the arrows illustrate the magnitudes of the velocities. All six satellites are moving away from the LMC. Phoenix 2 (Phx2), the most bound satellite, has the lowest relative radial velocity ($v_r\sim 6$ km/s). All satellites are currently located outside the distribution of bound DM particles (colored contours), but consistent with the distribution of LMC DM debris (shown with white dots).}
    \label{fig:LMC_satellites_rho}
\end{figure*}

We show the present day location and velocity vectors of the six most bound satellites 
of the LMC, in the LMC's reference frame, in Figure \ref{fig:LMC_satellites_rho}. The locations of the proposed Magellanic satellites are largely outside the currently bound DM distribution of the LMC, but are coincident with the distribution of the LMC's DM Debris \citep{Sales11, Kallivayalil18}. This suggests that some of these satellites could been bound to the LMC in the past \citep{Patel20}.  

Interestingly, all six satellites are currently 
moving away from the LMC.
Phoenix 2 has the lowest relative radial velocity, $v_r\sim 6$ km/s. \citet{Patel20} conclude that Phoenix 2 is a 
recently captured satellite and \cite{Jerjen18} identified the presence of tidal arms, possibly from the interaction with the LMC.
Our results are consistent with these findings, but new orbital calculations are needed that account for the LMC's time evolving potential.

We conclude that analytic models of the LMC that ignore mass loss will overestimate the 
number of satellites presently bound to the LMC. However, this does not mean that these 
satellites were not bound to the LMC in the past and hence a time-dependent analysis remains to be done.

\subsection{BFE applications in astrophysics}\label{sec:bfe_applications}

    Within the context of gravitational dynamics, BFEs are functional expansions that are used to compactly represent the density and potential fields of collisionless gravitational systems. BFEs could be used within gravity solvers to improve the performance of galactic N-body simulations \citep[e.g.,][]{Clutton-Brock73, Weinberg89, Hernquist92, Weinberg94, Weinberg95} or to analyze the output from N-body simulations  \citep[e.g.,][]{Lowing11}. Hence there is plenty of room for using BFEs in various astrophysical regimes, including: quantifying DM halo shapes, quantifying the morphology of galactic disks and the bar \citep[e.g.,][]{H-B05, Petersen19-disk}, building time-dependent potentials for galaxies and galaxy clusters,  computing fast orbit integration of tracers in time-dependent potentials (Garavito-Camargo, et al. in prep), and building strong gravitational lensing maps. 

    There are two important aspects that one needs to take into account when using BFEs: 1) choosing 
    the right BFE; and 2) choosing the length of the expansion (this will be discussed in detail in section \ref{sec:bfe_length}). 
    Choosing the appropriate BFE depends on the dynamical state of the system to be studied. If the system does not drastically change in morphology within a dynamical time, one can use the BFE that is built on an analytical basis \citep[e.g., this study, ][]{Clutton-Brock73, Hernquist92, Lilley18a, Lilley18b}. However, if the system is changing rapidly, the chosen BFE will not represent the system closely and hence the convergence of the expansion is not guaranteed. To alleviate this obstacle, \cite{Weinberg99} applied
    BFE to directly solve numerically the Sturm-Liouville equation in which a pair of bi-orthogonal functions are found at every 
    time-step. Hence the system can be represented accurately at every time-step. As such, BFEs provide a unique way to characterize 
    and simulate potentials of galaxies  accurately and efficiently, enabling broad applications.

\section{Conclusions}\label{sec:conclusions}

We have applied Basis Function Expansions (BFE) to N--body simulations of the MW--LMC system that were presented in \cite{garavito-camargo19a} (G19). These simulations follow the first infall of an LMC ($M_{\rm LMC}=$8,10, 18, 25 $\times 10^{10}$ M$_\odot$ at infall) towards the MW ($M_{\rm MW}=$1.2  $\times 10^{12}$ M$_\odot$), where the MW's velocity anisotropy profile can be isotropic (Model 1) or radially biased (Model 2). Our fiducial model is an LMC with an infall mass of 18 $\times 10^{10}$ M$_\odot$, about a MW Model 2, which generates the strongest wake \citep{garavito-camargo19a}. This fiducial LMC model has the mean expected infall mass for the LMC from abundance matching \citep[e.g][]{Moster13}.

In this study, we have quantified
the density, potential, and acceleration field of the combined MW--LMC system at the present time,
using a total of $\sim$ 236 coefficients in the BFE to describe the distortions to the MW, 20 coefficients for the bound LMC and a total of 1831 coefficients for the bound LMC with unbound DM debris. We find that the LMC does impact the density and potential of the MW in all simulations, resulting in an asymmetric DM distribution that deviates from common axisymmetric descriptions for DM halos, such as oblate, prolate, and triaxial.

We have also presented new methodology for reducing the noise in the BFE and for choosing the length of the expansion, which builds on work by \cite{Weinberg96}. This significantly reduces the number of coefficients in the expansion (by factors of 10), allowing us to identify coefficients that actually contain information. Furthermore, we have
illustrated how to use BFEs to model a host--satellite system, such as the MW--LMC. We used
multiple BFEs, one centered on the MW's cusp and the other on the LMC's cusp (see \S Appendix
\ref{sec:bfe_length} and \S \ref{sec:mwlmc_bfe}). 

This methodology is generic and can be applied to any 
simulated host galaxy and its satellites. For example, to model the M33-M31 system. Where M33 should induce a DM dynamical friction wake in M31's halo, which, if found, could also help constrain its orbit. 

Our main findings regarding the density and potential fields of the combined MW--LMC system at the present time as follows:

\begin{enumerate}

    \item \textbf{A dipole, $l=1$ mode, dominates the Collective Response, including the barycenter motion of the inner halo}.
    The gravitational energy of the coefficients in the BFE can be used to 
    identify the modes that dominate the structure of the perturbed MW halo.
    While the monopole term contributes the most, with 99\% of the total gravitational energy, dipole terms ($l=1$) are required to explain the North/South asymmetry induced by the Collective Response in agreement with recent studies \citep{Cunningham20, tamfal20}. These terms characterize the barycenter motion. 
    
    \item \textbf{The radial extent of the halo response is set by the $n$ modes.} After the monopole, the top ten most energetic terms have $l=m=0$. These $n$ modes
    build the radial extent of the halo response, starting at $\sim$30 kpc and extending to the virial radius. 
    
    \item \textbf{The asymmetric nature of the halo response (Collective Response and dynamical friction wake) is built by contributions from the angular terms.} 
    Asymmetric structures like the dynamical friction wake start to be reconstructed by $l,m >4$ modes. 

    Note that isolating the terms that contribute specifically to the dynamical friction wake vs. the Collective 
    Response will require a deeper analysis of the coefficients that is beyond the scope of this paper. However, this methodology does allow us to model the acceleration field from both these perturbations to model orbits (work in preparation).

    \item \textbf{The shape of the MW's DM halo response to the LMC is not well described as
    triaxial, prolate or oblate.}
    Cosmologically motivated triaxial, oblate and prolate halos are necessarily symmetric. 
    These halos do not have contributions from odd $l,m$ terms. In contrast, the MW halo response is dominated by odd terms, where
    the ratio of gravitational energy of the odd/even terms is 1-2.5 (excluding the monopole). 

    The inclusion of the bound LMC itself only exacerbates these deviations, as the bulk of its bound mass is in the South.
    As such, we find
    that none of the standard models accurately describe the present day shape of the MW--LMC DM halo.

    \item \textbf{Efforts to recover the original structure of the MW's halo should take into account the MW halo's response to the LMC.} The energy in high order, even, $l$ modes ($>l=2$) are of the same magnitude as those of the mean triaxial halo found in Illustris ($s_{\rho}=0.7,q_{\rho}=0.9$) \citep{Chua19}. As such, the halo response of the MW to the LMC must be accounted for in efforts to recover the initial shape of the MW halo and constrain the pre-LMC cosmological assembly history of the MW or the impact of different DM particle models on halo shape. 
    
    \item \textbf{Extreme values of triaxiality cannot be generated by perturbations to the halo from the LMC alone}. Even though the LMC can boost modes that mimic triaxiality, extreme halo shapes, such as those measured by \cite{Law10} or \citep{Erkal18b}, can not be induced in the halo by the LMC in our simulations. This motivates the study of the MW's DM halo response beyond spherical models prior the infall of the LMC, and also places importance on understanding the role of direct torques imparted by the LMC on halo tracers.

    \item \textbf{The boundary between the inner and outer MW halo occurs at 30 kpc and is set by the LMC's inner mass profile, which is observationally constrained and therefore the same in all LMC mass models.} Within 30 kpc, the DM halo moves coherently with the disk COM. As such, the orbital planes of objects within 30 kpc are expected to remain centered on the COM of the MW disk, consistent with results from previous studies \citep[e.g.,][]{petersen20}. 
    This result is not sensitive to the assumed LMC infall mass, because the inner mass profile is constrained by observations of the rotation curve, but may be impacted by different assumptions for the MW's initial mass profile.

    \item \textbf{The COM of the outer Halo ($>$30 kpc) can deviate from the COM of the MW's disk by as much as 15 kpc.} At radii larger than 30 kpc, the COM deviates from the COM of the disk, ranging from 2-15 kpc for the fiducial model. 
    The COM motion is the result of both the halo response and torques from
    the distorted LMC and cannot be properly accounted for in static models of the MW--LMC system.

     \item \textbf{The barycenter motion of the inner halo introduces a bias in observations of halo tracers at distances $>$ 30 kpc, where the velocity difference between the Northern and Southern Hemispheres can be as large as 100 km/s.}
     The shift in COM described above, is expected to manifest as radial velocity offsets between stars in the outer vs. the inner halo, which is referred to as the ``reflex motion" \citep{Gomez15, garavito-camargo19a, petersen20, Petersen21, Boubert20, Erkal20c}. 

    As a result, a dipolar pattern is expected in the radial velocities of stars beyond 30 kpc \citep{Cunningham20}. The velocity difference between stars in the South vs. stars in the North will vary as a function of Galactocentric distance, but can be as large as 50-120 km/s (see Figure~\ref{fig:vr_dipole}), depending on the mass of the LMC.

   \item \textbf{The magnitude of the barycenter motion scales with the LMC's infall mass. But, the location of the minimum of the combined MW+LMC acceleration field is constant with LMC's infall mass.} The amplitude of the displacement of the MW's COM, the radial velocity dipole, and the gravitational energy of the $m=1$, $l=1$ modes, all increase with the LMC's infall mass. However, the largest differences between LMC mass models appear at radii $>$ 70 kpc, and the increase in energy is not linear. The minimum of the combined acceleration field is roughly 30 kpc from the MW's center along the line of sight towards the LMC.

    \item \textbf{In our simulations, the bound mass of the LMC at present time is 30--50\% of its infall mass.}
  
    The bound LMC is tidally distorted, forming an S-shape represented by a BFE with 20 coefficients. 

    For all four LMC mass models ($M_{\rm LMC, infall}=8-25\times 10^{10}$ M$_{\odot}$), 
    the bound LMC DM mass is currently 
     $M_{\rm LMC, bound}=7-8\times 10^{10}$M$_{\odot}$. These bound particles are 30--50\% of the LMC's infall mass and can extend as far as 60 kpc 
    from the LMC COM, where the distribution is highly asymmetric.

    \item \textbf{In our simulations, the unbound DM debris from the LMC is expected to be the most massive inflow of DM particles experienced by the MW over the last 2 Gyrs}. 
    We have presented a methodology to accurately quantify the mass loss of the LMC (unbound DM particles) as it orbits the MW, using BFEs. 
    We found that over the last 2 Gyrs, in all mass models, the LMC has lost 50--70\% of its mass in the interaction 
    with the MW, which we call the LMC's DM debris. Some of this debris intersects with the MW's disk at high relative speeds, with interesting consequences for direct-detection experiments \citep{Besla19}. This debris has a leading and trailing component. The debris is very extended, spanning from $\hat{z}=-150$ to $150$ kpc.
    In addition to the dynamical friction wake, the LMC's DM debris is one of the causes for the simulated asymmetric shape of the MW's DM halo at present day.

    \item \textbf{The LMC has 1--2 bound satellites at the present-day (Phoenix 2 and the SMC).} We have found that at the present time, the only satellite that is bound to the LMC in all mass models is Phoenix 2. This stresses the importance of considering LMC mass loss when assessing the current dynamic state of the system. In our simulations, the SMC is bound today in only the fiducial LMC model. We note that while our simulations predict that only 1--2 satellites are likely currently bound to the LMC, this does not mean that the other satellites were not bound in the past \citep{Patel20,Sales11, Jethwa16, Sales17, Kallivayalil18, Erkal20b}, particularly if the gravitational influence of the SMC is also included \citep{Patel20}.

    \item \textbf{We define the sources of uncertainty in BFE --- \textit{bias} and \textit{noise} in the estimates of each coefficient, as well as \textit{bias} introduced by truncation of the series --- and present methods to reduce their influence, reducing the number of terms in the BFE by factors of 10}. In Appendix \ref{sec:bfe_length} we discuss how to reduce the bias and noise by smoothing the coefficients in a principal basis as described in \cite{Weinberg96}. 

    Finally, we illustrate that by sampling the halo randomly, we can reduce the variance and the number of coefficients needed to describe the system.

    \end{enumerate}

With 6D phase space information from upcoming surveys, the field will have an exciting opportunity to measure the impact of the LMC on the structure of the MW's DM halo. 
The first steps have already been taken through the identification of the reflex motion of the outer halo \citep{Erkal20c, Petersen21} and the tentative discovery of the stellar counterpart to the DM dynamical friction wake at distances of 60-100 kpc \citep{Conroy21}.
With the presented models, we now have a path forwards to not only identify the MW halo's response to the LMC, but also disentangle this effect from the initial structure of the MW halo created by the combination of its cosmological assembly history and the properties of the DM particle.

\section*{Acknowledgements}

This paper has benefited from conversations with 
Ekta Patel, Mike Petersen, Dennis Zaritsky, Jerry Sellwood, Ana Bonaca, and Peter Behroozi.
NG-C and GB are supported by the HST grant AR 15004, NASA ATP grant
17-ATP17-0006, NSF CAREER AST-1941096, and the Vatican Observatory Stoeger-McCarthy fellowship.
All the simulations where run on \textit{ElGato} HPC computing cluster at UArizona, which 
was supported by the National Science Foundation under Grant No. 1228509.

ECC is supported by a Flatiron Research Fellowship at the Flatiron Institute. The Flatiron Institute is supported by the Simons Foundation.
KVJ's contributions were supported by NSF grant Chaos c, GG013913, AST-1715582. This work was supported in part by World Premier International Research Center Initiative (WPI Initiative), MEXT, Japan
FAG acknowledges support from Fondecyt Regular 1181264, and funding from the Max
Planck Society through a Partner Group grant

\textit{Software:} Astropy \citep{astropy:2013, astropy:2018},  pygadgetreader
\citep{pygadgetreader}, matplotlib \citep{Hunter:2007}, numpy \citep{numpy}, scipy \citep{scipy},
ipython \citep{ipython}, scikit-learn \citep{scikit-learn, sklearn_api}, 
gala \citep{gala}, jupyter \citep{jupyter}, h5py \href{http://depsy.org/package/python/h5py}{http://depsy.org/package/python/h5py}. healpy \citep{zonca19}. This research has made use of NASA’s Astrophysics Data System (ADS), and the curated research-sharing platform arXiv.

\bibliography{references} 

\appendix

\section{BFE derivation of the Hernquist basis:}\label{sec:scf_derivation}
For completeness with here we briefly derived the main equations of the Hernquist Basis function expansion \citep{Hernquist92}. For a detailed and comprehensive derivation of this expansion we refer the reader to sections 2.2 and 3.1 in \cite{Hernquist92} or section 2.1 in \cite{Lowing11}.
Both the radial and angular contributions to the expansion are assumed to be separable. The radial part is expanded in Gegenbauer polynomials $C_n^{\alpha}(\xi)$, while the angular part is expanded in Spherical Harmonics. In this paper we follow the notation
of \cite{Lowing11}. The resulting expansion in density and potential is the product of 
both the radial and angular expansions as shown in Equations \ref{eq:rho} and \ref{eq:phi}.

\begin{equation}\label{eq:rho}
  \rho(r, \theta, \phi) = \sum_{n,l,m}^{\infty} A_{nlm} Y_{lm}(\theta, \phi) \rho_{nl}(r)
 \end{equation}

\begin{equation}\label{eq:phi}
  \Phi(r, \theta, \phi) = \sum_{n,l,m} A_{nlm} Y_{lm}(\theta, \phi) \Phi_{nl}(r)
\end{equation}

$A_{nlm}$ are the coefficients of the expansion.

The indices $n,l,m$ denote the order of the expansion in the radial, azimuthal and polar 
components respectively. $\Phi_{nl}(r)$ and $\rho_{nl}$ are 
expressed in terms of $C_n^{\alpha}(\xi)$ where $\xi = \dfrac{r-1}{r+1}$ as follows:

\begin{equation}\label{eq:rhonl}
\rho_{nl}(r) = \dfrac{K_{nl}}{2\pi}\dfrac{r^l}{(1+r)^{2l+3}}C_{n}^{2l+3/2}(\xi)\sqrt{4\pi}
\end{equation}

\begin{equation}\label{eq:phinl}
\Phi_{nl}(r) = - \dfrac{r^l}{r(1+r)^{2l+1}}C_{n}^{2l+3/2}(\xi)\sqrt{4\pi}
\end{equation}

Note that the lowest order term ($l=0$ and $n=0$) in equations \ref{eq:rhonl} and \ref{eq:phinl} is the Hernquist profile
where the gravitational constant and total mass of the halo are defined as $G$=1, $M$=1 respectively. $K_{nl}$ is defined as:

\begin{equation}
K_{nl}=\dfrac{1}{2}n(n+4l+3) +(l+1)(2l+1)
\end{equation}

The coefficients of the expansion can be found using the bi-orthogonal properties of $\rho_{nlm}$ and $\phi_{nlm}$.
The basis is build by design bi-orthogonal, that is: 

\begin{equation}
  I^{n'l'm'}_{nlm} = \int \rho_{nlm}({\bf{r}})[\Phi_{n'l'm'}({\bf{r}})]^*
  d{\bf{r}} =
  I_{nl} \delta_{ll'}\delta_{mm'}\delta_{nn'}
\end{equation}

\noindent Where the orthonormal properties of the spherical harmonics and
the ultraspherical harmonics where used to find that (see \cite{Hernquist92}
for a detailed derivation): 

\begin{equation}\label{eq:Inl}
  I_{nl} = -K_{nl}
  \dfrac{4\pi}{2^{8l+6}}\dfrac{\Gamma(n+4l+3)}{n!(n+2l+3/2)[\Gamma(2l+3/2)]^2}
\end{equation}

With this bi-orthonormal property the coefficients can be computed as
follows:

\begin{equation}\label{eq:Anlm}
  A_{nlm} = \dfrac{1}{I_{nl}} \int \rho({\bf{r}})[\Phi_{nl}(r)Y_{lm}(\theta, \phi)]^*
  d{\bf{r}}
\end{equation}

The basis can be re-written in only real quantities by replacing Equation
\ref{eq:Anlm} in Equation \ref{eq:rho}. The imaginary quantities will
cancel out when performing the sum over $m$ as a result the coefficients
$A_{nlm}$ are splitted into cosine and sine contributions defined as:

\begin{equation}\label{eq:coeff}
\begin{aligned}
  S_{nlm} = \dfrac{(2-\delta_{m0})}{I_{nl}} \sum_k^{N} m_k
         \Phi_{nl}(r_k)Y_{lm}(\theta_k) \cos{m\phi_k} \\
         T_{nlm} = \dfrac{(2-\delta_{m0})}{I_{nl}} \sum_k^N m_k 
         \Phi_{nl}(r_k)Y_{lm}(\theta_k) \sin{m\phi_k} 
\end{aligned}
\end{equation}

The sum in equations \ref{eq:coeff} is computed over all the particles, $N$, in the halo. Hence the basis contains information for all halo particles
in the halo. In practice, the number of particles is finite, and hence the limits of the sums in Equations \ref{eq:rho}
\ref{eq:phi} cannot be infinite. Instead, one has to truncate the expansion at a given value of $n_{max}$ and $l_{max}$.
As such the final expression for the Hernquist BFE is:

\begin{equation}\label{eq:rho_bfe}
 \begin{split}
  \rho(r, \theta, \phi) =  \sum_{n}^{n_{max}} \sum_l^{l_{max}} \sum_m^l  Y_{lm}(\theta) \rho_{nl}(r)
  (S_{nlm} \cos{m\phi}+ T_{nlm} \sin{m\phi})
\end{split}
\end{equation}

\begin{equation}\label{eq:phi_bfe}
\begin{split}
 \Phi(r, \theta, \phi) = \sum_{n}^{n_{max}} \sum_l^{l_{max}} \sum_m^l Y_{lm}(\theta) \Phi_{nl}(r)
 (S_{nlm} \cos{m\phi}+ T_{nlm} \sin{m\phi})
\end{split}
\end{equation}

\section{Choosing the coefficients of a BFE to minimize noise and truncate the expansion}\label{sec:bfe_length}

An accurate estimate of the force field of the halo using BFE relies on finding 
the least uncertain estimator (analytic or numerical form of the BFE) and therefore it 
relies on the information content of the coefficients.

The length of the expansion is one the free parameters in BFEs methods, along with the scale length of the halo. Choosing too few terms
in the expansion will cause a poor representation of the system, while too many terms
will increase the noise caused by the discrete nature of the simulation. High order terms in the expansion
describe small-scale fluctuations of the system. However, some of these small-scale features might be
artificial due to sources of noise. One can interpret the number of terms in the expansion 
as degrees of freedom, and hence it motivates the need 
to identify, discard and correct the coefficients in the expansion that are biased or noisy and maximizing the signal \citep[e.g][]{Weinberg96} by looking at correlations in the coefficients.
Here we start by describing the sources and types of noise uncertainty presented in BFEs
as follows:

\textit{BFE truncation bias:}
BFE uses a set of predetermined bi-orthogonal functions that approximates the density and potential of the halo.
If the series were infinitely long, the underlying physical forms could be represented exactly.
The representation embodied in a finite number of coefficients will be \textit{biased} towards the lower order terms function and away from the `truth'.
The amount of bias in each coefficient in the expansion depends on the 
functional form of the BFE.
The expansion will get closer to the `true' halo phase-space as more coefficients are included, even if they are biased. However, truncating the expansion too early causes
a poor representation of the force field that cannot be improved by the traditional means of increasing particle number of the N--body simulation.

\begin{figure}
    \centering
    \includegraphics[scale=0.3]{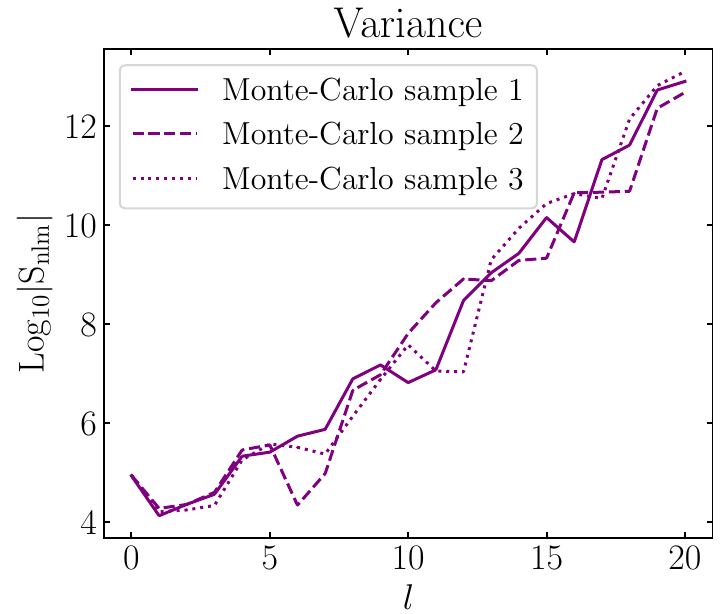}
    \caption{Effect of the variance noise in the coefficients of a BFE expansion of a DM halo. We show the coefficients corresponding to $n=m=0$ as a function of $l$.  
    The variance noise in the amplitude of the coefficients is caused by the random Monte Carlo sampling of a halo.
    The number of particles is the same for the three halos shown. Both the number of particles in the halo and its random Monte Carlo sampling produce noise that is captured in the 
    amplitude of the coefficients. High order coefficients ($l>5$) are more sensitive to noise than low order coefficients.}
    \label{fig:noise_bfe}
\end{figure}

\textit{Variance noise:}
The second source of uncertainty comes from the discrete representation of the phase-space. The particles represent a sampling of the phase space, 
as such they are a random, imperfect representation of the density distribution. If the phase-space is 
sampled with different random realizations, the amplitude of the coefficients will be slightly different. 
This manifests as \textit{variance} when computing the coefficients, this is illustrated in Figure \ref{fig:noise_bfe}.

\textit{Bias in individual coefficient evaluation:} The third source of uncertainty is that any estimate of the coefficient
comes from taking the mean over the estimate from each individual particle.
This assumes that the \textit{mean} is a good representation of the coefficients. However, the underlying distribution of these estimates is unknown and thus the mean may bias the evaluation of each coefficient.

To identify the coefficients that are sensitive to noise, and hence the ones to exclude from the expansion, we follow the method explained in \cite{Weinberg96}. This method characterizes the bias noise in each coefficient and reduces the degrees of freedom by decreasing the number of coefficients needed in the expansion. We review the main aspects of the method here and refer the reader to \cite{Weinberg96} for a detailed conceptual and mathematical explanation.
 
\textbf{Addressing Bias:} To quantify the noise in each expansion term, we compute the `signal-to-noise' of a given coefficient $A_{nlm}$ as: 

\begin{equation}
\Gamma=\dfrac{\rm{A}_{nlm}}{\sqrt{\rm{var}({\rm{A}_{nlm}})}}    
\end{equation}

\cite{Weinberg96} showed that the variance of the coefficients $\rm{var}(\rm{A}_{nlm})$ is: 

\begin{equation}\label{eq:var}
 \rm{var}(\rm{A}_{nlm}) = \sum_{i}^{N} \Psi_{nlm}^2(x_i)
   - \dfrac{1}{N}\left(\sum_{i}^{N} \Psi_{nlm}(x_i)\right)^2,
\end{equation}

where $N$ is the number of particles and $\Psi_{nlm}$ for the BFE expansion is defined as:

\begin{gather}\label{eq:covar}
  \Psi_{nlm}(x_i) = (2-\delta_{m,0}) m_{i}
  \tilde{\rm{A}}_{nl}\Phi_{nl}(r_i)Y_{l,m}(\theta_i) 
  \begin{pmatrix}
  \cos{m \phi_i} \\
  \sin{m \phi_i}
  \end{pmatrix}
\end{gather}

Where the sine and cosine contribution comes from the definitions in Equation \ref{eq:coeff}. Once the coefficients $A_{nlm}$ and its variances are computed, the bias in each coefficient (in the principal basis, see below) can be corrected by multiplying it with the `smoothing' factor:

\begin{equation}
    b_{nlm} = \left[1 + \dfrac{1}{\Gamma^2}\right]^{-1}
\end{equation}

In order to choose the coefficients that are less affected by the bias one has to find the optimal
value of $\Gamma_{opt}$ that reduces the bias. Coefficients with a value of $\Gamma > \Gamma_{opt}$
are included in the expansion, while those that don't satisfy this condition are the terms that represent `noise' and thus discarded, this is discussed in detail Section \ref{sec:bfe_length}.

\textbf{Addressing truncation noise using principal basis of the coefficients:} 
For the BFE the coefficients $A_{nlm}$ are split into 
two correlated coefficients $S_{nlm}$ and $T_{nlm}$  (following \citet{Lowing11} notation) that account for the real and imaginary part of the $A_{nlm}$ (see Equation \ref{eq:coeff}). 
Therefore, one needs 
to find first the \textit{principal basis} to un-correlate these coefficients. This is done following 
section 2.2 in \cite{Weinberg96}. In practice, for the BFE expansion one needs to diagonalize the
2$\times$2 covariance matrix $S_{S,T}$:
\begin{gather}\label{eq:covar}
  S_{S,T} = 
  \begin{bmatrix}
  \sum_{i}^{N} \Psi_{c, nlm}^2 & \sum_{i}^{N} \Psi_{c, nlm} \Psi_{s, nlm} \\
  \sum_{i}^{N} \Psi_{s, nlm}\Psi_{c, nlm} & \sum_{i}^{N} \Psi_{s, nlm}^2 \\
  \end{bmatrix}
\end{gather}
where $\Psi_{s}$ and $\Psi_{c}$ refer to each component in equation \ref{eq:covar}. 
The diagonalization of the covariance matrix to $S_{S,T}$ results in the rotation matrix
used to rotate the coefficients $S_{nlm}$ and $T_{nlm}$ into the principal basis.
Both the principal basis transformation and the smoothing allows to reduce the number 
of coefficients since the `signal' of each coefficient is identified and corrected by bias.

\textbf{Addressing Variance:} The above method however, does not address the variance caused by the random sampling.
In order to compute the mean value of the coefficients, we do $M$ random samples of the halo particles and compute the BFE expansion 
for all of the samples. The coefficients $\rm{A}^{opt}_{nlm}$ corrected by the variance are computed using the mean of the
$M$ coefficients $\rm{A}_{nlm, i}$ in each expansion:
\begin{equation}
    \rm{A}_{opt} = \dfrac{1}{M} \sum_{i}^{M=\sqrt{N}} \rm{A}_i
\end{equation}
where $N$ is the total number of particles in the simulation.
Similarly, the optimal variance of each coefficients is the mean of all the variances:
\begin{equation}    
    \rm{var}({A}_{opt}) = \dfrac{1}{\sqrt{N}}\sum_{i}^{\sqrt{N}} var(A)_i
\end{equation}

Where we have assumed that the variance noise behaves as shot noise whose standard deviation is $\sqrt{N}$. 

In Section~\ref{sec:gof_example} we illustrate how by truncating the expansion using allows to 
reduce the noise in the BFE density representation of a low-resolution halo. It is however, 
not obvious how to chose the value of $\Gamma_{opt}$ needed to truncate the expansion. In this
work, we have chosen a value of $\Gamma_{opt}=5$ by visual inspection of noise signatures in the density field. Methods to find $\Gamma_{opt}$ analytically would be presented in future work.

\subsection{Example of the resulting BFE after selecting the `non-noisy' coefficients}\label{sec:gof_example}

\begin{figure}
    \centering
    \includegraphics[scale=0.6]{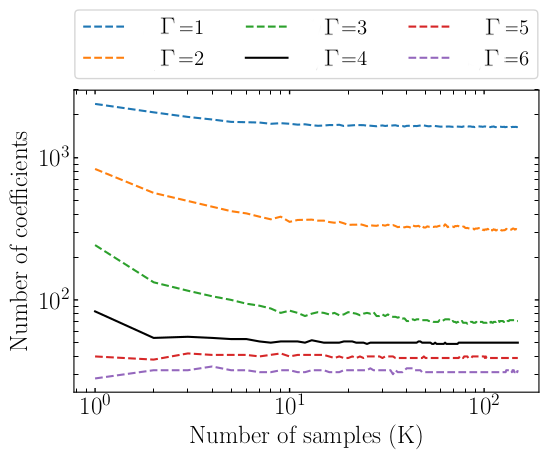}
    \caption{Number of coefficients in the BFE as a function of the number of random realizations (samples) used to compute the amplitude of the coefficients. Different lines represent the values of signal-to-noise threshold ($\Gamma$) used to select the coefficients. Sampling the halo and choosing an appropriate signal to noise are critical in choosing the length of the BFE. Sampling the halo reduces the noise due to the random distribution of the particles. Smoothing the halo reduces the noise due to the discrete nature of the system.}
    \label{fig:N_coeff_sn}
\end{figure}

\begin{figure*}
    \centering
    \includegraphics[scale=0.53]{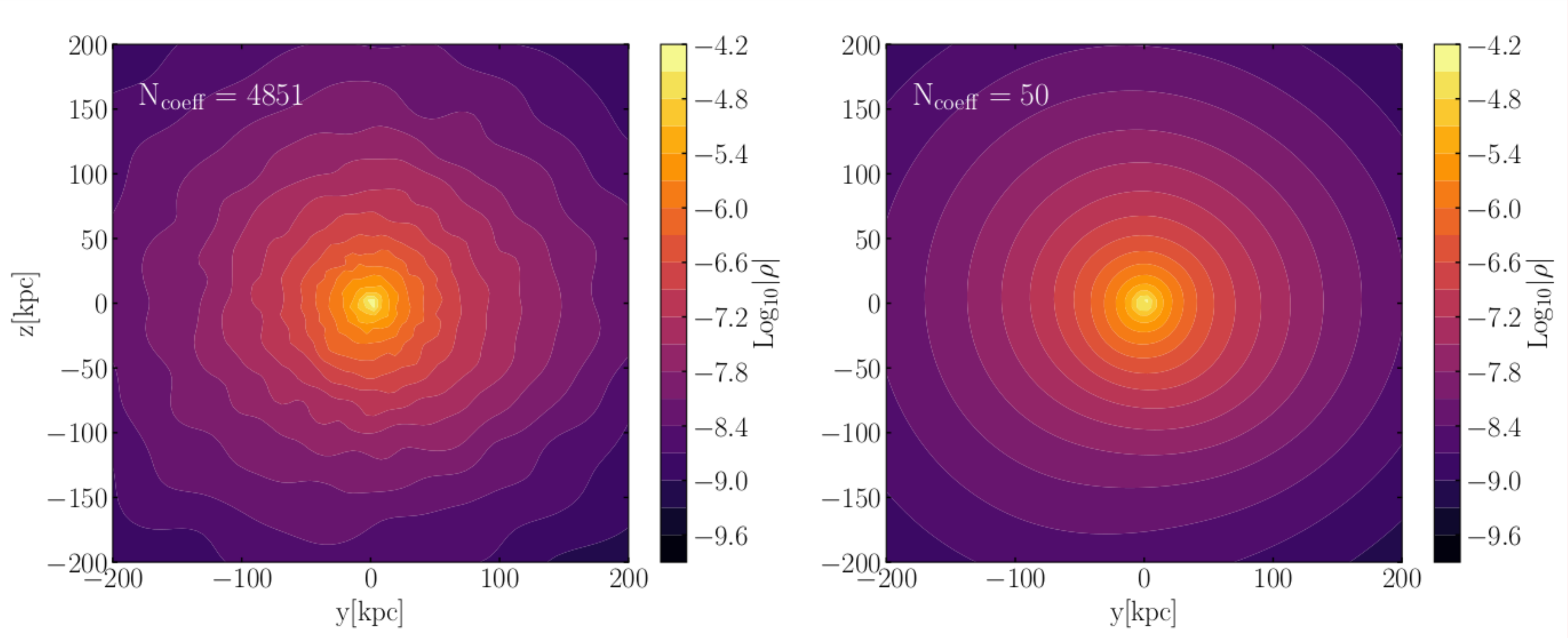}
    \caption{Contour maps of the reconstructed density of a DM halo sampled randomly with $10^6$ particles. \textit{Left panel:} Reconstructed density using the BFE expansion using all of the 4851 coefficients computed using $n_{max}=20$ and $l_{max}=20$. \textit{Right panel:} Reconstructed density field using the 50 smoothed coefficients with $\Gamma>5$, which contain most of the `information.' Smoothing the coefficients not only decreases the number of coefficients needed in the BFE, but also decreases the variance and the truncation noise.}
    \label{fig:N_coeff_sn_rho}
\end{figure*}

We apply the methods described in section \ref{sec:bfe_length} 
a low resolution MW-like halo of $10^5$ particles. We compute the coefficients using Equation~\ref{eq:coeff} and then the density field using Equation~\ref{eq:rho_bfe} and shown in 
the left-hand side panel of Figure~\ref{fig:N_coeff_sn_rho}. After correcting for bias, truncation error
and variance we select 50 coefficients that contain all of the information of the phase-space. The
density field is show in the right-hand side panel of Figure~\ref{fig:N_coeff_sn_rho}. 
We illustrate the effect of the variance in Figure~\ref{fig:N_coeff_sn}, where we select the coefficients 
for six $\Gamma$ values as a function $K$ random samples. We find convergence in the number of coefficients
for $K>10$. For $\Gamma$=4 we found the 50 coefficients used to compute the density field in Figure \ref{fig:N_coeff_sn_rho}.  

\section{Connecting the terms in the BFE to the nature of the system}\label{ref:app_intuition}

In this section, we discuss the contribution of the radial ($n$) and  angular $(l,m)$ 
terms of the BFE to the density, potential or acceleration fields of DM halos. 

We start by discussing the radial modes. The halo density profiles are defined in terms of the $n$ and $l$, as follows:

\begin{equation}\label{eq:bfe_r}
    \rho_{nl}=\dfrac{K_{nl}}{2\pi}\dfrac{r^l}{r(1+r)^{2l+3}} C_n^{(2l+3/2)}(\xi) \sqrt{4\pi},
\end{equation}

\noindent where $\xi=(r-1)/(r+1)$ and $K_{nl}=\dfrac{1}{2}n(n+4l+3) + (l-1)(2l+1)$. Hence

\begin{equation} 
\rho \sim n l r^{-(l+4)} C_{n}^{(2l+3/2)}(\xi)
\end{equation}

\noindent as such, the amplitude of $\rho$ increases proportionally to $n$ and $l$. However, note that as $l$ increases the 
density profile also decreases more sharply as a function of $r$.

\begin{figure}[h]
    \centering
    \includegraphics[scale=0.5]{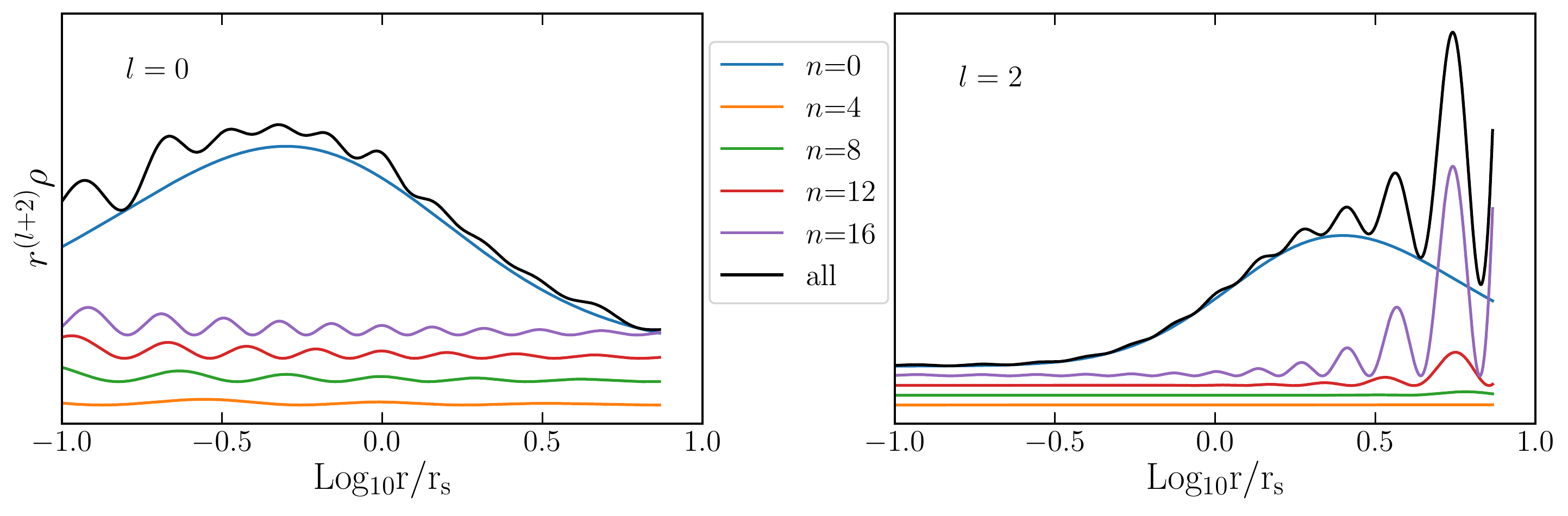}
    \caption{Density profiles of the radial components of the BFE defined in equation \ref{eq:bfe_r} as a function of $n$ and $l$. Colored lines indicate the contribution of each mode and the black line indicates the total contribution from all the modes. Note, that the y-axis has a different scale in all the panels. As $n$ increases, the number of local minima in the density distribution increases, since this is modulated by the ultraspherical polynomials $C_{n}^{\alpha}$. On the other hand, as we increase both $l$ (right panel) and $n$ (different lines), the amplitude of each mode increases.
    At higher values of $l$ and $n$ the density profiles do not decrease since the contributions from both $l$ and $n$ from the ultraspherical polynomials are equivalent to the $-(l+4)$ exponent in equation \ref{eq:bfe_r}.}
    \label{fig:bfe_r2}
\end{figure}

In Figure \ref{fig:bfe_r2}, we plot the density profiles of each of the 
$n$ terms (solid color lines), with  $l=0$ and $l=2$, left and right panel
respectively. We have scaled the amplitude of each term differently for visualization purposes. In this example, the $n=0$ term dominates the density
profile and the high order radial terms correspond to perturbations. The order of $n$ is proportional to the local maxima in the density, these will
correspond to thin spherical shells. The locations
of the maxima are different for each term and hence multiple terms can amplify
or decrease a particular perturbation at a given radius. 

For the $l=2$ modes,
the radial terms have the same behaviour with $n$, however the radial profile decreases faster than in the $l=0$ case. Overall, this radial behaviour is expected from the solution to Liouville's equation, to which the Poisson equation is a particular case. 

Given the above, we expect contributions from higher order $n$ terms in the BFE if:
\begin{enumerate}
    \item There are perturbations at different radii, for example the DM wake induced by the LMC;
    \item  The halo is not spherical. As shown in the next section \ref{sec:ideal_halos}, the BFE for Triaxial, Oblate, and Prolate halos have important
contributions from radial terms. Since these halos are elongated in specific
directions, the superposition of radial terms will elongate the halos
in those direction;
\item The DM halo's radial dependency is different from the zeroth order BFE term. In this case  
we expect contributions from radial terms with $l>0$, e.g. if using the \cite{Hernquist92} BFE to describe a NFW halo.
\item  If the scale length of the halo is not chosen accurately, radial terms will also appear.
\end{enumerate}

Overall, we can think of the Radial terms ($n$) as the ones that set the radial extent of the perturbations, where larger values of $n$ represent perturbations at larger radii. 

\begin{figure*}[h]
    \centering
    \includegraphics[scale=0.5]{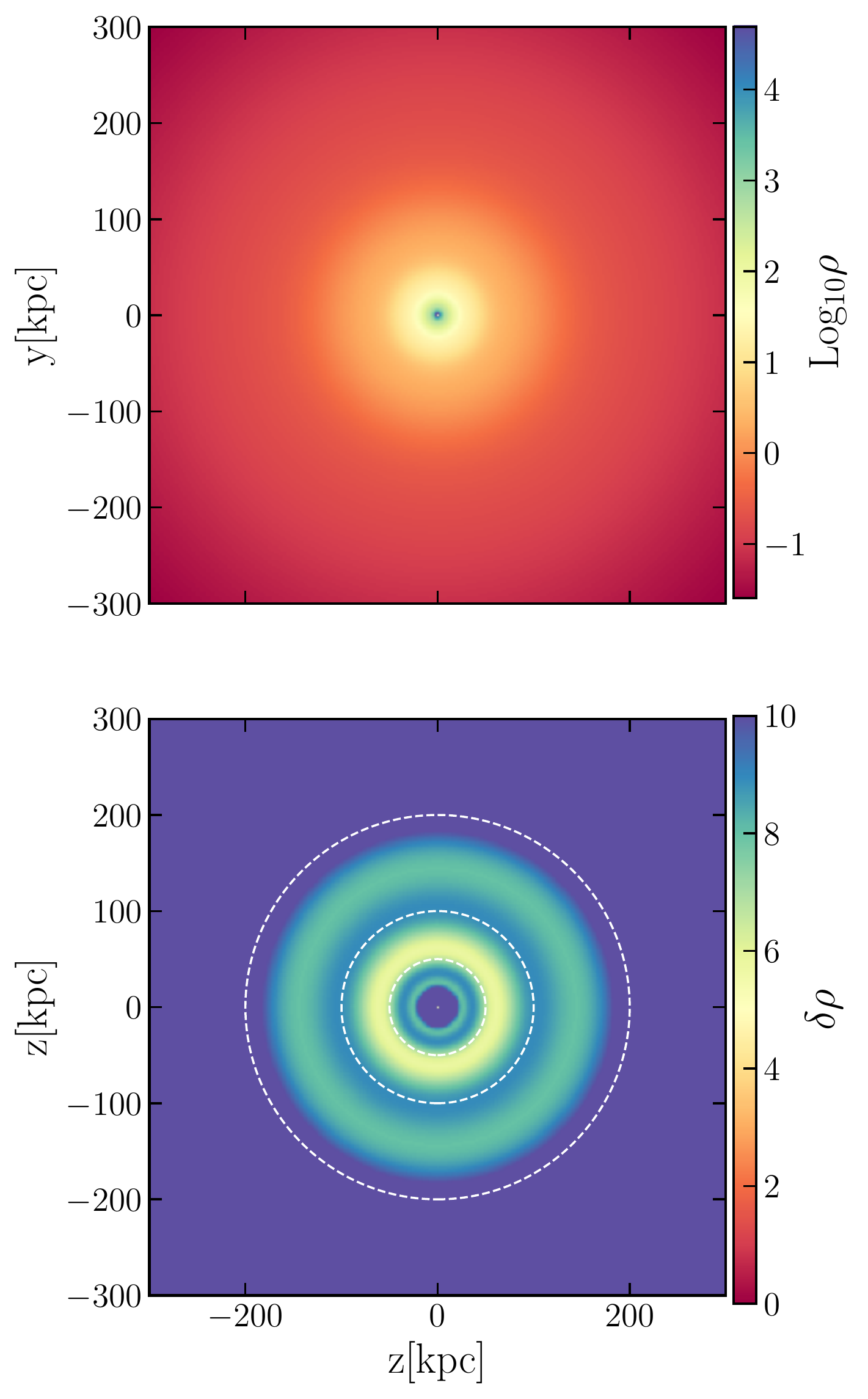}
    \includegraphics[scale=0.5]{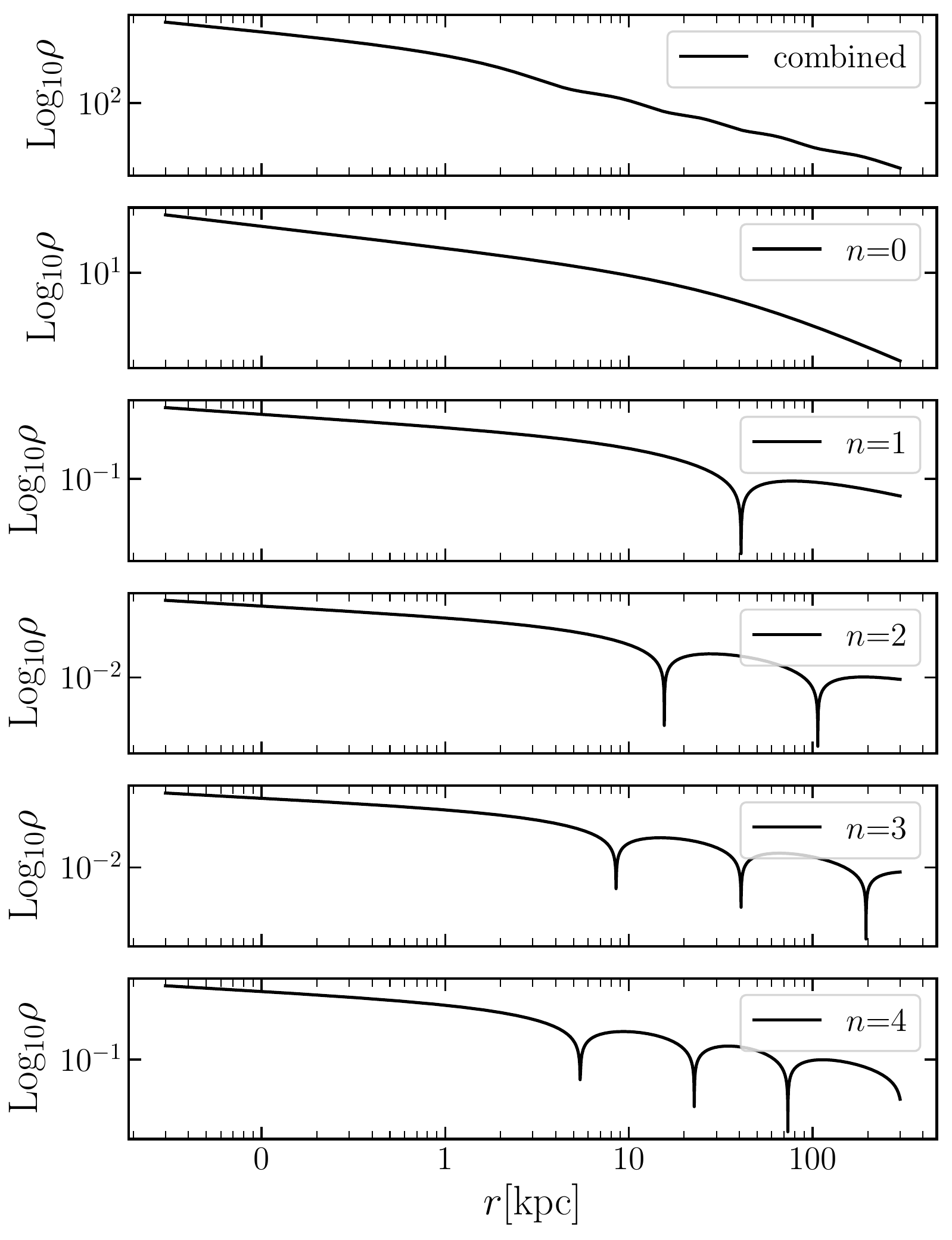}

    \caption{The most energetic terms in this system are the ones with $l=m=0$. These radial terms set the scale of where the perturbations in the halo are the strongest. These five terms set perturbations at several radii as shwon in the the right panel.}
    \label{fig:bfe_r}
\end{figure*}

For example, in the case of
$l=m=0$, the radial $n$ terms describe very thin spherical shells located at
different radii. Thicker shells can be created by combining multiple $n$-modes as shown in Figure \ref{fig:bfe_r}.

The angular contribution in the BFE is contained in the spherical harmonics. For example, a 
dipole in the density field, such as that generated by the reflex motion of the MW to the LMC \citep{Gomez15, garavito-camargo19a, petersen20, tamfal20}, is captured in the $l=1$ coefficients. High order coefficients appear if small-scale perturbations are
present.For example, if one computes an expansion 
in a DM halo including its subhalos one should fine that high order $l, m$
coefficients appear. 

In the case of the DM wake produced by the LMC,
Figure \ref{fig:wake_l_terms} shows how the wake is reconstructed as
we increase the order of the $l$ terms. Since the wake is a large scale perturbation 
in the halo, terms up to $l=3$ are sufficient to describe the wake. 
For further applications we refer the reader to
\cite{Cunningham20}where spherical harmonic are used to described the perturbations
induced by the wake in the velocity field of the MW's DM halo.

BFEs are a powerful tool to characterize asymmetric and radially varying DM halo shapes, owing to the ability of the method
to characterize perturbations in halos. As discussed in this section, deviations in the density profile from spherical halos can be captured by adding terms to the
expansion. Low order terms describe large scale perturbations, while 
high order terms describe small scale perturbations.

\section{Idealized halos:}\label{sec:ideal_halos}

Here we compute the BFE of idealized halos, whose properties are summarized in Table~\ref{tab:ideal_halos}, in order to gain intuition for which coefficients will contribute
the most in each of these halos. We build three idealized
halos whose main properties are summarized in table 
\ref{tab:ideal_halos}. We choose some extreme values 
of $s_{\rho}$ and $q_{\rho}$ for the oblate and prolate in order 
to maximize the signal in the BFE coefficients. The triaxial halo, on the other hand is consistent with mean values from
the Illustris
cosmological simulation \citep{Chua19}.

\begin{figure}
    \centering
    \includegraphics[scale=0.4]{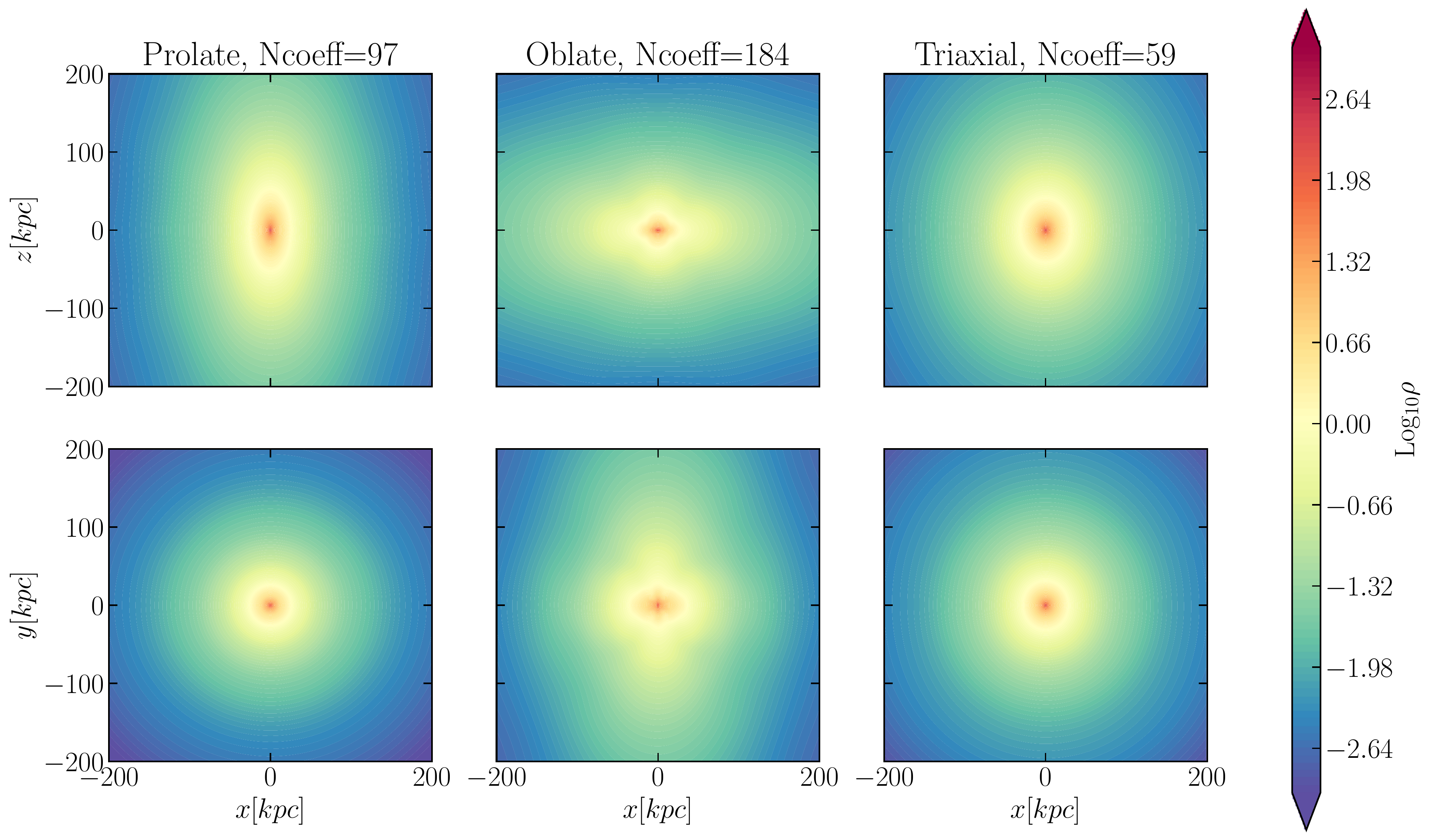}
    \caption{Density contours in the $\hat{y}-\hat{z}$ and $\hat{x}-\hat{y}$ planes of the prolate, oblate and triaxial halos are shown. 
    The coefficients were computed up to $n_{max}=l_{max}=20$ and the $\Gamma$=5 was used to truncate the expansion. For the oblate halo the expansion needs high order terms to reproduce the density field.}
    \label{fig:density_contours}
\end{figure}

We compute the length of the BFE expansion following the method described in section \ref{sec:bfe_length}. 
The halo density field computed with the BFE is shown in Figure \ref{fig:density_contours}. For the oblate halo, we find that the expansion requires a larger number of terms than for the prolate halo. This is due
to the functional form of the spherical harmonics. 

In order to study the contribution of each coefficient, we compute 
the energy in the coefficients as described in section \ref{sec:power_shape}.
We show the energy $U_{lm}$ in the $l-m$ space in Figure 
\ref{fig:halos_m_l} where we have summed the energies from all the 
$n$ coefficients. That is, $U_{lm}=\sum_{n} U_{nlm}$. Our main findings are: 1) None of the halos have contribution from odd $m$ terms, since the corresponding spherical harmonics are not
radially symmetric. 2) Prolate halos have no contribution from the $m>0$ terms. The spherical harmonic lobes corresponding to the $m=0$ terms are perpendicular to the $x-y$ plane, which coincides with the major axis of the prolate halo.
3) Oblate halos, on the other hand, do have a significant contribution from $m>0$ terms. In fact, the most energetic coefficients are in the $m=2$ and $m=4$ term. 4) Triaxial halos lie in between prolate and oblate halos. There is contribution from $m>0$ terms, but the main contribution to the energy comes from the $m=0$ modes. However, in more extreme triaxial halos than considered here,
the energy of the $m>0$ terms can be much larger.

In addition to the energy in the coefficients, their sign 
also contains information of the halo. We study the sign of 
the $S_{nlm}$ coefficients (see equation \ref{eq:coeff}), since 
all the $T_{nl0}$ coefficients are zero. The main difference among 
the halos is that the $l=2$ and $l=6$ 
in the Oblate halos have negative signs since those spherical harmonics
are negative in the mid-plane and positive in the poles. For oblate halos,
the main contribution is in the mid-plane and hence the need to add energy
in the mid-plane and subtract towards the poles. For the triaxial halo, 
the $m>0$ terms are negative, note that the corresponding spherical harmonics are not axisymmetric and hence the subtraction of these terms are needed in order to build the asymmetries of triaxial halos.

\begin{figure}[h]
    \centering
    \includegraphics[scale=0.55]{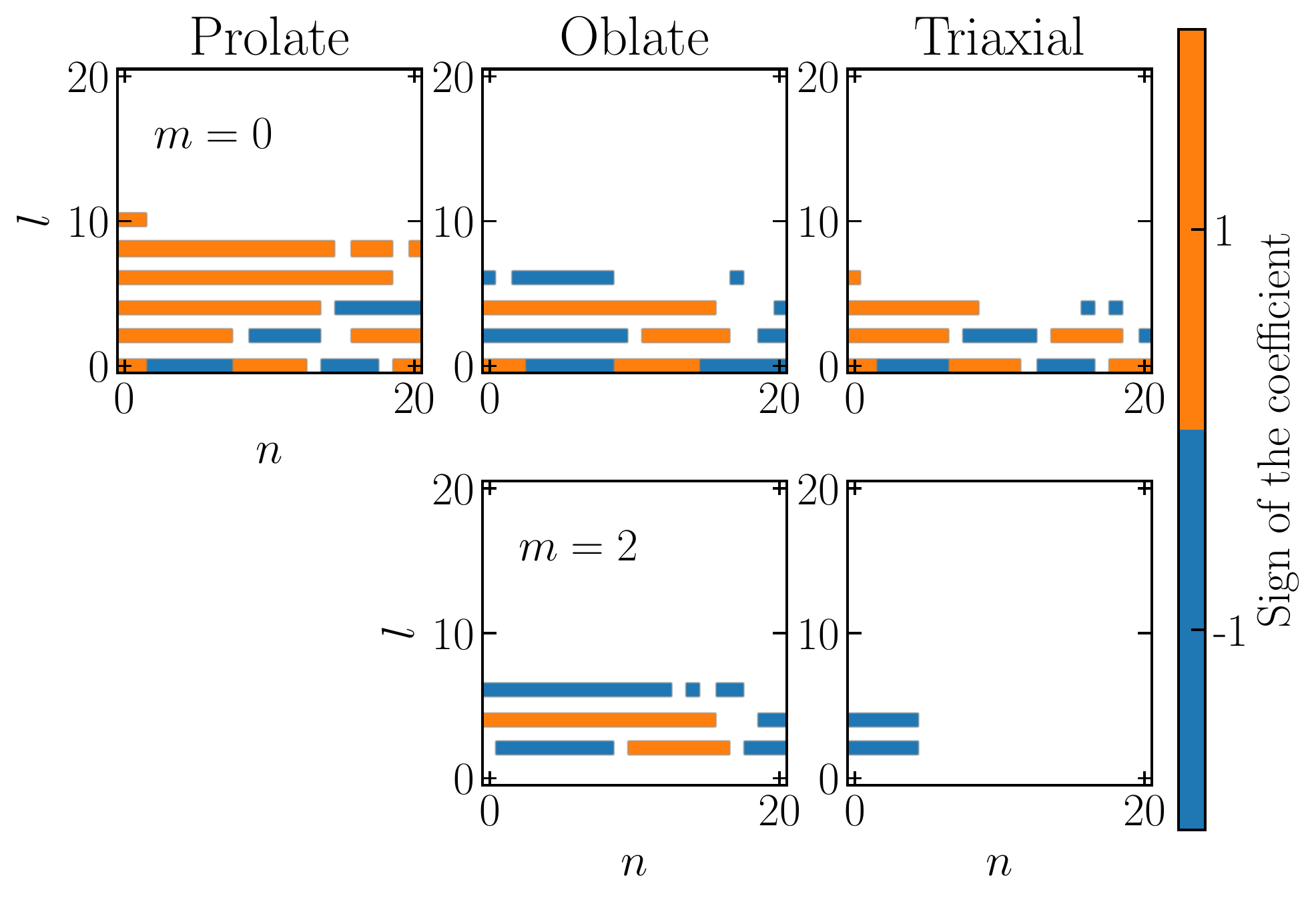}
    \caption{The sign of the coefficients also tells us about the nature of the system. Here we show the sign of the $S_{nlm}$ coefficients: orange represents positive and blue negative. We only plot results up to $m=2$ since this includes the majority of the coefficients. For the Oblate halos the $l=2$ and $l=6$ coefficients have negative signs, these terms are subtracted in order to create the oblate shape in the $x-y$ plane.  In the triaxial halo the m=2 modes are negative, these terms are non axisymetric whose help to build the axial asymmetries of triaxial halos. } 
    \label{fig:my_label}
\end{figure}

\end{document}